\documentclass[useAMS,usenatbib]{mn2e}

\usepackage{graphicx} \usepackage{aas_macros} \usepackage{lscape}
\usepackage{amsmath}

\newcommand{\gtsim}{\mbox{{\raisebox{-0.4ex}{$\stackrel{>}{{\scriptstyle\sim}}$}}}}
\newcommand{\ltsim}{\mbox{{\raisebox{-0.4ex}{$\stackrel{<}{{\scriptstyle\sim}}$}}}}

\title[Winds and streams in protogalaxies]{The impact of supernovae driven winds on
  stream-fed protogalaxies} 
\author[L. C. Powell, A. Slyz and J. Devriendt]{Leila
C. Powell$^{1,2}$\thanks{E-mail: leila.powell@cea.fr}, Adrianne
Slyz$^{2}$ and Julien Devriendt$^{2,3}$\\ 
$^{1}$Service d'Astrophysique, CEA-Saclay, Orme des Merisiers, 91191 Gif-sur-Yvette Cedex, France \\ 
$^{2}$Oxford Astrophysics, Denys Wilkinson Building, Keble Road, OX1 3RH, Oxford, UK \\ 
$^{3}$Universit\'e Claude Bernard Lyon I, CNRS UMR 5574, ENS-L, Observatoire de Lyon, \\
9 Avenue Charles Andr\'e, 69561 St-Genis-Laval Cedex,France}

\begin{document}

\date{Accepted . Received ; in original form }

\pagerange{\pageref{firstpage}--\pageref{lastpage}} \pubyear{2010}

\maketitle

\label{firstpage}

\begin{abstract}

Supernovae (SNe) driven winds are widely thought to be very
influential in the high-redshift Universe, shaping the
properties of the circum-galactic medium, enriching the intergalactic medium (IGM)
with metals and driving the evolution of low-mass galaxies. However, it is not yet fully understood how SNe driven
winds interact with their surroundings in a cosmological context,
nor is it clear whether they are able to significantly impact the evolution of low-mass galaxies from which they originate by altering the amount of cold
material these accrete from the cosmic web.
Indeed, due to the strong constraints on resolution imposed by limited
  computational power, all cosmological hydrodynamics simulations to
date resort to implementing more or less physically well motivated and complex subgrid models
to trigger galactic winds. To explore this issue, we implement a
standard Taylor-Sedov type solution, widely used in the community to depict the combined
action of many SN explosions, in a cosmological resimulation of a low
mass galaxy at $z \ge 9$ from the `{\sc nut}' suite. However, in contrast
with previous work, we achieve a resolution high enough  
to capture {\em individual} SN remnants in the Taylor-Sedov phase, for which the
Taylor-Sedov solution actually provides an accurate description of the
expansion. We report the
development of a high-velocity, far-reaching galactic wind
produced by the combined action of SNe in the main galaxy and its
satellites, which are located in the same or a neighbouring dark matter 
halo. Despite this, we find that (i) this wind carries out very little
mass (the measured outflow is of the order of a tenth of the
inflow/star formation rate) and (ii) the cold gas inflow rate remains
essentially unchanged from the run without SNe feedback. Moreover, 
there are epochs during which star formation is enhanced in the feedback run relative to its radiative
cooling only counterpart. We attribute this `positive' feedback to the metal enrichment
that is present only in the former. We conclude that at very high
redshift, efficient SNe feedback can drive large-scale galactic winds
but does not prevent massive cold gas inflow from fuelling
galaxies, resulting in long-lived episodes of intense star formation.

\end{abstract}

\begin{keywords}
methods:numerical--galaxies: high-redshift--galaxies: formation--galaxies: evolution--intergalactic medium--supernovae: general
\end{keywords}

\section{Introduction}

Significant outflows of gas from star-forming galaxies, known as
galactic winds, have been observed both in local galaxies
\citep[e.g.][]{lehnert_heckman_1996, cmartin99_outflows} and high
redshift ($z \sim3$) Lyman break galaxies (LBGs)
\citep[e.g.][]{pettini_etal_2001, shapley_etal_2003}. These winds are
usually associated with starbursting galaxies (for example LBGs have
star formation rates (SFRs) of $\sim 100 {\rm M}_{\odot}/{\rm yr}$)
and in this case are often referred to as `superwinds'.  These winds
can reach large distances from the source galaxy and are typically
observed to have temperatures greater than the `escape temperature' of
the potential well of their host halo e.g. M82, a classic example of a
superwind \citep[][]{lehnert_heckman_weaver_1999}. This suggests that
at least some of their metal enriched gas may be able to escape into the IGM.

The production of these superwinds is attributed to simultaneous
SNe explosions whose remnants can overlap producing a bubble of
hot, low density gas that can `blowout' of the galaxy and become a
wind \citep{mckee_ostriker_1977}. The main requirement for `blowout'
is that the rate of SNe is high enough, such that the remnants overlap
before they can cool radiatively \citep{heckman_armus_miley_1990,
david_forman_jones_1990}. Observations of outflows from M82 are
consistent with multiple SNe supplying energy to drive the wind
\citep{heckman_armus_miley_1990}. Alternatively, it has been
advocated that multiple coherent SNe, as seen in the superwind
scenario, are not the only mechanism able to produce outflows from
dwarf galaxies. Dwarf galaxies can lose a significant fraction of
their mass on timescales of a Gyr, in a more quiescent manner. Fuelled
only by an average SFR, cold gas clouds in the interstellar medium (ISM) are evaporated by
SNe remnants and form a wind if the evaporated gas reaches high enough
temperatures \citep{efstathiou_2000}.

Due to their ability to eject material from galaxies, galactic winds
are considered a potential solution to several outstanding problems in
the field of structure formation. They are proposed to play a role in
the pollution of the IGM and intracluster medium with metals, the
occurrence of metallicity gradients within individual galaxies
\citep[e.g.][]{heckman_etal_2000}, the properties of dwarf galaxies
\citep[][]{dekel_silk_1986} and even to provide a reservoir of hot gas
for re-accretion by galaxies at lower redshifts
\citep{oppenheimer_etal_2010}. 

The most influential role of galactic winds, however, is arguably
their impact on the IGM. \citet{furlanetto_loeb_2003} demonstrate
analytically that LBGs can propagate $\sim 100$kpc into the IGM, which
is compatible with the sizes of observed HI-deficient regions around
LBGs \citep{adelberger_etal_2003}. Based on cosmological simulations,
\citet{aguirre_etal_2001} argue that SNe driven winds from massive
galaxies alone ($M_{\rm baryon}>10^{10.5} {\rm M}_{\odot}$) could
pollute the entire IGM to its observed level. It has also been proposed, however,
that protogalaxies (with masses comparable to present day dwarf
galaxies) could be better candidates for polluting the IGM,  since the
lower velocity winds they would produce at $z=9$ would not  perturb
the IGM in the way that superwinds at $z=3$ would
\citep[][]{madau_ferrara_rees_2001}. There is also evidence for metals
in the IGM by  $z=5$, requiring pollution at an earlier epoch than
that probed by observations of LBGs. Simulations of an isolated dwarf
by \citet{maclow_ferrara_1999} show that almost all metals are ejected
from dwarf galaxies  (with $M_{\rm gas}<10^9 {\rm M}_{\odot}$), even though the ejection of gas mass is surprisingly
inefficient.

It is clear that galactic winds have the potential to impact vast
regions of space and influence structure formation on large
scales. Therefore, the importance of studying this phenomenon in its
full cosmological context cannot be overestimated. However, while there have been many
successful numerical simulation studies of galactic winds in
individual galaxies
\citep{maclow_ferrara_1999,mori_ferrara_madau_2002,
scannapieco_bruggen_2010}, capturing the production of a galactic wind
in cosmological simulations remains a challenge. In such simulations,
galactic winds tend to be imposed directly, rather than allowed
to develop naturally as a result of overlapping SNe bubbles
\citep{springelhernquist03_sf,
scannapieco_etal_2006,oppenheimer_dave_08}. We recall that an
important ingredient of analytical studies of SNe driven winds is a
multiphase ISM i.e. a hot, rarefied atmosphere, containing cold dense
gas clouds \citep[e.g.][]{mckee_ostriker_1977,efstathiou_2000}. If the
resolution of a cosmological simulation is not sufficient, the
multiphase nature of the ISM is not captured and it is not possible to
accurately model SNe remnants. In fact, the inclusion of a multiphase ISM `by
hand' is a common feature of galactic wind models
\citep[e.g.][]{springelhernquist03_sf, scannapieco_etal_2006}. 

\begin{figure*}
  \includegraphics[width=0.31\textwidth,trim = 0mm 0mm 0mm 0mm,
clip]{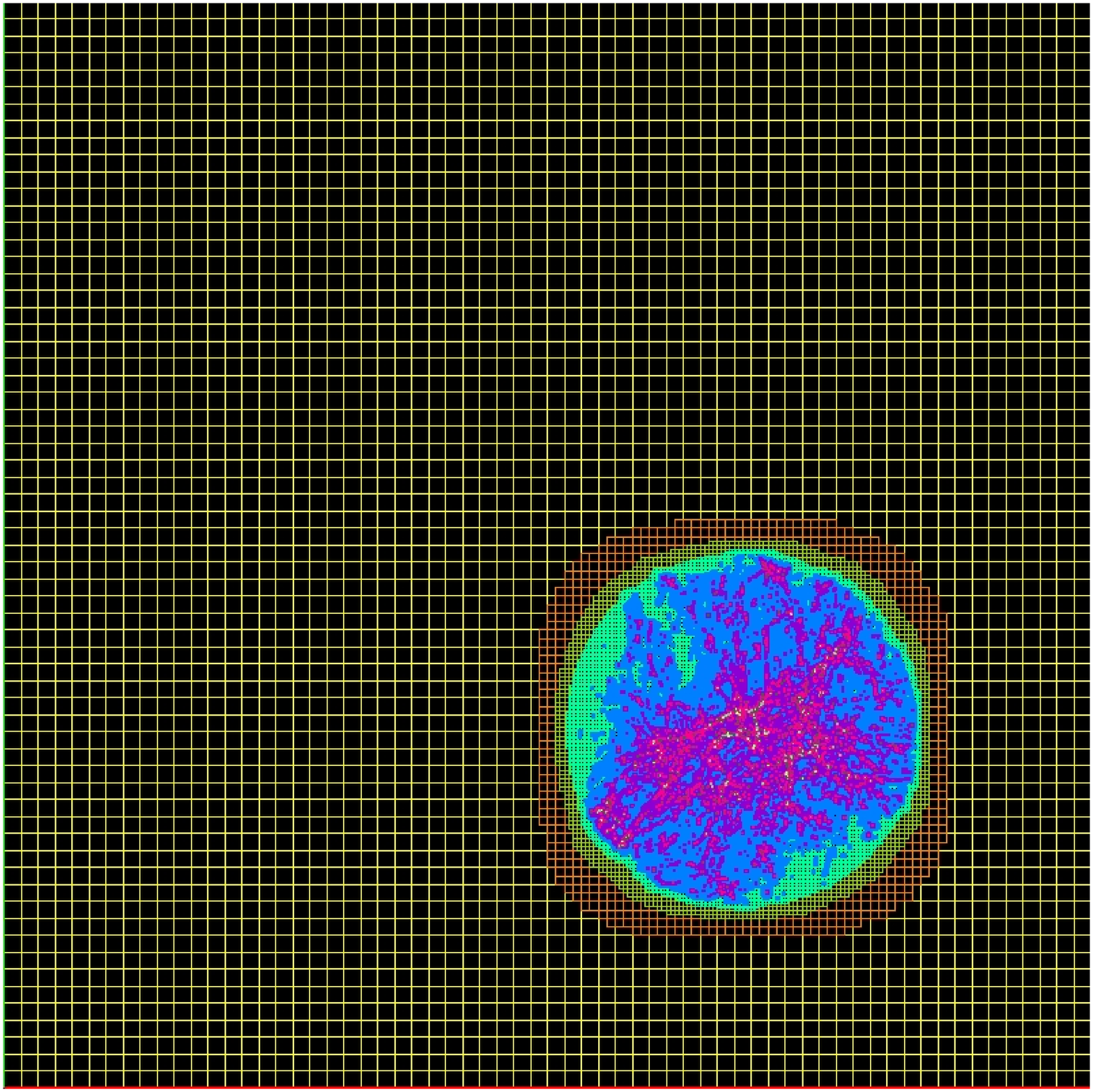}
   \includegraphics[width=0.31\textwidth,trim = 0mm 0mm 0mm 0mm,
clip]{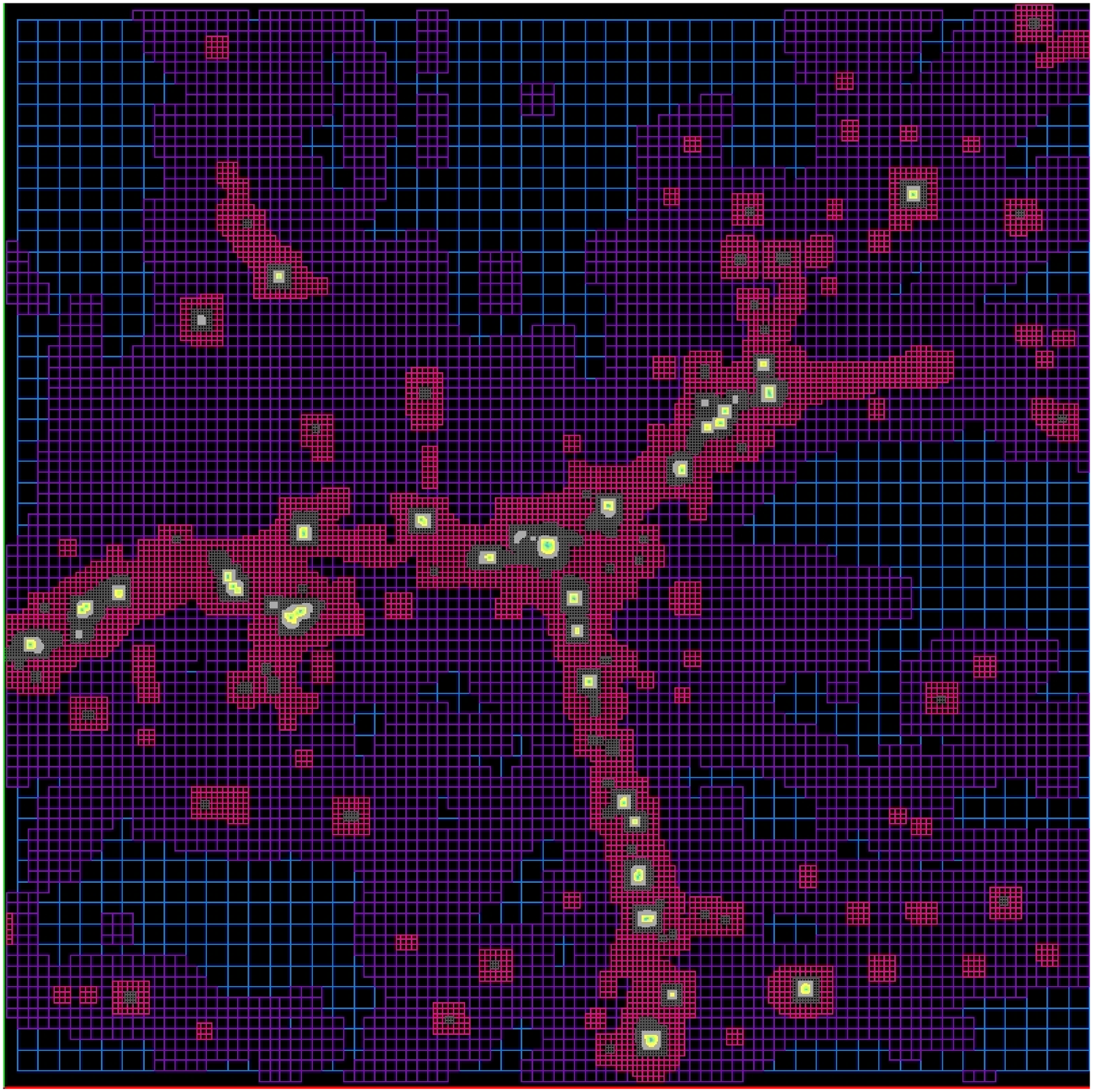}
    \includegraphics[width=0.31\textwidth,trim = 0mm 0mm 0mm 0mm,
clip]{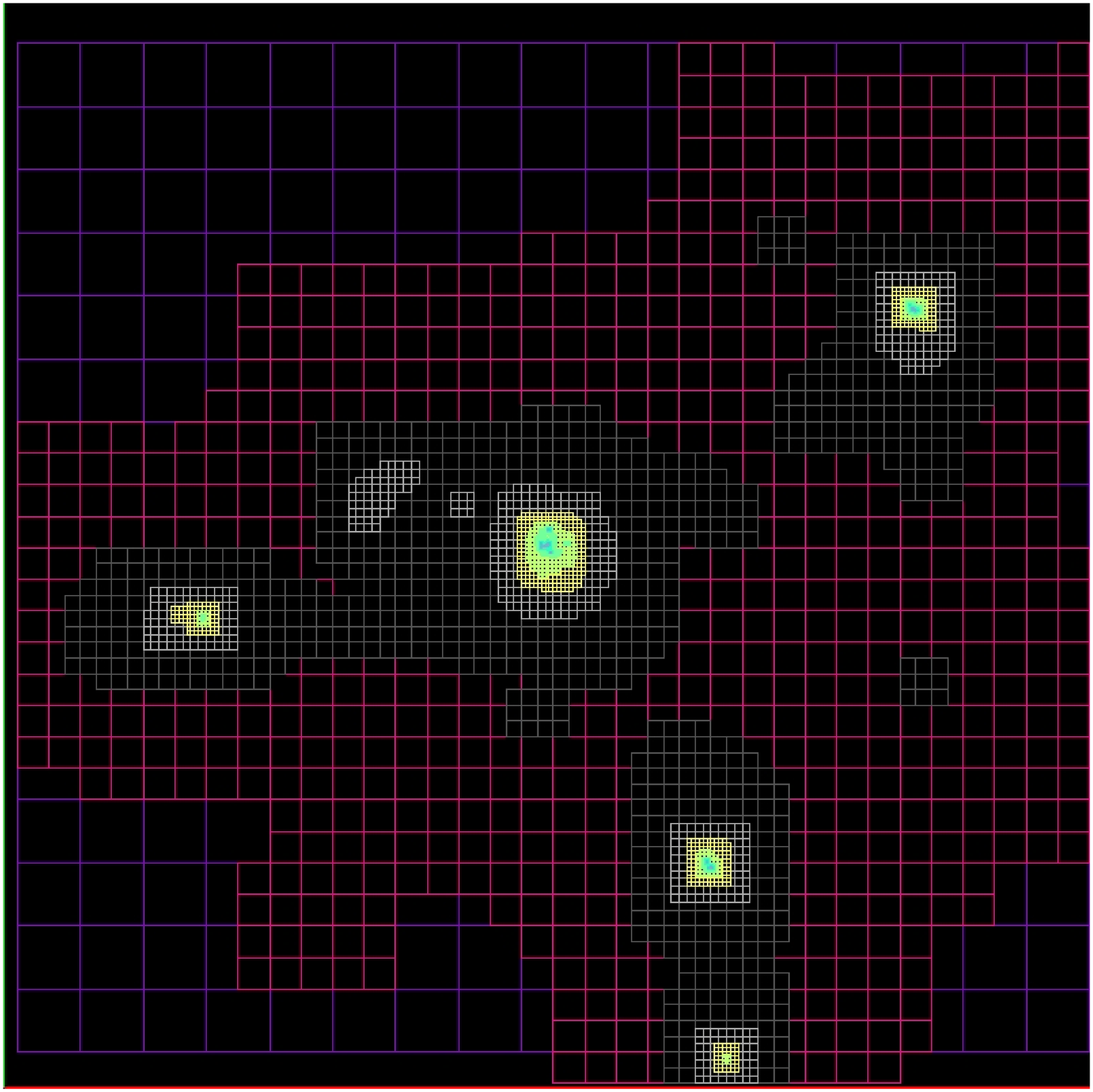}\\
     \includegraphics[width=0.31\textwidth,trim = 0mm 0mm 0mm 0mm,
clip]{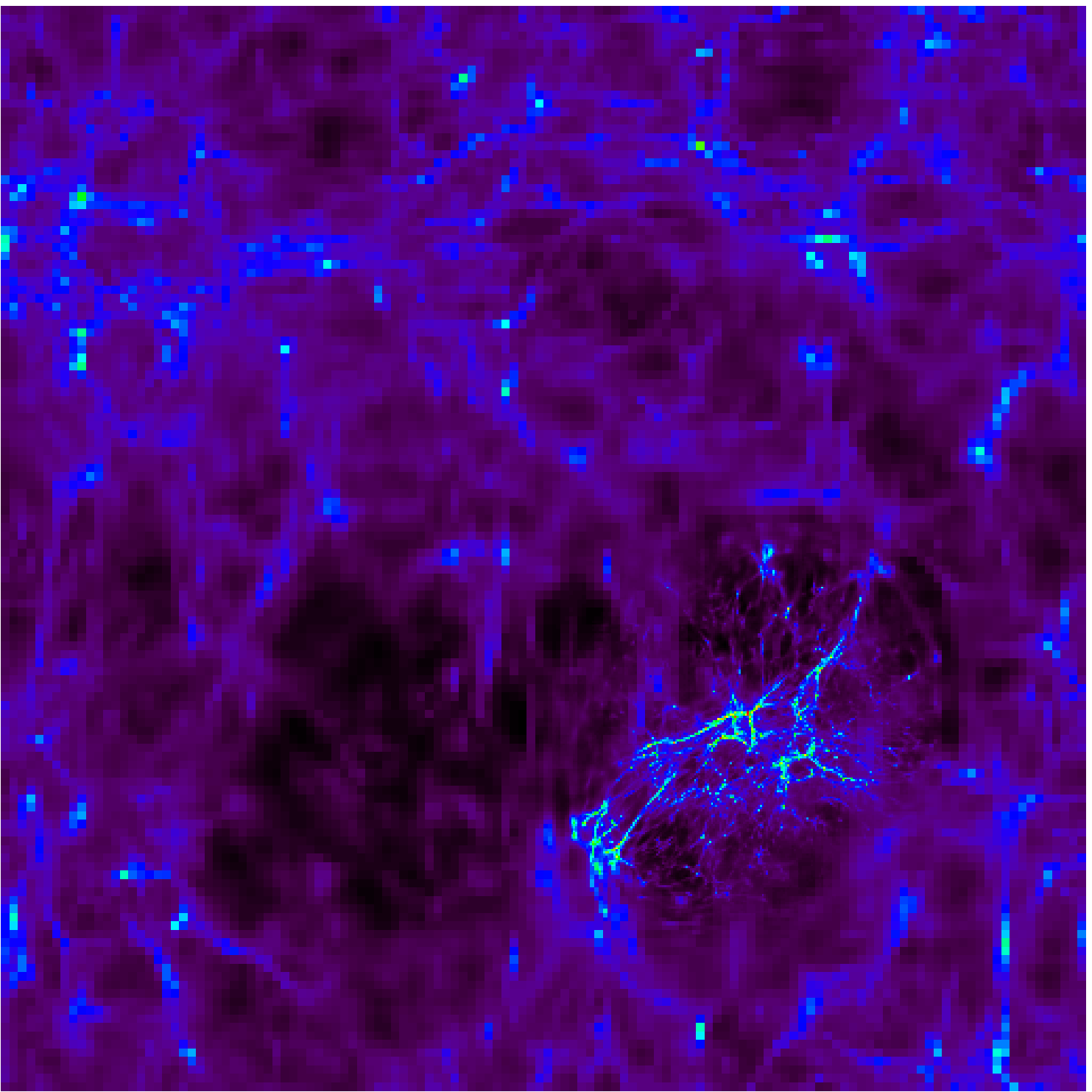}
            \includegraphics[width=0.31\textwidth,trim = 0mm 0mm 0mm
0mm, clip]{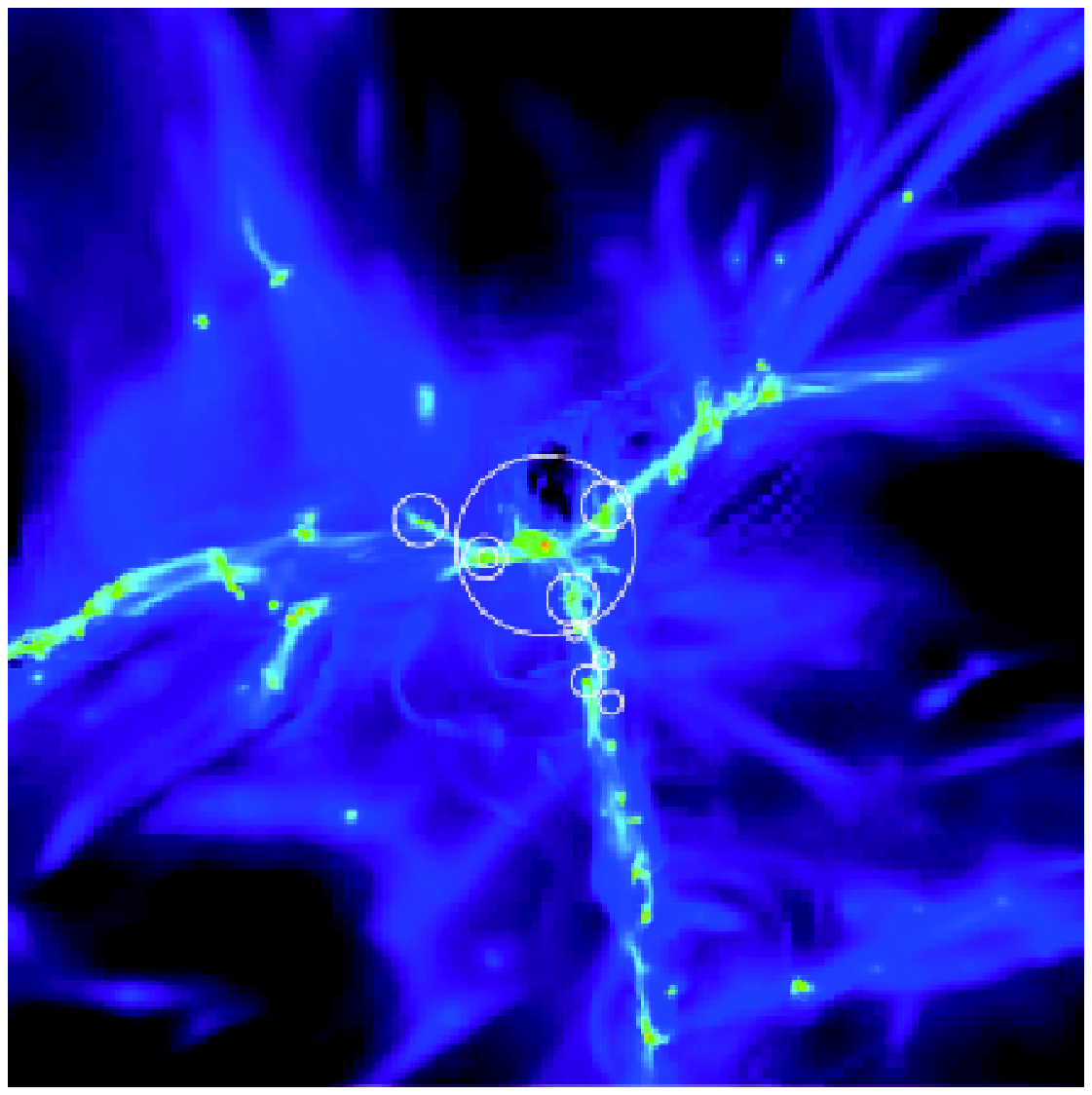}
              \includegraphics[width=0.31\textwidth,trim = 0mm 0mm 0mm
0mm, clip]{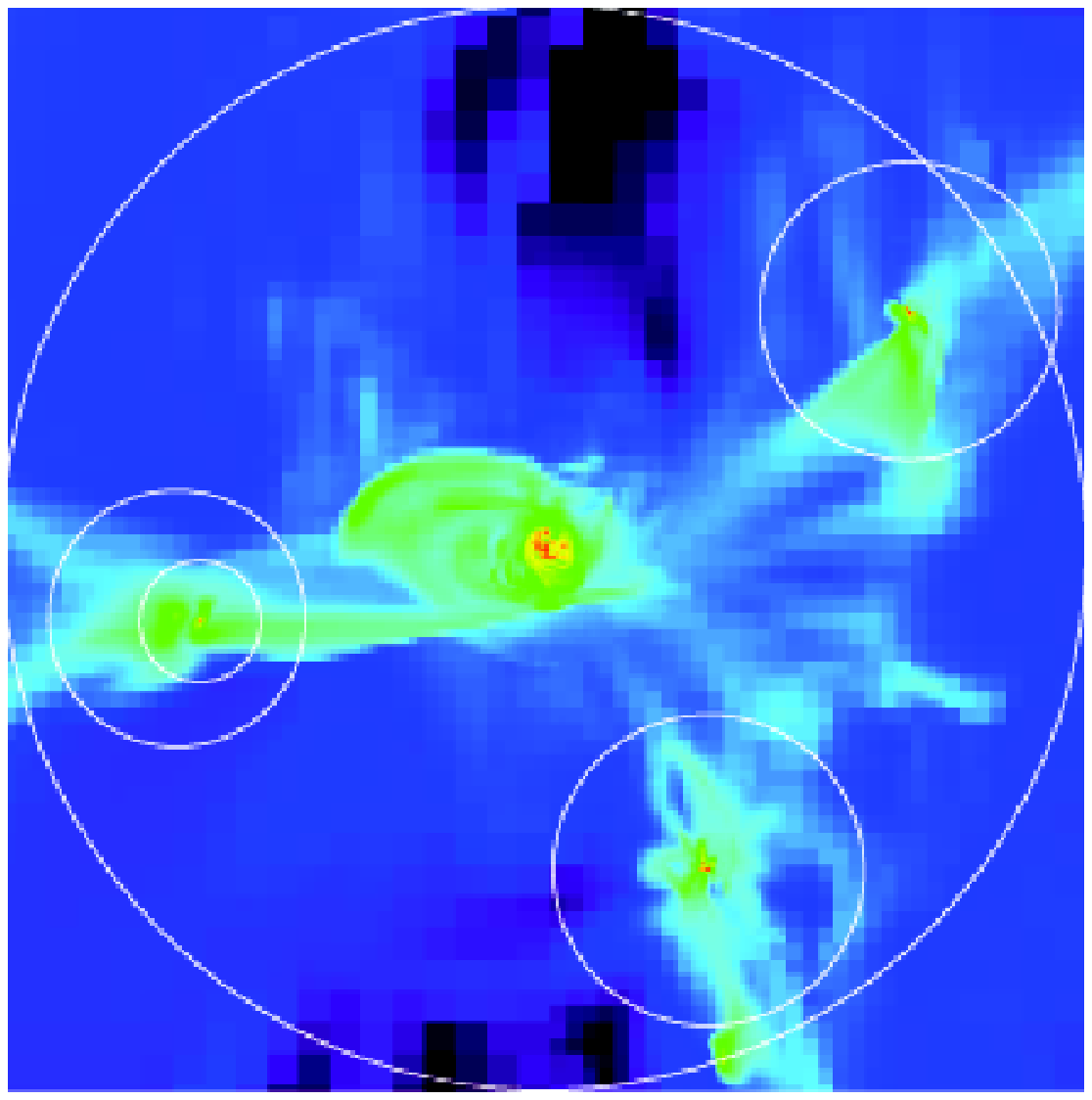}

       \caption{Maps showing a z-projection of the hierarchy of grids
(first row) and corresponding gas density (second row) in the feedback
run at $z=9$. From left to right: The whole box, a cube of side
$12r_{\rm vir} \approx 60$kpc centred on the main halo and a cube of side $2r_{\rm
vir}$ also centred on the main halo respectively. The white circles in the middle
and right gas density images (second row) indicate the virial radius
of the main halo and its subhaloes. In the top left panel showing the
whole box, we can see the $128^{3}$ grid covering the whole image and
the 3 nested grids of equivalent resolution $256^{3}$ (orange), $512^{3}$ (yellow) and
$1024^{3}$ (green) centred on our main halo. Additional grids can also be seen
(blue,pink) indicating where additional levels of AMR have been
triggered in the $1024^{3}$ grid. The middle and right panels show the
triggering of additional levels of AMR (up to a maximum of 15) in
dense structures, such as the main galaxy disc and subhaloes.} \label{grids} 
\end{figure*}

Examining galactic winds in a cosmological context, rather than in
isolation, introduces further complications. Galaxies are now embedded
in the cosmic web and are subject to inflows as well as outflows.  The
mode of inflow will be dependent on the galaxy halo mass; 
low-mass galaxy haloes ($M_{\rm vir} < M_{\rm shock} \approx 10^{12} M_\odot$ ) 
cannot host a stable shock at the virial radius and so primarily accrete cold ($T < T_{\rm vir}$) gas
whereas high-mass galaxy haloes ($M_{\rm vir} > M_{\rm shock}$) primarily accrete
shock-heated, hot ($T \sim T_{\rm vir}$) gas \citep{shock_1d}.  It has been suggested that this
framework, when coupled with active galactic nuclei (AGN) and SNe
feedback, could naturally give rise to some of the bimodality in
galaxy properties we observe but usually fail to reproduce with
simulations \citep{mshock}. Indeed, semi-analytic modelling has shown
that incorporating such processes results in an improved fit to most
galaxy property trends e.g. colour bimodality, luminosity function etc
\citep{cattaneo}.  \citet{keres2b} have shown, however, that AGN
radio-mode feedback (which can prevent hot gas from being accreted)
has little effect on high-mass galaxies in cosmological simulations
because most of these were built hierarchically from lower mass
galaxies which gained their baryons via cold accretion. They suggest
it is feedback in low mass galaxies, such as SNe driven winds, that is
of greatest importance. It seems that studying accretion and
feedback in low mass protogalaxies is, therefore, vital for our understanding of
galaxy evolution as a whole.

Recent cosmological simulations are in agreement that for low mass
galaxies (with $M_{\rm vir} < M_{\rm shock}$) most inflowing gas is
cold \citep[e.g.][]{keres,bimodal_marenostrum,brooks}. All of these studies also demonstrate qualitatively 
that much of this cold inflow occurs along filaments (e.g. Fig.~17 of \citet{keres}, Fig.~5 of  
\citet{bimodal_marenostrum} and Fig.~5 of \citet{brooks}). What has not been quantified, however, is how 
much cold accretion occurs in a spherically symmetric way i.e. with the same 
geometry as assumed for hot accretion.  While it is reasonable to assume that this spherical cold accretion 
can be safely ignored when the galaxy is surrounded by a halo of shock-heated gas (i.e. it has $M_{\rm vir} 
\sim M_{\rm shock}$), this is not the case when the halo is below the shock mass threshold and there is not necessarily anything 
to impede the spherical accretion of cold material from the IGM. Indeed, most of the visual demonstrations that 
cold accretion takes place along filaments in these previous studies focus on examples when the galaxy has a hot
 gas halo and cold accretion is necessarily confined to the filaments anyway. This paper thus complements 
such studies by measuring how cold gas accretion is divided between the filamentary and diffuse (spherically 
symmetric) components in low mass ($M_{\rm vir} < M_{\rm shock}$) haloes, in simulations both with and without SNe feedback.

Most of the aforementioned studies also incorporate SNe feedback, although it is unclear whether a substantial galactic 
wind ever develops and they do not address the specific impact of this on the accretion processes. Recent work 
by \citet{vandevoort_etal_2010} attempts to tackle this issue with a suite of cosmological simulations employing 
different physics. We note, however, that their SNe feedback is implemented by giving gas particles a velocity 
kick and so the galactic wind is an input rather than a result that arises naturally. They conclude that the 
relative importance of cold accretion onto the galaxy halo is robust to changes to the feedback recipes, but that 
the relative importance of cold accretion onto the galaxy itself is decreased in the presence of SNe-driven 
winds. We revisit this question in this paper, but at much higher
resolution and redshift ($z \ge 9$), where SNe blastwaves 
are individually resolved and so it is not necessary to put in a galactic wind by hand. We also emphasize that our
 high spatial and mass resolution allow us to fully resolve the
 filaments at very high redshift ($z \ge 9$), contrary to previous studies in which the
 spatial resolution, in particular, is lacking in this redshift range. These resolution effects potentially enhance the
 likelihood of the filaments being destroyed by the SNe feedback, changing the balance between
inflow and outflow and thus the subsequent evolution of the galaxy.

A few high resolution studies of cosmological accretion in
individual galaxies have been performed, but these tend to focus either on more massive
galaxies (i.e.~ with  $M_{\rm vir} > M_{\rm shock}$) at lower redshift
($z \sim 3$) \citep{agertz_clumpygal,ceverino_clumpygal} or on the
early formation stages of the first galaxies ($M_{\rm vir} \approx 5 \times 10^{7} M_\odot$) 
at $z>10$ where direct accretion of gas from the IGM briefly precedes cold filamentary  
inflows \citep{greif_etal_2008}. We are primarily interested in studying the
epoch bracketed by these studies; that in which
filamentary accretion is important for low mass haloes, during which
galactic winds may also develop.

In summary, for a low mass protogalaxy (i.e. with  $M_{\rm vir} <
M_{\rm shock}$) at high redshift ($\approx z=9-10$), there are two
main predictions about its interaction with the cosmic web  : 1) a galactic wind will develop and extend far into the IGM,
polluting it with metals and 2) cold gas will flow rapidly into the
host halo via the web's filaments.  What happens when these inflow and
outflow mechanisms occur simultaneously? We  tackle this question with
a suite of cosmological resimulations (incorporating different
physics) with $0.5$pc physical resolution in the densest regions, of a galaxy with $M_{\rm
vir} \sim 5 \times 10^{9} M_\odot$ at $z=9$.  At such high resolution, we can model
individual SNe with a Sedov blastwave allowing a galactic wind to
arise naturally. We examine the mechanism via which the wind develops
and measure its properties.  Furthermore, by measuring the inflow and
outflow rates and comparing these with a control run (without SNe) we
investigate whether the presence of a hot galactic wind can alter the
accretion processes in a protogalaxy in such a way as to significantly
impact its evolution.

\begin{figure*} \centering
  \includegraphics[width=0.4\textwidth,trim = 0mm 0mm 0mm 0mm,
clip]{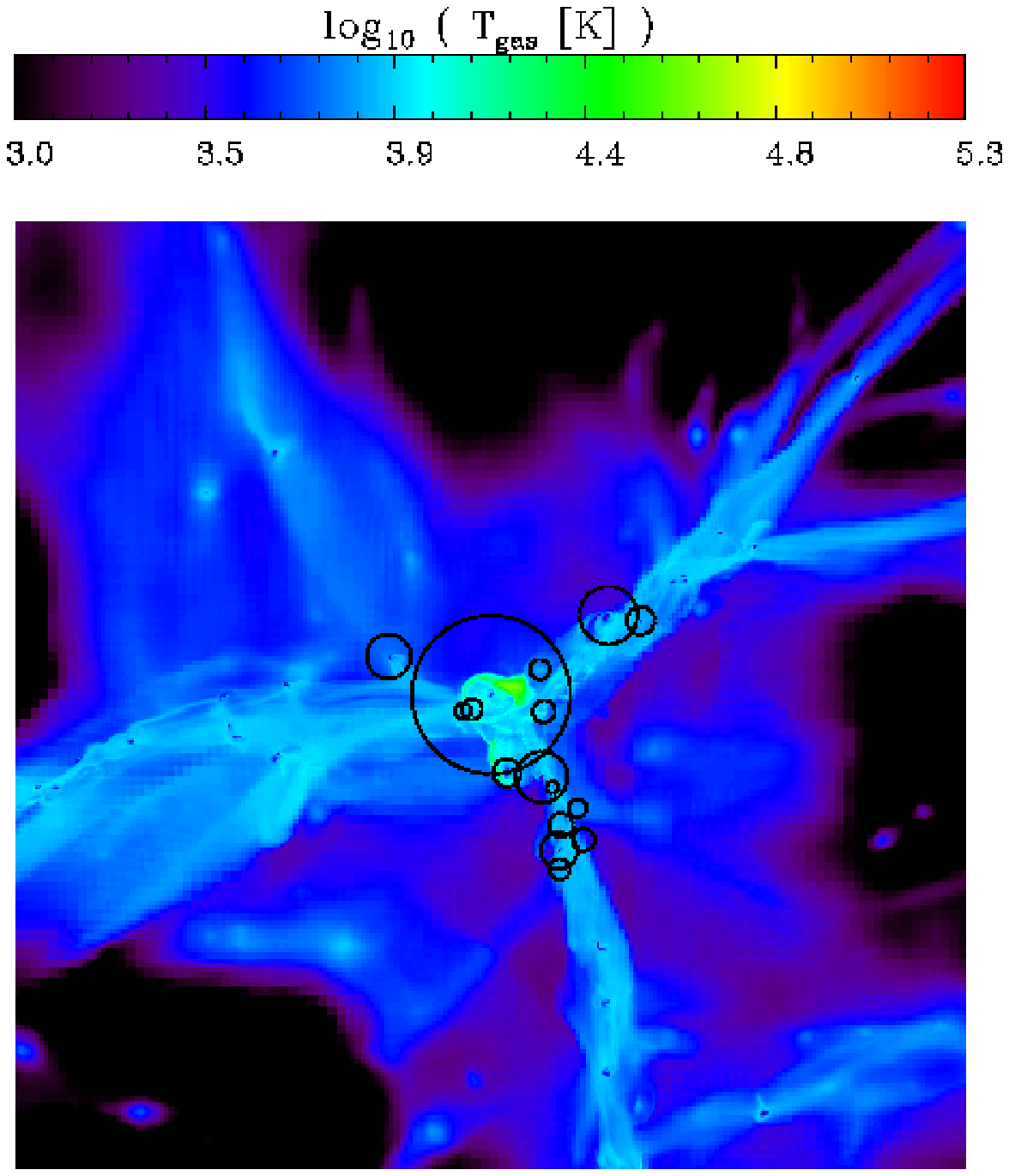}
   \includegraphics[width=0.4\textwidth,trim = 0mm 0mm 0mm 0mm,
clip]{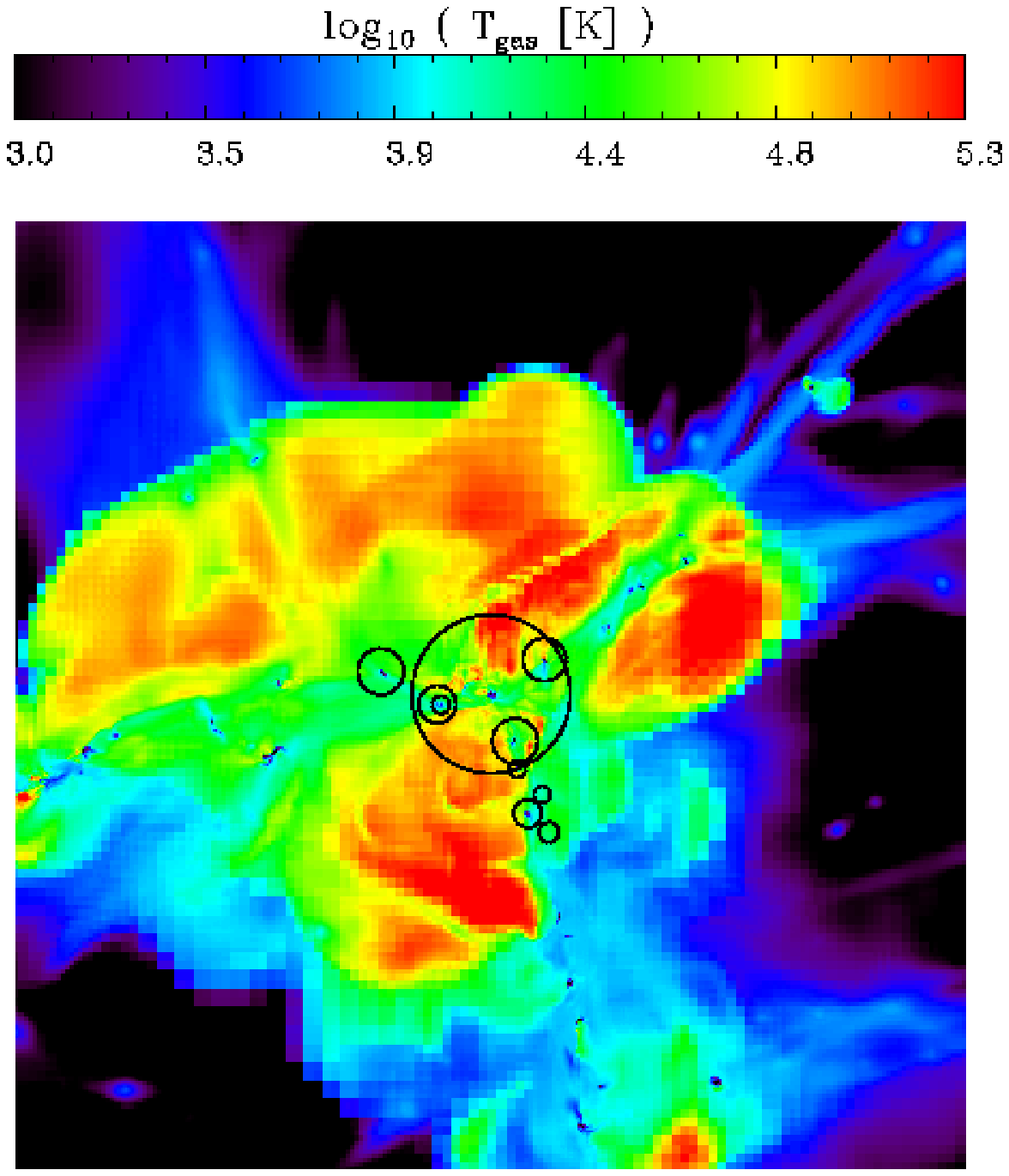}\\
  \includegraphics[width=0.4\textwidth,trim = 0mm 0mm 0mm 0mm,
clip]{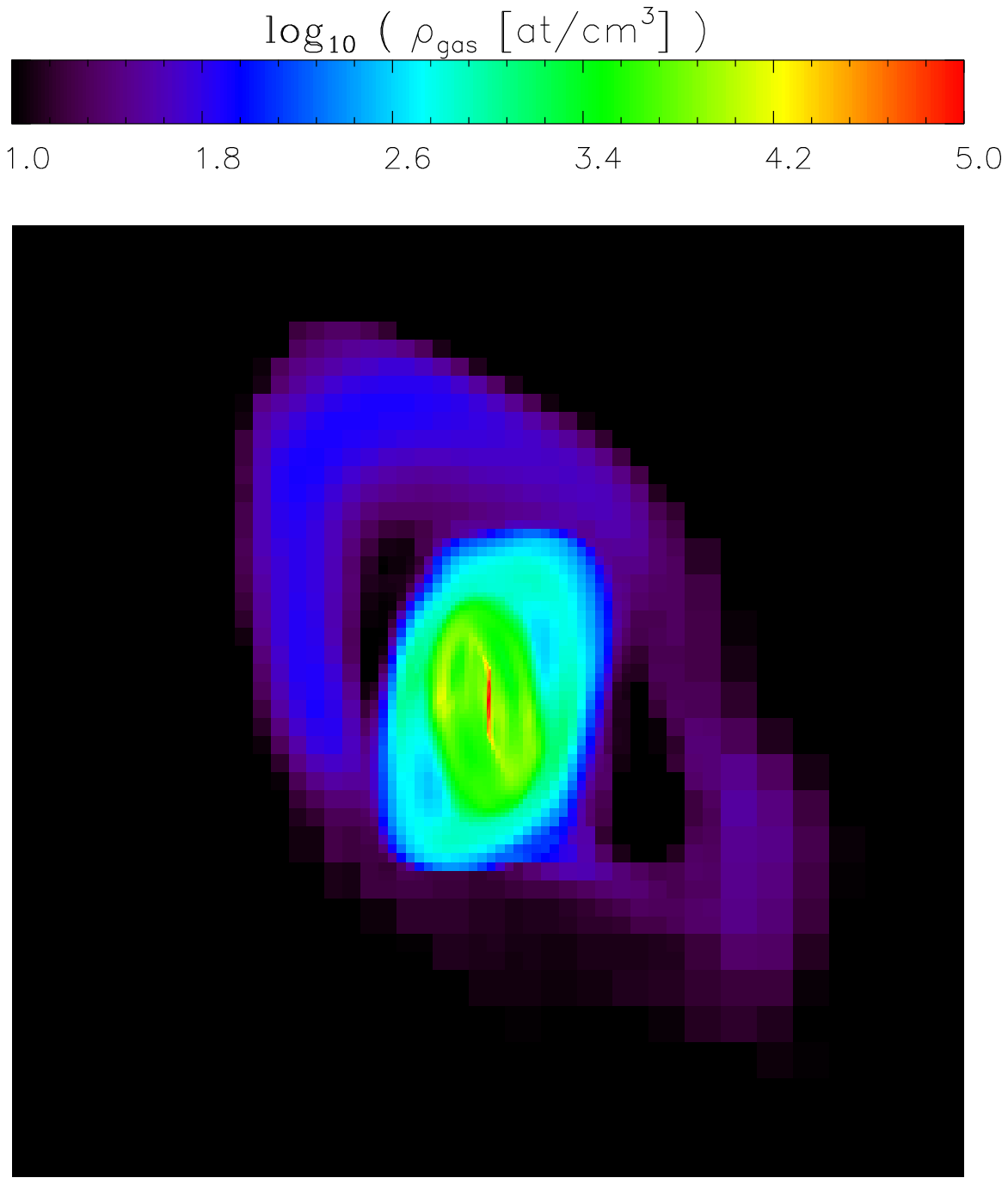}
   \includegraphics[width=0.4\textwidth,trim = 0mm 0mm 0mm 0mm,
clip]{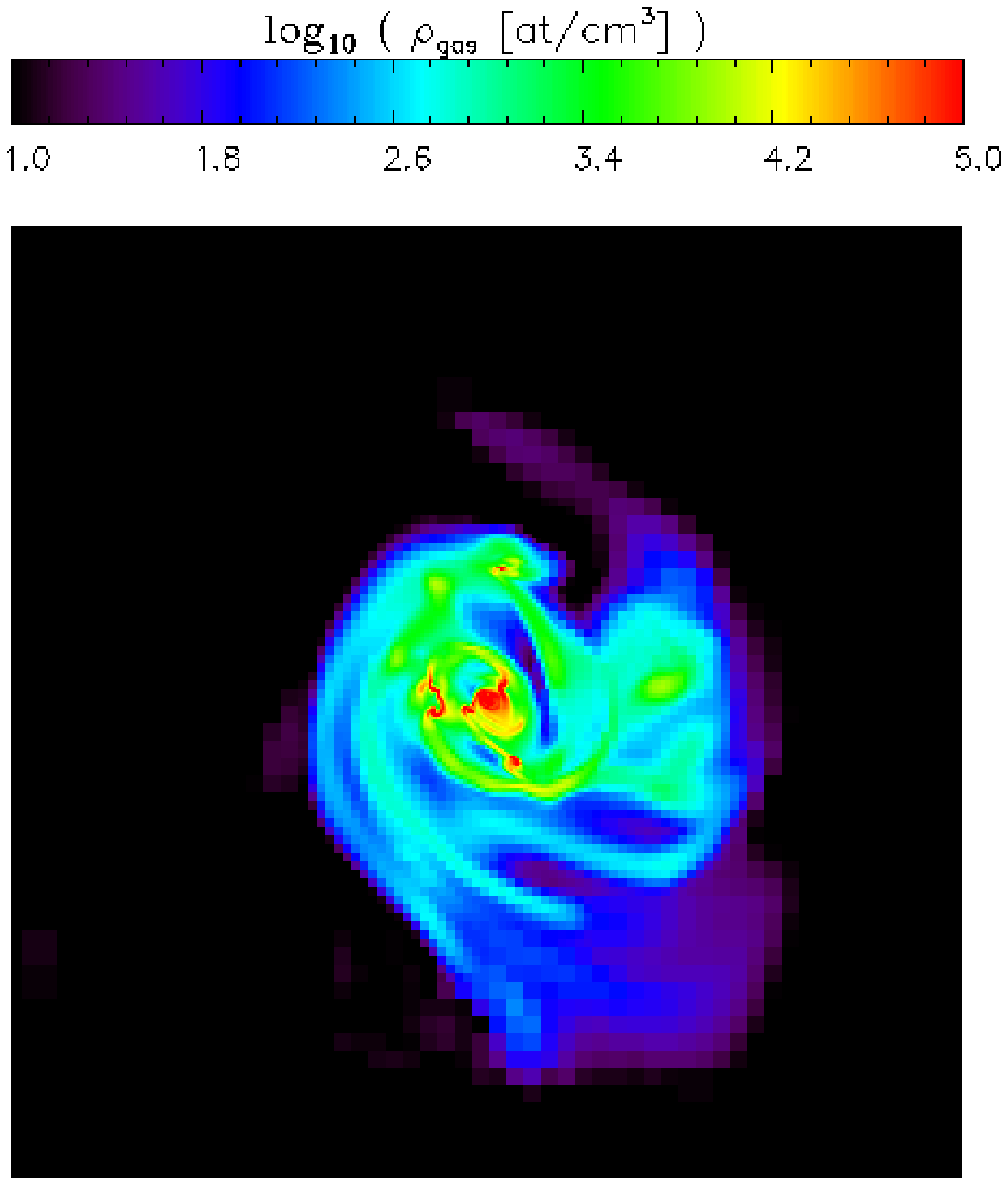}
   \caption{{\bf Top row: }Density weighted temperature for a cube of
$12 r_{\rm vir} \approx 60$ kpc on a side centred on the main halo for the cooling (left)
and feedback (right) runs at $z=9$. Black circles in the top images indicate the
position and virial radius of the main halo (the largest circle) and
its subhaloes. The colour scale has been fixed to $3.0 \le \log T \le
5.3$; $\log T=5.3$ is the threshold for gas to be considered `hot'
(see Table \ref{catsummary}) and the lower limit of $\log T=3$ has
been chosen such that the filaments can be seen clearly since a
significant mass fraction of the gas is in the range $3.3 \le \log T
\le 4.3$. {\bf Bottom row: } Projected density for a cube of
$0.2 r_{\rm vir}$ on a side centred on the main halo, showing the disc approximately face-on in the cooling (left)
and feedback (right) runs at $z=9$. Note the clumps visible in the feedback run (right).}
    \label{tempmaps} \end{figure*}

This paper is structured as follows. In Section \ref{sec:sim} we
describe the suite of simulations we have performed and outline the
gas physics employed (cooling, star formation, SNe feedback), further
details of which are provided in the Appendix (Section
\ref{sec:appendix}). In Section \ref{sec:filwind} we distinguish
between different gas phases in both runs, relating these to physical
structures e.g. the filaments, satellite galaxies, the hot galactic
wind (feedback run only). In Section \ref{sec:hotgas}, we explore
the mechanism responsible for the development of the galactic wind in
the feedback run and measure the wind properties. We measure the
inflow and outflow in all gas phases in Section \ref{sec:acc} and then
relate the dominant accretion modes and outflows to the star formation
rate in Section \ref{sec:sfr}, comparing the cooling and feedback runs
in order to isolate the impact of the SNe feedback on the
formation and evolution of this galaxy at high-redshift. Finally, we
outline our conclusions in Section \ref{sec:conc}.

\section{The {\sc nut} simulations}\label{sec:sim}

The {\sc nut} Simulations \footnote{Nut is the Egyptian goddess of the night
sky} are a suite of ultra-high resolution (maximum physical spatial
resolution of $\approx 0.5$pc at all times) cosmological resimulations
of a Milky Way like galaxy forming at the intersection of 3 filaments
in a $\Lambda$CDM cosmology. To date
we have performed 3 simulations, with different gas physics, with the
AMR code, {\sc ramses} \citep{ramses}: i) adiabatic with a uniform UV
background \citep{uvbackground} turned on instantaneously at
z=8.5 ii) cooling, star formation and UV background and iii) as ii)
but with the addition of supernova (SN) feedback, including metal
enrichment and its subsequent changes to the cooling
function. Simulations i) and ii) are presented elsewhere (Powell et
al, in prep);  for the sake of completeness we summarise the technical
details below.

The galaxy resides in a halo with a virial mass of 1.05$\times10^{11}
M_{\odot}$ at $z \approx3 $ and has been selected such that it lies in
an isolated environment (not in the centre of a cluster or group at
$z=0$). The cosmology employed in the simulation {\tt is}
$\Omega_{\lambda}=0.742, \Omega_{\rm m}=0.258$, $\Omega_{\rm b}=0.045$
and $\sigma_{8}=0.8$, with comoving box-size $= 9$ Mpc h$^{-1}$ and
the simulations start at $z \approx 499$ (parameters are consistent
with {\it WMAP}$5$ \citep{wmap5}). There are three nested grids
located in the region where the galaxy halo will form which
corresponds to a volume of approximately (2.7 comoving Mpc
h$^{-1})^{3}$. This has the equivalent resolution of 1024$^{3}$ dark
matter particles which each have a mass of ${\rm M}_{\rm
DM}=5.4\times$$10^{4}$ $M_{\odot}$.  Further levels of refinement, up
to a maximum of $15$, are applied to the grid in this region in order
to keep the spatial resolution, in the densest region, below $\sim$ 1
pc physical at all times. We use a quasi-lagrangian refinement
strategy (i.e. the gas mass of cells stays roughly constant) in which
the next level of AMR is triggered when the baryonic mass in a cell
reaches $8\times m_{\rm sph}$ (where $m_{\rm sph}=9.4 \times10^{3}
M_{\odot}$), or the number of dark matter particles in the cell
reaches $8$. Fig.~\ref{grids} illustrates the grid structure (top row)
and corresponding gas density with halo and subhaloes indicated with
white circles (bottom row), showing the set-up of the nested grids and
the locations where additional levels of AMR are triggered (see
caption for details).  Additional technical details, including
descriptions of the star formation and SN feedback prescriptions, are
given in the Appendix.

Fig.~\ref{tempmaps} shows temperature maps of a region of size $12 r_{\rm
vir}$ (top row) and density maps of a region of size $0.2 r_{\rm vir}$
(bottom row) centred on the resimulated galaxy at $z=9$ for the cooling (left column) and feedback
(right column) runs. Since cooling is present in both simulations, 
 a thin, rapidly rotating disc forms,
which is continuously fed by cold gas that streams along the filaments
at $\sim$Mach $5$. The disc is gravitationally unstable and
occasionally fragments into star-forming clumps of several parsecs in radius (e.g. bottom
right panel of Fig.~\ref{tempmaps}) which our resolution allows us to resolve along 
with the filaments and the scale-height of the thin disc (this is demonstrated 
in Powell et al, in prep). In our simulation with SN feedback a far-reaching galactic
wind develops at very high-redshift. Star formation in the galaxy, its
satellites and haloes embedded in the filaments results in many SNe
explosions, which are individually resolved. The numerous bubbles
overlap, creating an extended cavity through which the hot gas escapes
the galactic potential in the form of a wind.  By $z=9$, when the dark matter halo mass has reached
$\approx 5 \times 10^9 {\rm M}_{\odot}$, the wind extends to $\approx 6 r_{\rm
vir}$ i.e. $\approx 30$kpc, filling the virial sphere with hot, diffuse
gas. 

  \begin{figure} \centering
  \includegraphics[width=0.4\textwidth,trim = 0mm 0mm 35mm 5mm,
clip]{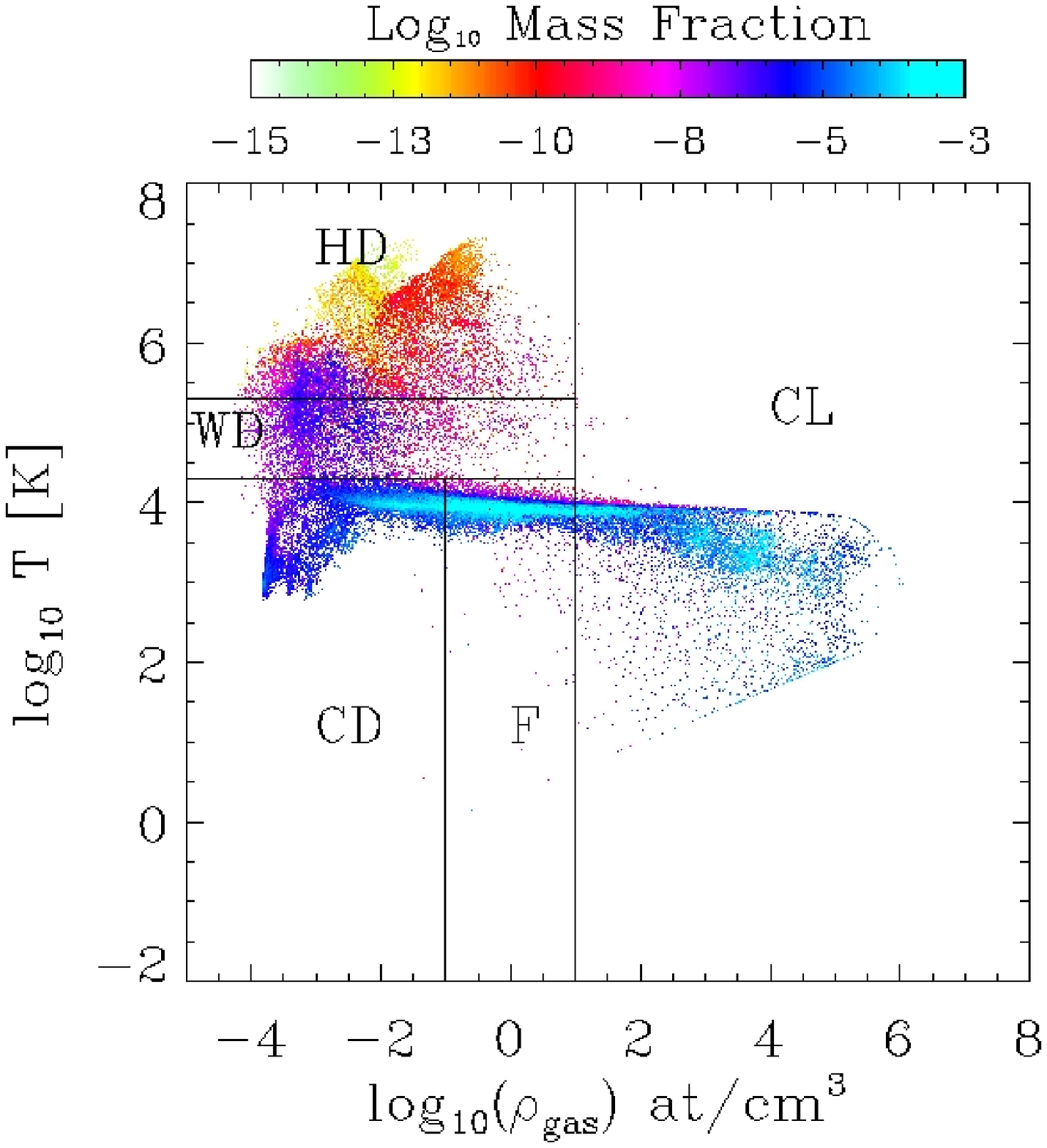}
   \includegraphics[width=0.4\textwidth,trim = 0mm 0mm 35mm 5mm,
clip]{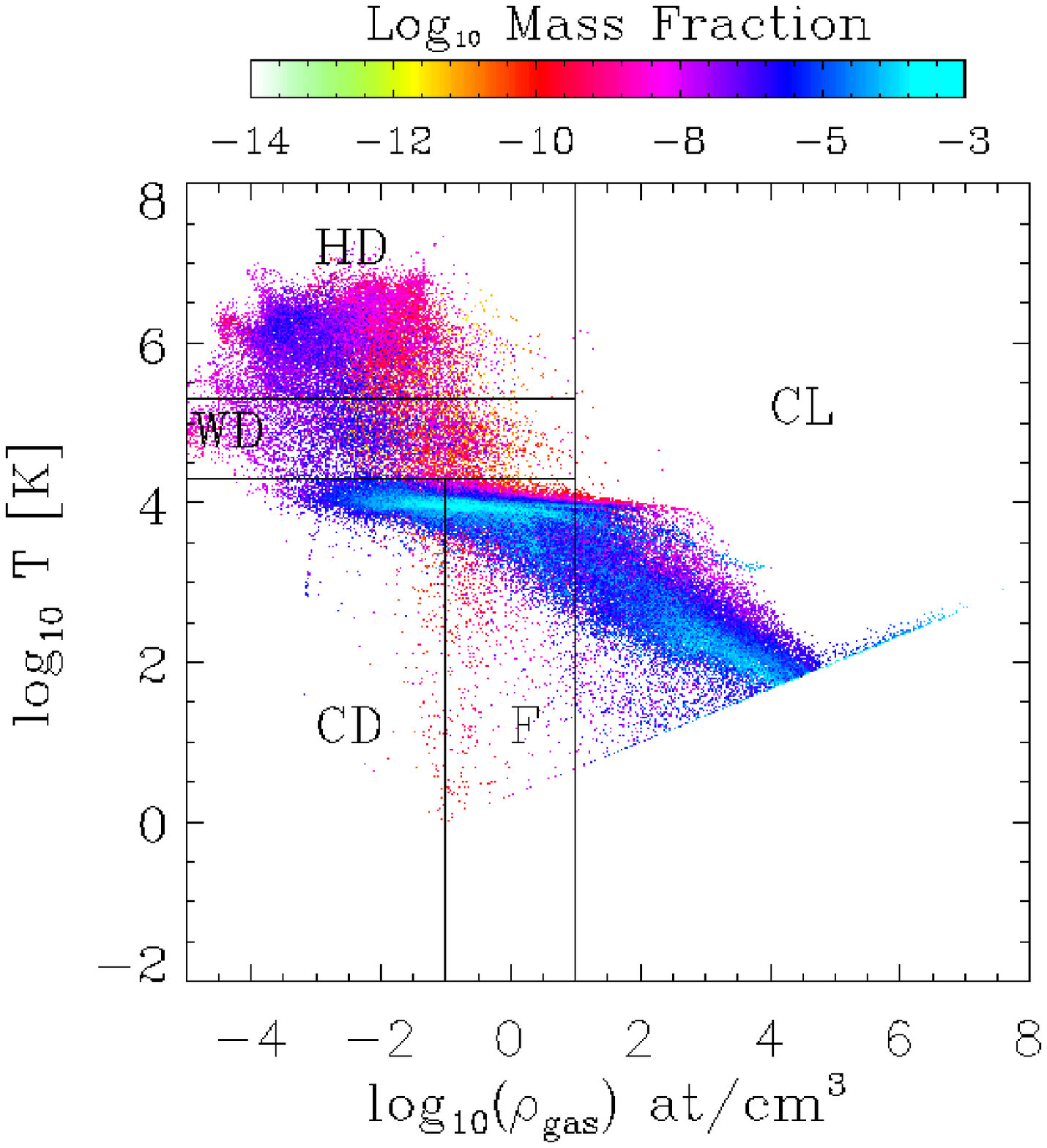}
   \caption{Temperature-density histograms for all gas within the
virial radius of the main progenitor in the cooling (top) and feedback (bottom) run at
$z=9$. The
colours show the fraction of the total gas mass within each bin. Solid
horizontal lines show the temperature cuts at $T=2\times10^{4}$K and
$T=2\times10^{5}$K and vertical lines show the density cuts at $0.1$
atoms/cc and $10$ atoms/cc which we have chosen to distinguish the
different accretion modes, labelled: clumpy (`CL'), filaments (`F'),
hot diffuse (`HD'), warm diffuse (`WD') and cold diffuse (`CD'). The
virial temperature for this mass halo (M $\sim 5 \times 10^9 {\rm M}_{\odot}$)
is $1.3 \times 10^{5}$K and $1.6 \times 10^{5}$K for the cooling and
feedback runs, respectively.}
    \label{histofluxcuts} \end{figure}

\section{Identifying the wind and filaments} \label{sec:filwind}

In order to quantify the relationship between filamentary accretion and galactic winds, 
we define 5 different gas phases, using temperature and density cuts and measure
mass inflow and outflow rates in each of these phases.

\begin{figure*} \centering \fbox{
    \includegraphics[width=0.45\textwidth,trim = 0mm 0mm 0mm 0mm,
clip]{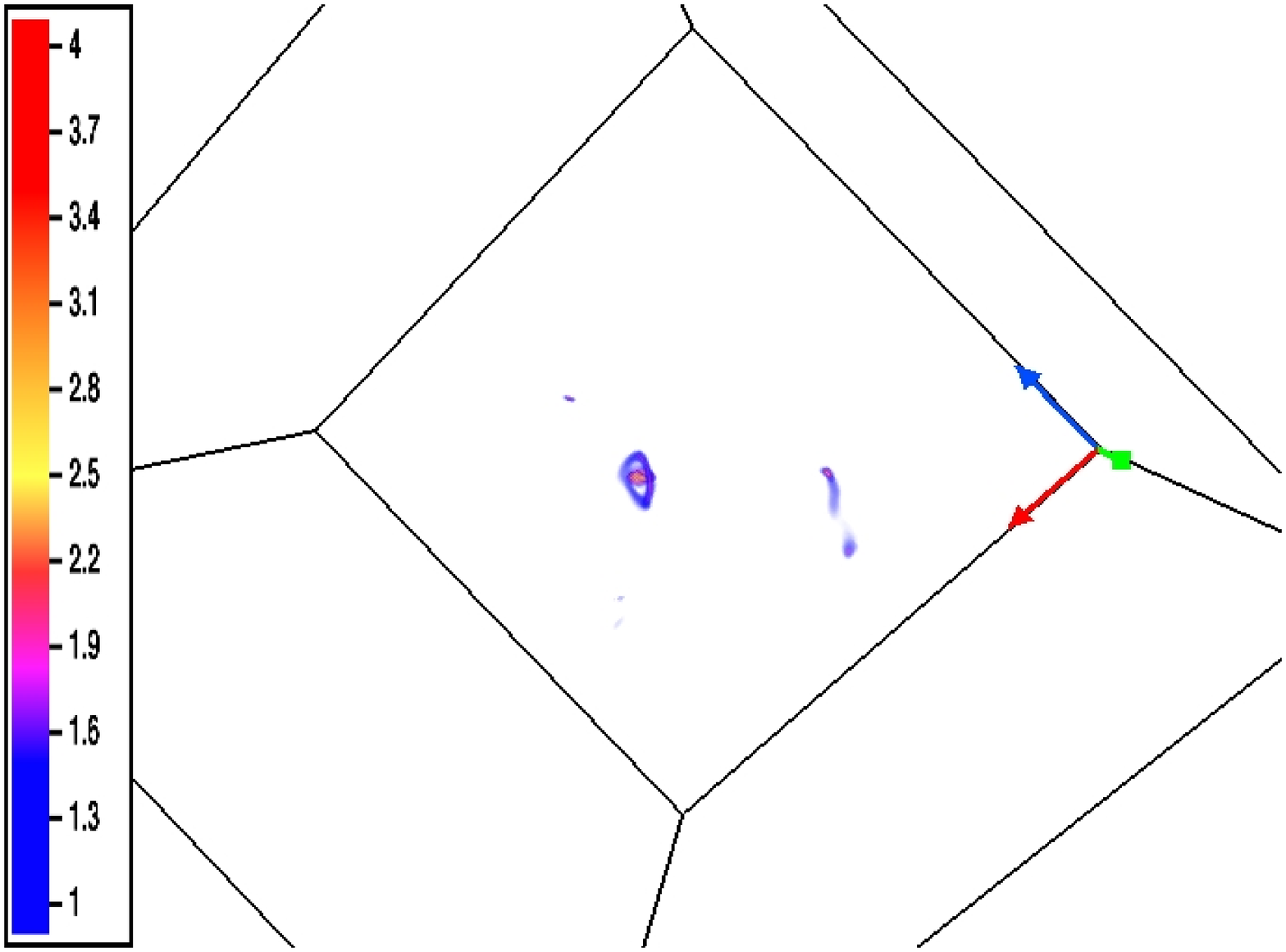}} \fbox{
      \includegraphics[width=0.45\textwidth,trim = 0mm 0mm 0mm 0mm,
clip]{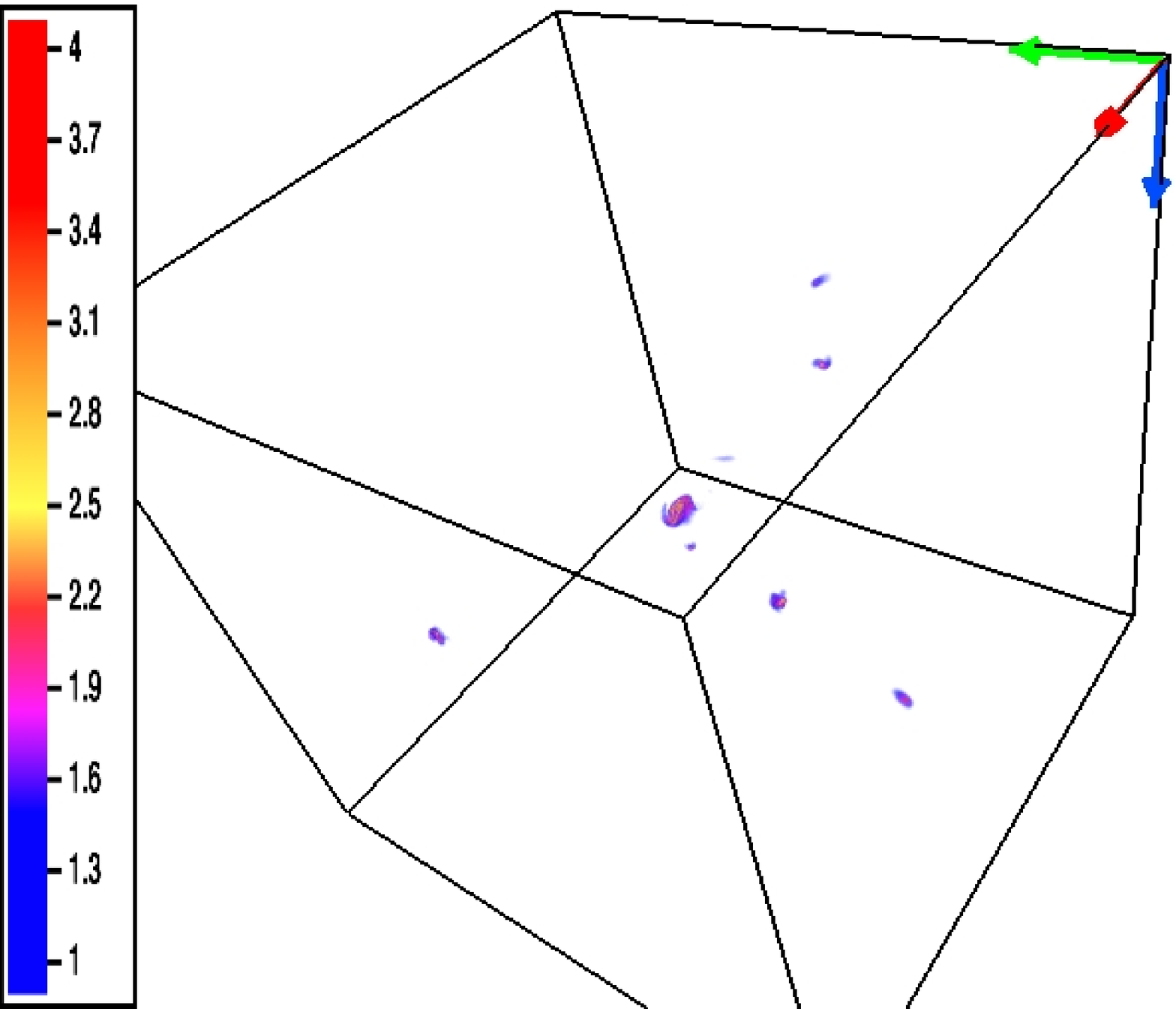}}\\ \fbox{
  \includegraphics[width=0.45\textwidth,trim = 0mm 0mm 0mm 0mm,
clip]{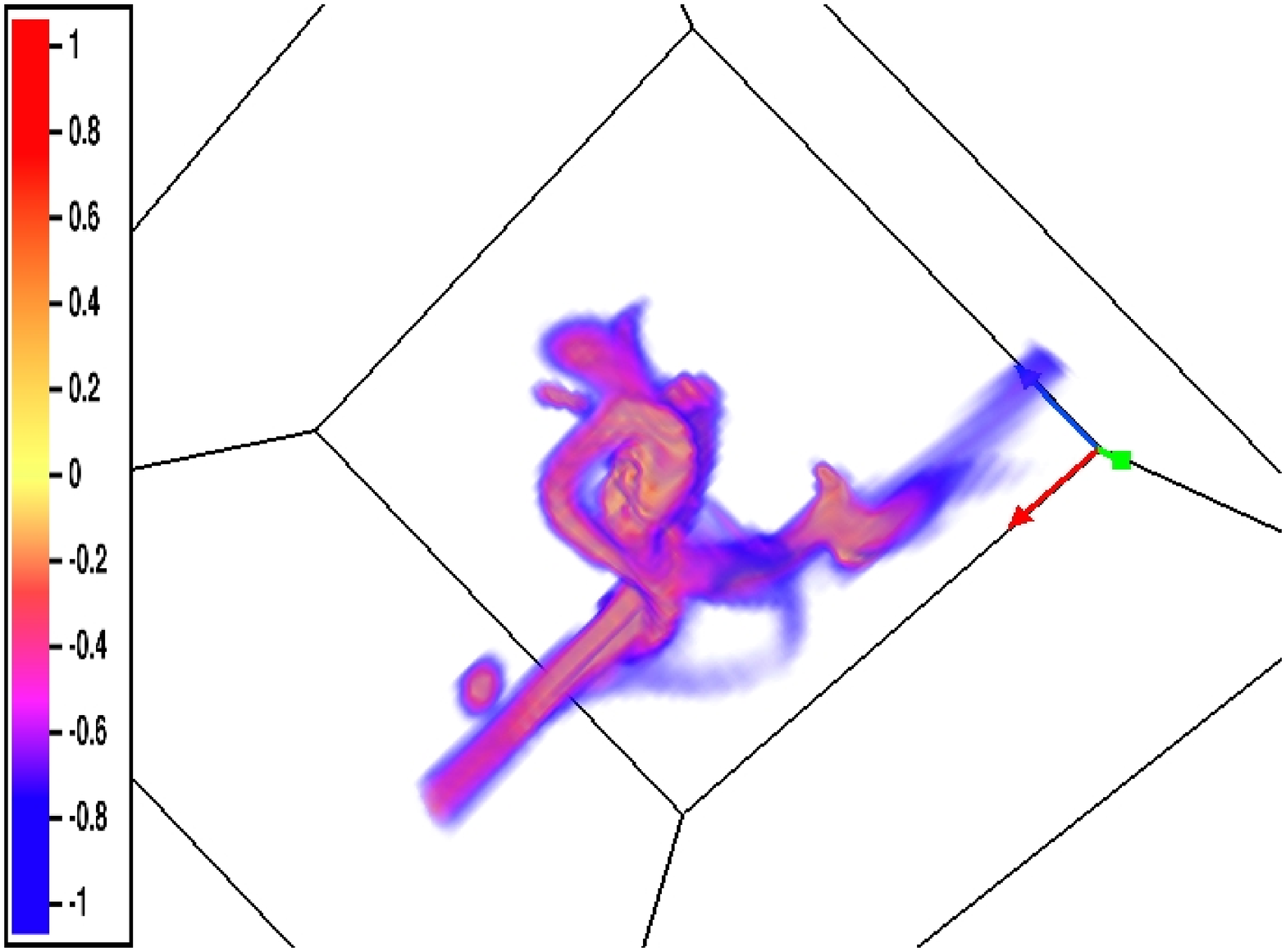}} \fbox{
  \includegraphics[width=0.45\textwidth,trim = 0mm 0mm 0mm 0mm,
clip]{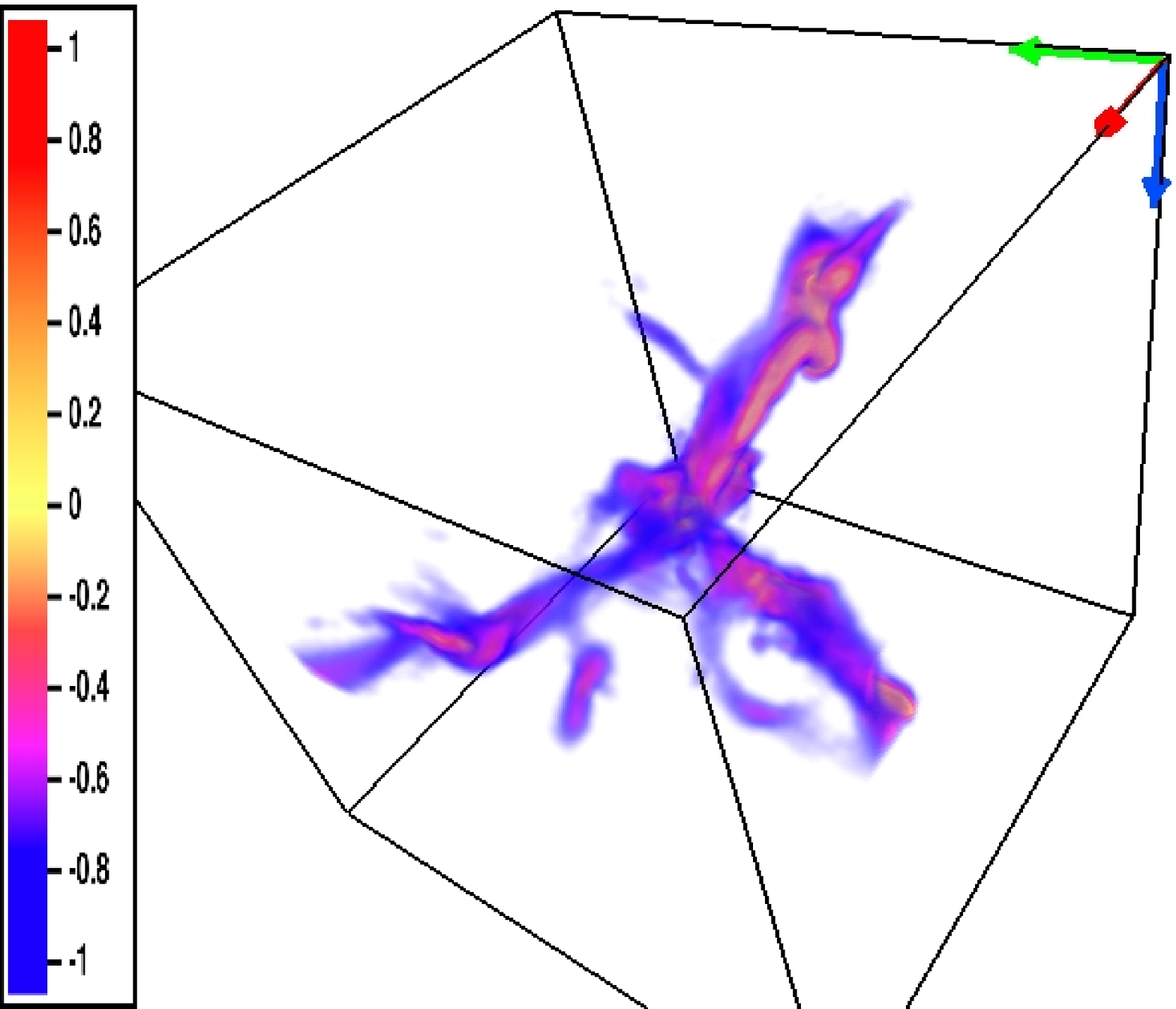}}\\ \fbox{
  \includegraphics[width=0.45\textwidth,trim = 0mm 0mm 0mm 0mm,
clip]{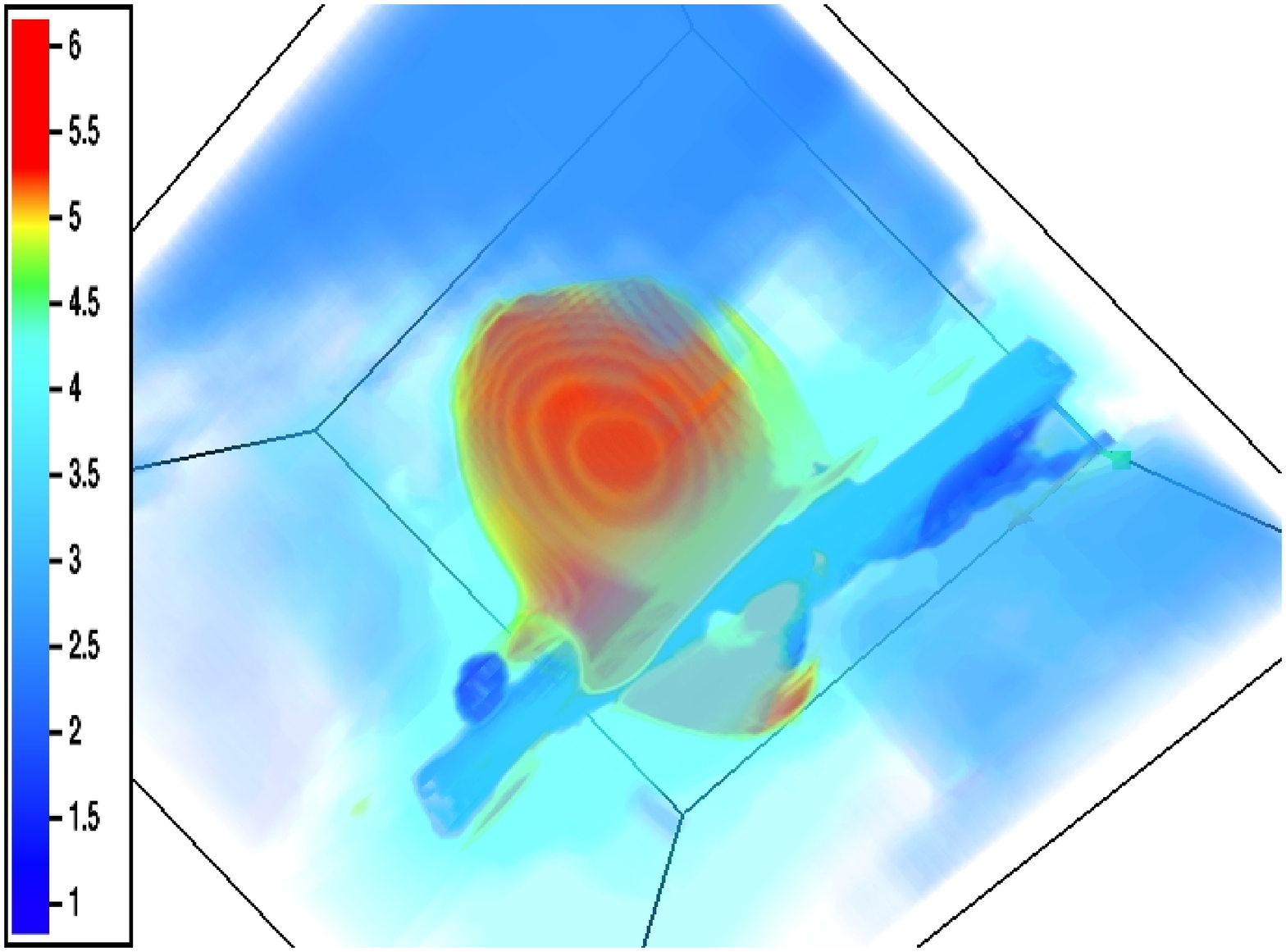}} \fbox{
    \includegraphics[width=0.45\textwidth,trim = 0mm 0mm 0mm 0mm,
clip]{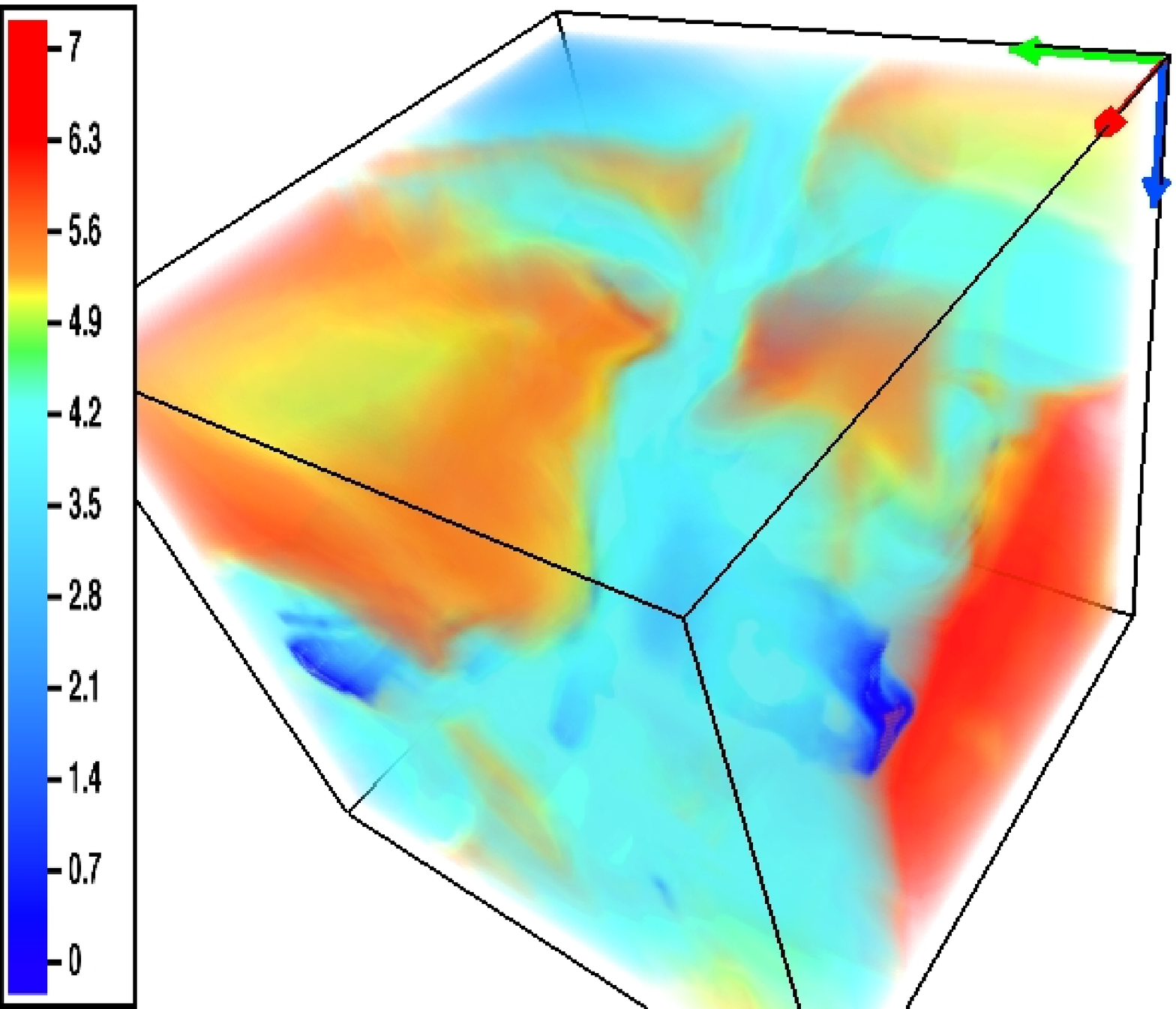}}
 \caption{3D visualisations (made with {\sc vapor}) at $z = 9$ to
illustrate the different gas phases in which we measure the different
modes of accretion and the outflows in the cooling (left) and feedback (right)
runs. The cube is centred on the main progenitor halo, extends out
to $r_{\rm vir}$ and was extracted on level 15 (equivalent resolution
= $40$pc). Top row: Rendering of the density for $\rho > 10$
atoms/cc showing the main galaxy and some satellites.  Middle row:
Rendering of the density in the range $0.1 < \rho <10$ atoms/cc for which we define our
filamentary accretion. Bottom row: Rendering of the temperature for gas with
$\rho < 0.1$ atoms/cc and $T<2\times10^{4}$K representing the cold diffuse phase (light-dark blue),
 and gas with $\rho < 10$ atoms/cc and $T>2\times10^{4}$K representing the warm and hot diffuse phases
(yellow-red). Note that gas with $T>2\times10^{5}$K
is shown in red for the cooling run (left) and orange in the feedback
run (right). Colour bars show density scale in $\log$ atoms/cc (top
and middle rows) and the temperature scale in $\log$ K (bottom
row). The $x$, $y$ and $z$ unit vectors are shown with red, green and
blue arrows, respectively.} \label{3dfluxcuts} \end{figure*}

Fig.~\ref{histofluxcuts} shows density-temperature histograms for gas
inside the virial radius in the cooling (top) and feedback (bottom)
runs.  The region of cold ($T<10^{4}$K), dense ($\rho \gtsim 10$
atoms/cc) gas in both phase diagrams can be attributed to gas that has
condensed into galaxies, either the main galaxy or its subhaloes. The
cooling is more efficient in the feedback run due to the increased
metallicity following enrichment from SNe. This accounts for
the greater mass fraction of gas in this region at the lowest
temperatures and highest densities. The {\it edge} apparent here is
due to the polytrope used to prevent artificial fragmentation below
the Jean's scale (more details in the Appendix). A striking feature,
common to both the cooling and feedback runs, is the prevalence of gas
with $T\sim10^{4}$K. We identify this with filaments (see Fig.~\ref{3dfluxcuts}) 
defined as containing all gas with $T<2\times10^{4}$K, $0.1\le \rho \le 10$ atoms ${\rm cm}^{-3}$.

Since we are analysing high redshift {\tt ($z \geq 9$)} outputs, the virial temperature
is always below $T = 2\times10^{5}$K. While this temperature threshold
has been employed in other studies to successfully separate
shocked-heated gas from cold gas
\citep[e.g.][]{keres,bimodal_marenostrum}, it does not isolate all the
gas heated to temperatures of the order of T$_{\rm vir}$ and higher in our simulations (as is clear from
Fig.~\ref{histofluxcuts}). To account for the gas which has
experienced heating (i.e. has $T>2 \times10^{4}$K and so is not
`cold'), but has not reached the `standard' definition of hot used in the aforementioned 
literature ($T>2 \times 10^5$K), we 
define an additional category of {\it warm} gas which has $2\times
10^{4}{\rm K} \leq T \leq 2\times 10^{5}{\rm K}$.  In the bottom panel
of Fig.~\ref{histofluxcuts} (feedback run) we can clearly see a cloud
of hot ($T>T_{\rm vir} \approx 1.5 \times 10^5$K), diffuse ($\rho \ltsim 10$ atoms/cc) gas
which contains a significant fraction of the total gas mass. Perhaps
surprisingly, we see a similar feature in the top panel of
Fig.~\ref{histofluxcuts} for the cooling run, albeit one that makes a
less significant contribution to the total gas mass. 
It is predicted that infalling gas will shock on impact with the dense ISM if it has not passed through a shock 
at the virial radius (and so was accreted cold) (e.g. Birnboim \& Dekel 2003). The hot phase in the cooling run 
must be the result of shock-heating of this type, since there are no SNe to heat the gas in that simulation. 
However, since the mass in the hot phase in the cooling run is tiny compared to that in the filaments (see e.g. 
colours for the hot diffuse gas component at z=9 in Fig.~\ref{histofluxcuts} (top panel)), and since the cooling time of metal poor material 
at 10$^6$ K with a density comparable to that of the cold filaments ($\sim$ 1 at/cc) is around $\sim$ 1 Myr,
i.e. much larger than the time resolution of the simulation on the level of refinement reached at these densities 
($\sim$ 0.1 Myr), we 
conclude it is primarily the cold diffuse gas that is being shock heated. Indeed, if the filamentary material 
were being continuously shocked one would expect to see a prominent hot phase with densities equaling or exceeding
 that of the filaments due to compression of the gas as it passes through the shock. This scenario is clearly 
excluded by Fig.~\ref{histofluxcuts}, which shows that only a tiny mass fraction of such high density hot material exists.  
We undertake a more detailed study of how and where shock-heating occurs in Powell et al, (in prep).

\begin{table}
\centering \begin{tabular}{|p{0.1\textwidth}|p{0.06\textwidth}|p{0.06\textwidth}|p{0.06\textwidth}|p{0.06\textwidth}|}
\hline {\bf Category} & {\bf $T_{\rm min}$ (K)} & {\bf $T_{\rm max}$
(K)} & {\bf $\rho_{\rm min}$ (at/cc)} & {\bf $\rho_{\rm max}$
(at/cc)}\\ &&&&\\ \hline clumpy &0 & $\infty$ & 10 & $\infty$\\ \hline
filaments & 0& $2\times10^{4}$ & 0.1 & 10\\ \hline cold diffuse & 0&
$2\times10^{4}$ &0 & 0.1\\ \hline warm diffuse & $2\times10^{4}$ &
$2\times10^{5}$ &0 & 10\\ \hline hot diffuse &
$2\times10^{5}$&$\infty$ & 0& 10\\
\hline \end{tabular} \caption{Summary of the temperature and density
thresholds used to distinguish different components of the gas inside
the virial radius of the main
progenitor.} \label{catsummary} \end{table}

The categories we have created are summarised in Table
\ref{catsummary} and in Fig.~\ref{histofluxcuts} we show how the thresholds divide up the $\rho-T$
phase diagrams at $z=9$ for the cooling (top) and feedback (bottom)
runs. Fig.~\ref{3dfluxcuts} shows visualisations of these categories
created using 3D visualisation software {\sc vapor}
(www.vapor.ucar.edu). The results of a $\rho > 10$ atoms ${\rm cm}^{-3}$ cut
on the density appears in the top row; the central galaxy and several satellites
are clearly defined by the clumpy category in both the cooling and
feedback runs. The gas that meets our definition of the filaments
($T<2\times10^{4}$K, $0.1\le \rho\le 10$ atoms ${\rm cm}^{-3}$) is
illustrated with a rendering of the density in Fig.~\ref{3dfluxcuts}
(middle row). The filamentary category is successful in isolating the
filaments in both the cooling (left) and feedback (right) runs. It
appears that the filaments are thinner and less well-defined in the
feedback run, so one already wonders whether this impacts 
the accretion rate. This image also shows an interesting feature of
the structure of the filaments; they become twisted and loop around
the central galaxy. This is particularly evident in the cooling run (middle row,
left) and is linked to the way these filaments dissipate their energy as they find
their way to the centre of the halo. 

Fig.~\ref{3dfluxcuts}, bottom row shows 3D renderings of the
temperature of the diffuse gas ($\rho \ltsim 10$ atoms/cc). It is
immediately apparent that the virial sphere contains significantly
more hot gas in the feedback run (right) than the cooling run
(left). Approximately $50$ per cent of the virial sphere is filled
with hot gas in the feedback run, yet only $5$ per cent in the cooling
run. This large covering factor of hot gas even in haloes below the
shock mass threshold could explain recent observations of
high-redshift galaxies in which outflows, rather than cold inflows (as
expected for cold-accretion dominated haloes) are observed \citep{steidel_etal_2010,FG_keres_2010}. 
The hot gas in the cooling run (bottom row, left) surrounds the wound up filaments 
and originates from the filaments dissipating their kinetic energy
as they stream through the cold, diffuse medium on their approach to the centre.
We believe this loss of energy in a shock in combination with gravitational focusing
causes the filaments to wrap around the galactic disk. We explore the details of
this in Powell et al, in prep.

\begin{figure*}
  \includegraphics[angle=90,scale=0.5,trim = 0mm 0mm 0mm 0mm,
clip]{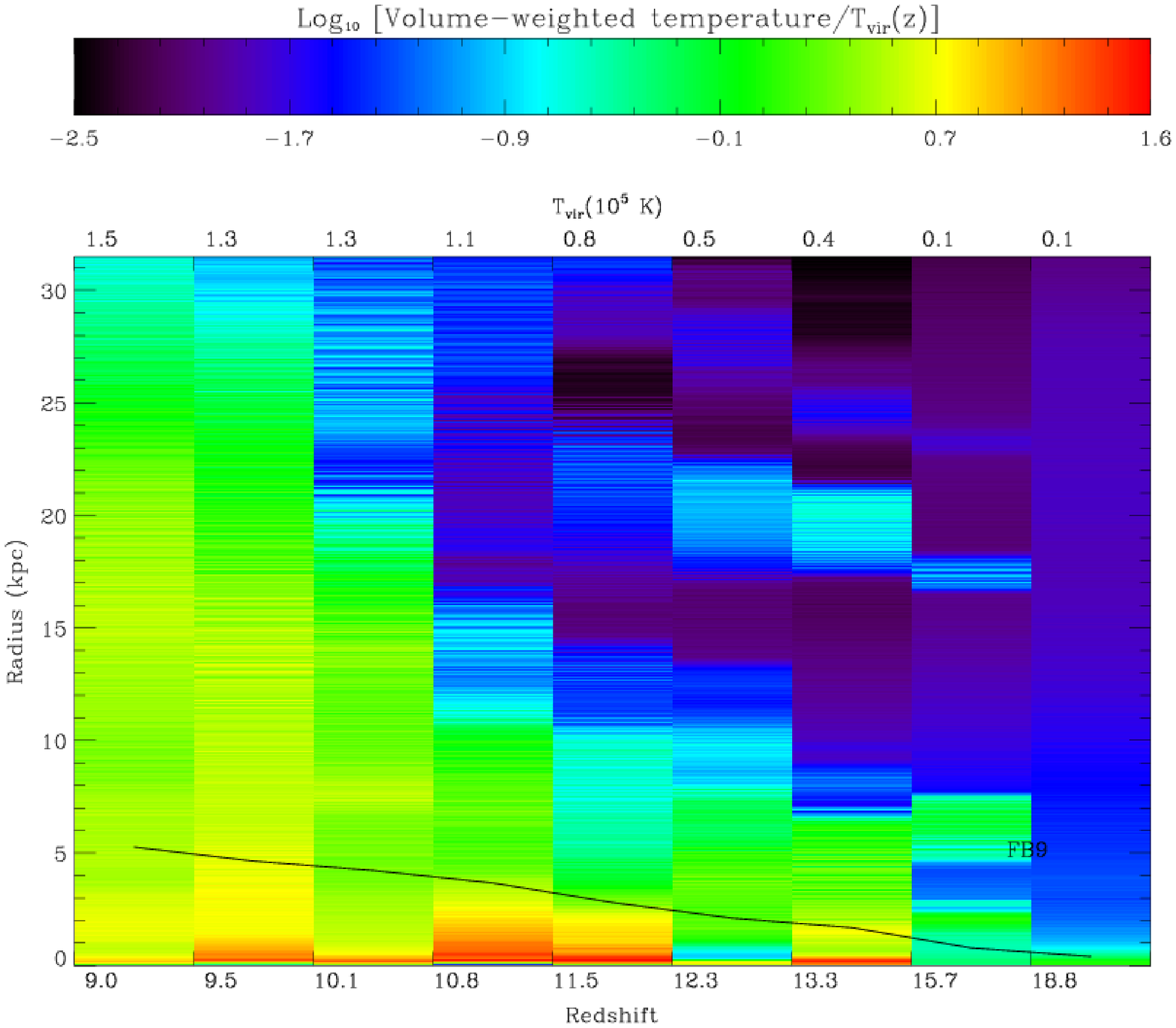}
  \caption{Volume-weighted temperature as a fraction of $T_{\rm vir}$ in
10 pc radial bins in a sphere of radius $31.8$kpc physical ($6r_{\rm
vir}$ at $z=9$) centred on the halo of the main progenitor, versus redshift for the
feedback run. The solid line shows $r_{\rm vir}$ of the main halo at
each redshift. The volume-weighted temperature is above $T_{\rm vir}$
out to $\sim 5-6r_{\rm vir} $ at $z < 14$ indicating that the wind has escaped the
potential well of the main halo.}
  \label{volradhisto_fb9_1d} \end{figure*}

\section{The galactic wind}\label{sec:hotgas}

We identify the galactic wind illustrated in Fig.~\ref{tempmaps} (top, right) with
both the warm and hot diffuse phases defined in the previous
section. It is clear from the phase diagrams in Fig.~\ref{histofluxcuts} that these phases contain significantly more mass
in the feedback run than in the cooling run and the only possible culprit is SNe feedback. We quantify the
extent of the galactic wind by examining the time evolution of the
volume-weighted temperature (represented by the colour-coding) as a function of radius in
Fig.~\ref{volradhisto_fb9_1d}. The temperature is calculated in 10 pc physical radial bins within a sphere of radius 
$31.8$ kpc physical ($6r_{\rm vir}$ at $z=9$) centred on the main
progenitor halo at each redshift.  By $z=9$ the hot gas in the
feedback run already extends out to $\sim 5-6 r_{\rm vir}$.
We note that such a wind `bubble' extent is in fair agreement 
 with the estimate derived semi-analytically by \citet{furlanetto_loeb_2003} (their Fig 1, top right
  panel) using a spherical thin-shell approximation. This is rather
  surprising in light of all the simplifying assumptions that go into
  their modelling. For instance, they assume that star formation and
  feedback only occur in galaxies when they cross the cooling
  threshold {\it and} are captured by the main 
 progenitor halo. In contrast star formation in our simulated galaxies
 has started earlier in most cases and winds have had time to develop prior to
mergers. The similar sizes of the wind in semi-analytic calculations
arise from the fact that Furlanetto and Loeb's key
(incorrect) assumption that all the gas has collapsed in a dense medium before triggering a
starburst and that therefore the wind's spherical thin shell expands into an
empty halo, actually has consequences very similar to those in the scenario
which unfolds in our feedback simulation. In the latter, most of the halo is
also devoid of dense material, but this is because this gas is
preferentially channelled along narrow filaments which continuously fuel 
star formation, regardless of whether galaxy mergers are taking place or
not. Winds from our simulated galaxies expand perpendicularly to this
filamentary network into the low pressure intrahalo medium and
depending on their sizes, galactic wind bubbles can merge before 
or after their host haloes merge.

\subsection{Development of the wind}

\begin{figure*} \centering
    \includegraphics[width=0.24\textwidth,trim = 0mm 0mm 34mm 5mm,
clip]{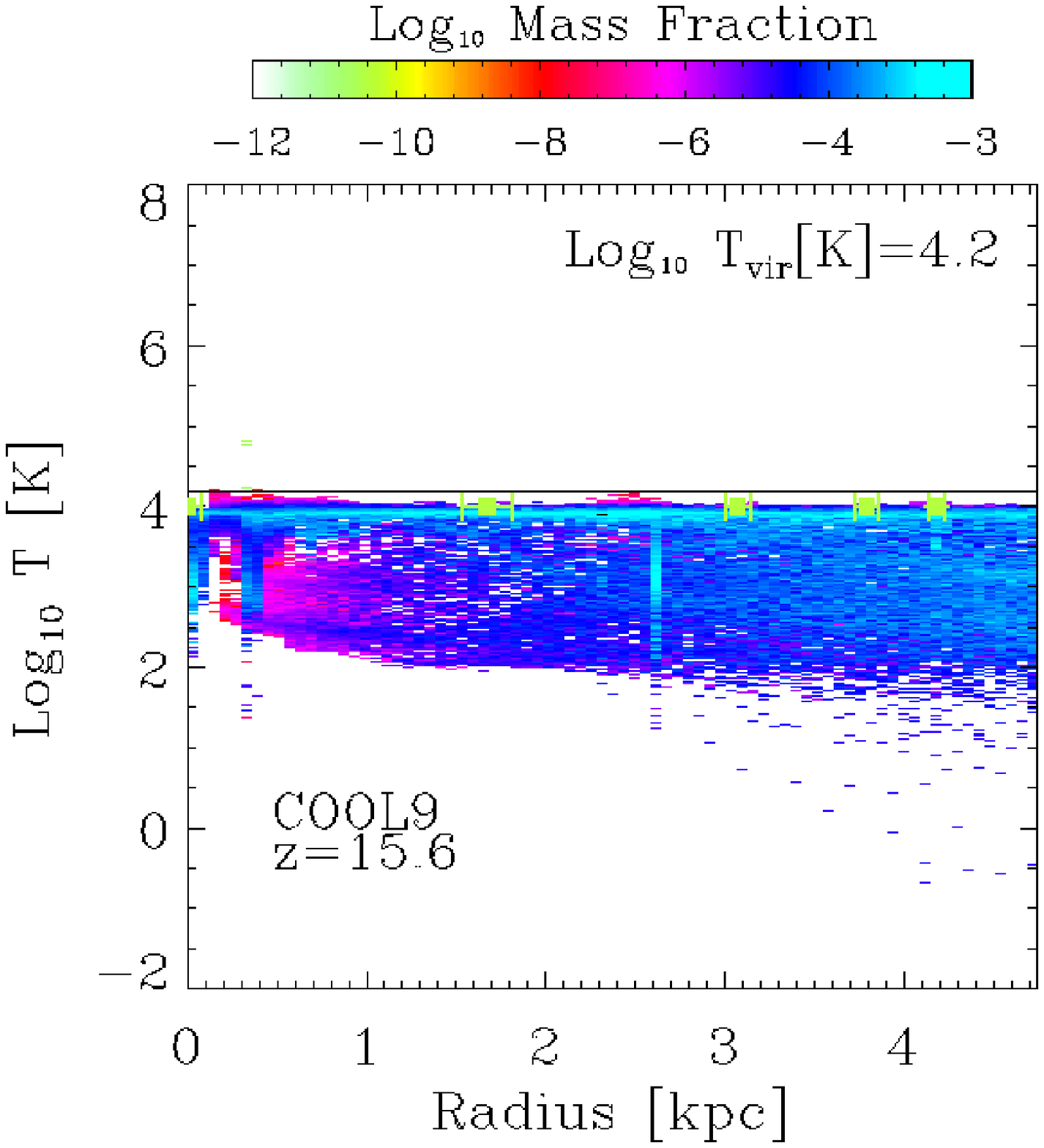}
         \includegraphics[width=0.24\textwidth,trim = 0mm 0mm 34mm
5mm, clip]{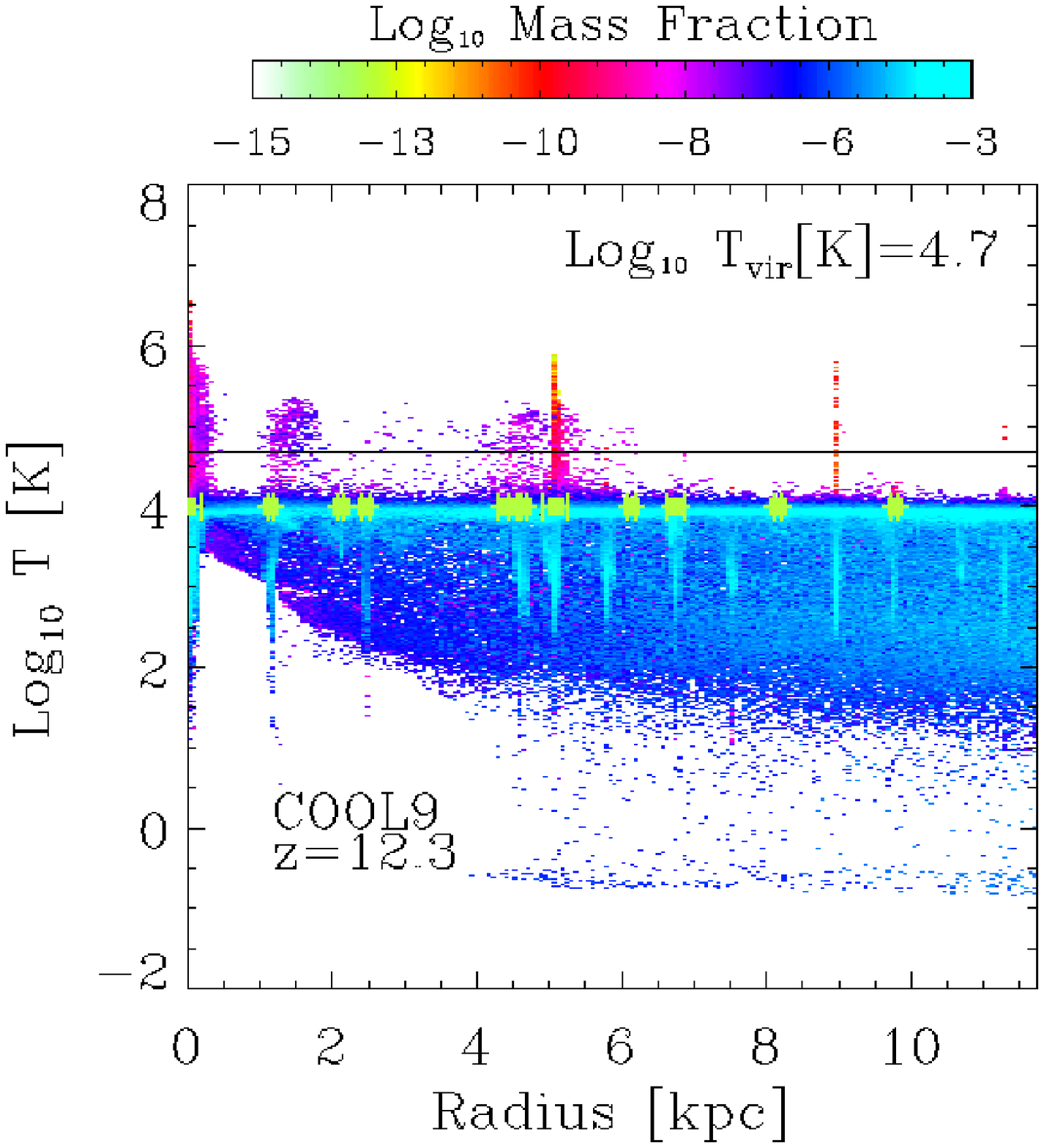}
             \includegraphics[width=0.24\textwidth,trim = 0mm 0mm 34mm
5mm, clip]{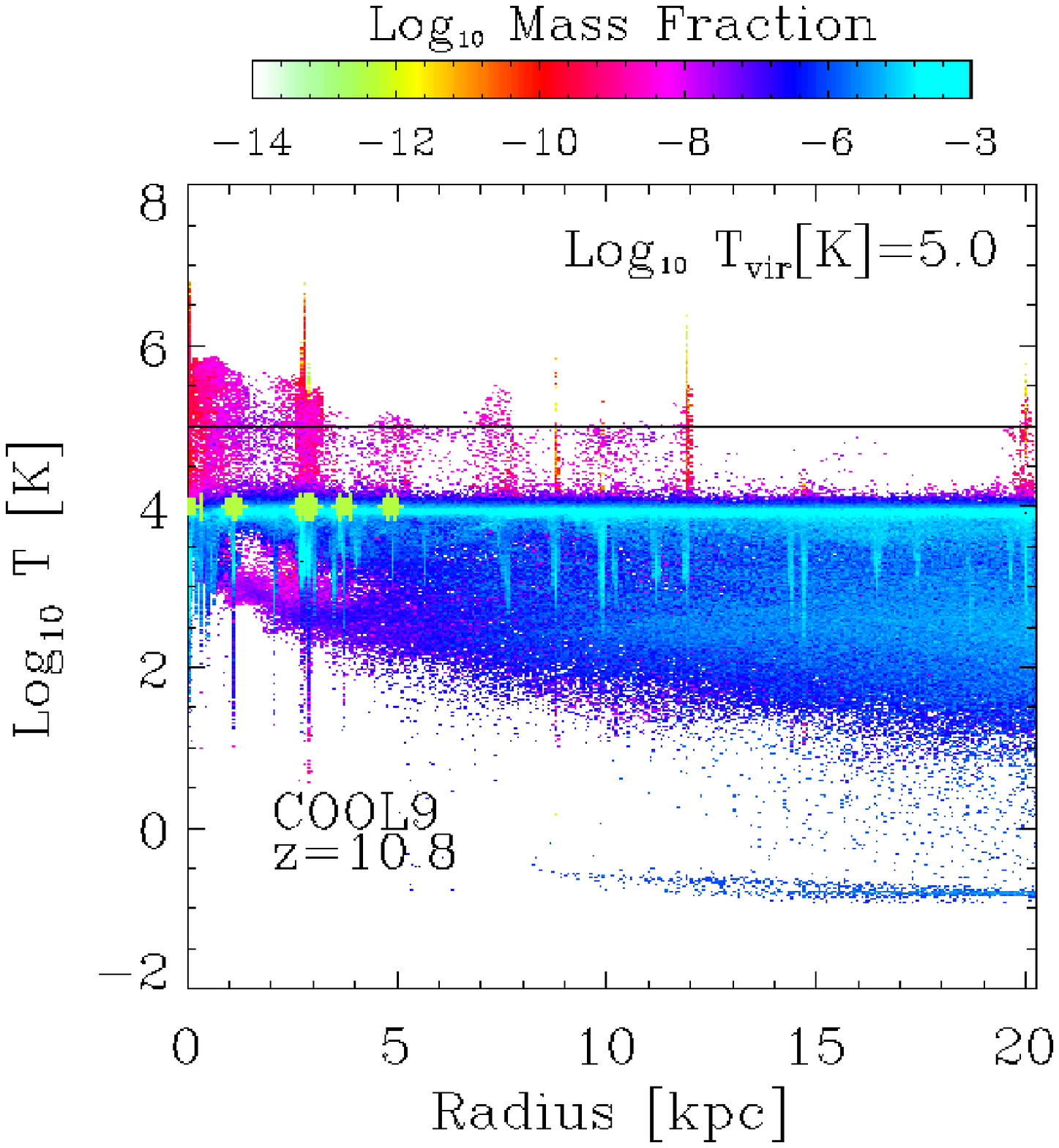}
                   \includegraphics[width=0.24\textwidth,trim = 0mm
0mm 34mm 5mm, clip]{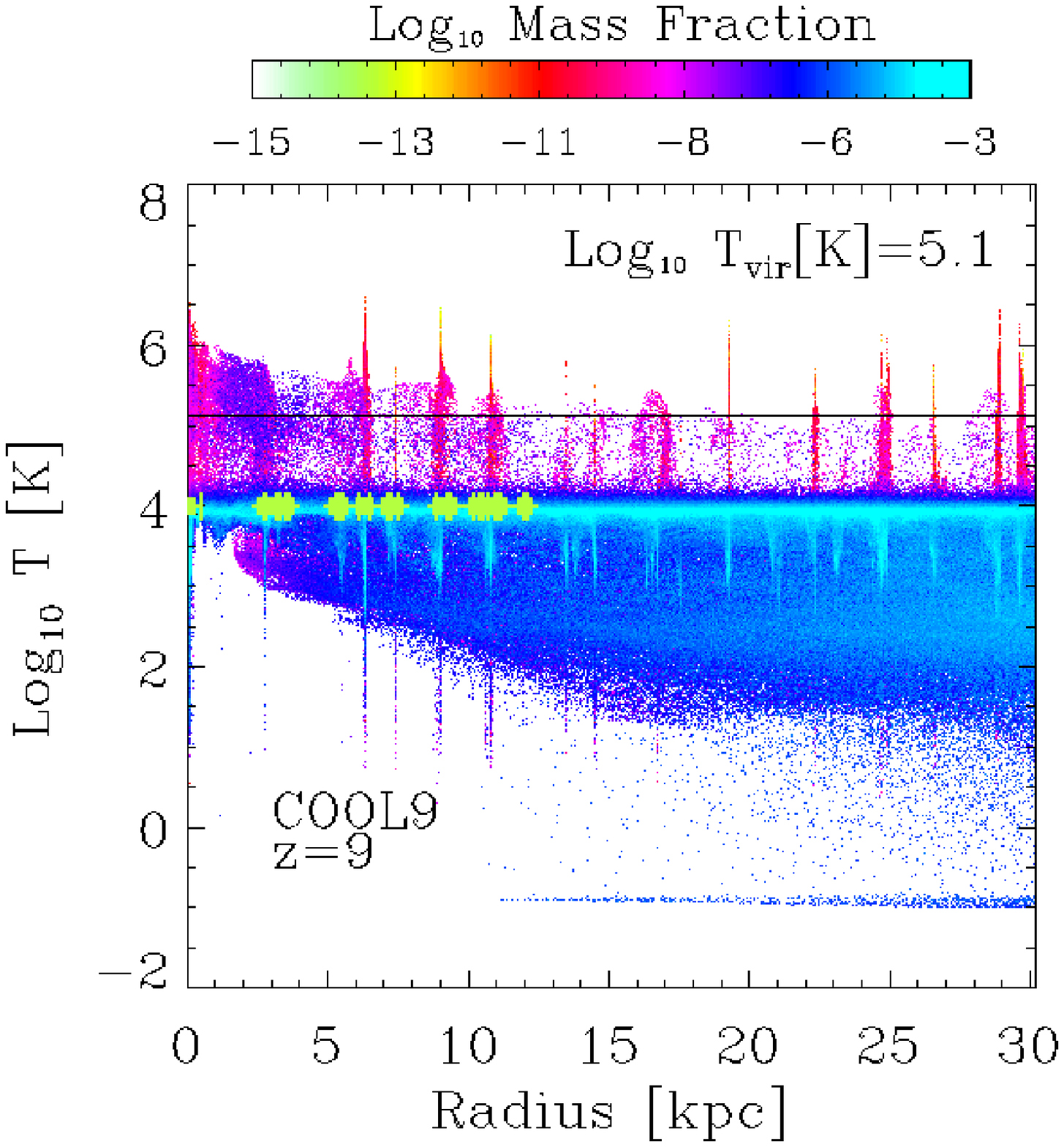}

    \includegraphics[width=0.24\textwidth,trim = 0mm 0mm 35mm 5mm,
clip]{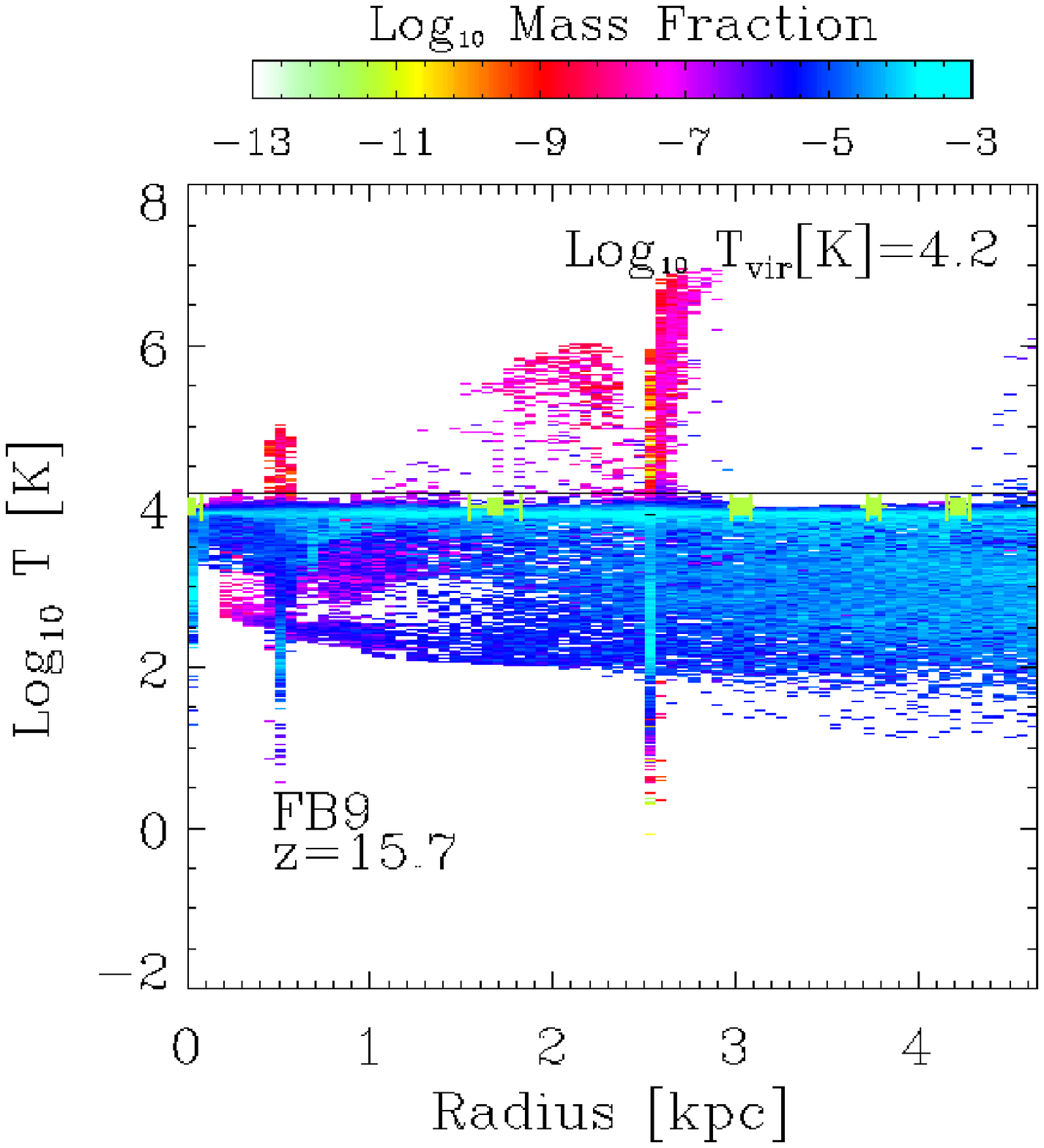}
        \includegraphics[width=0.24\textwidth,trim = 0mm 0mm 35mm 5mm,
clip]{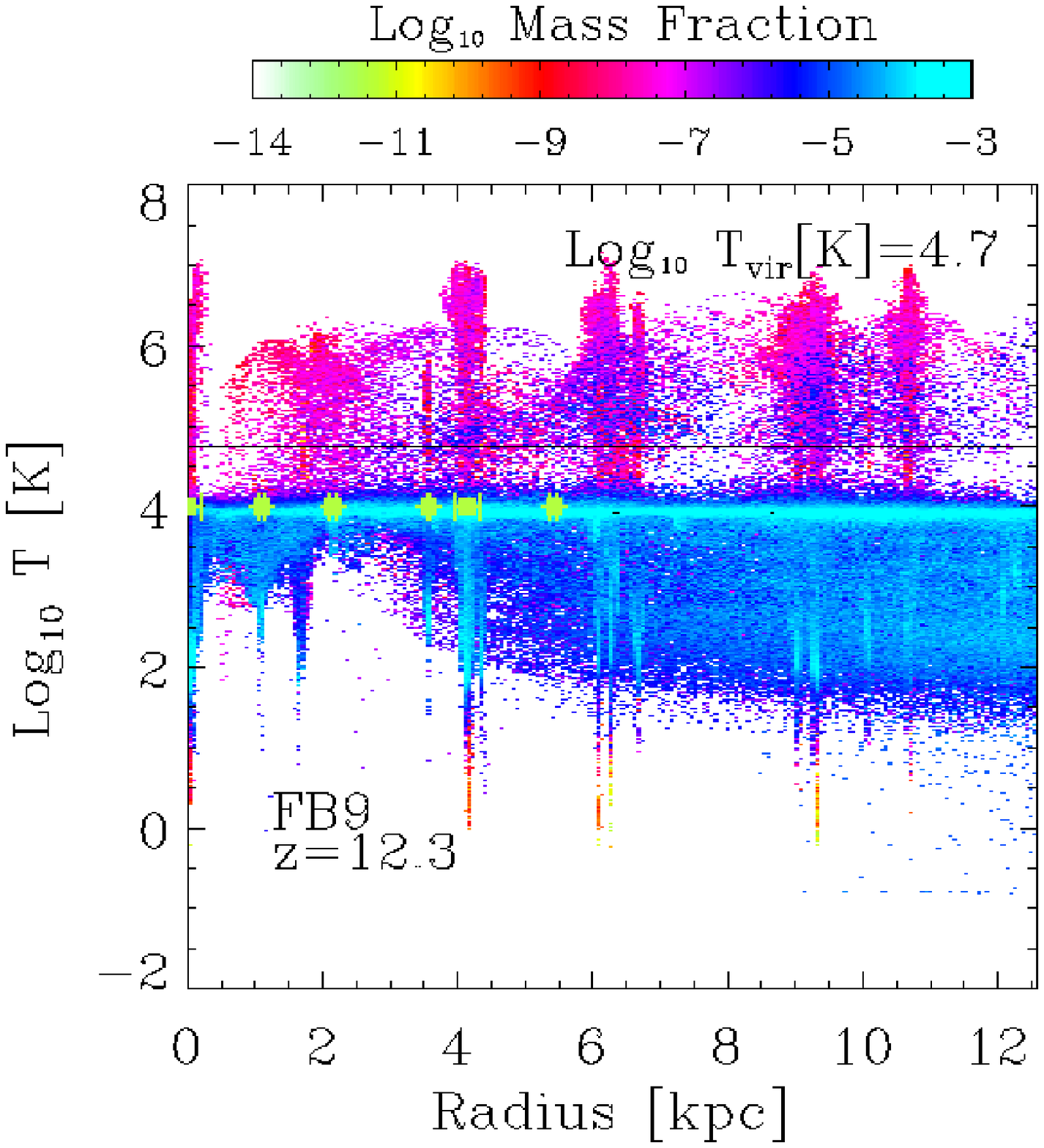}
            \includegraphics[width=0.24\textwidth,trim = 0mm 0mm 35mm
5mm, clip]{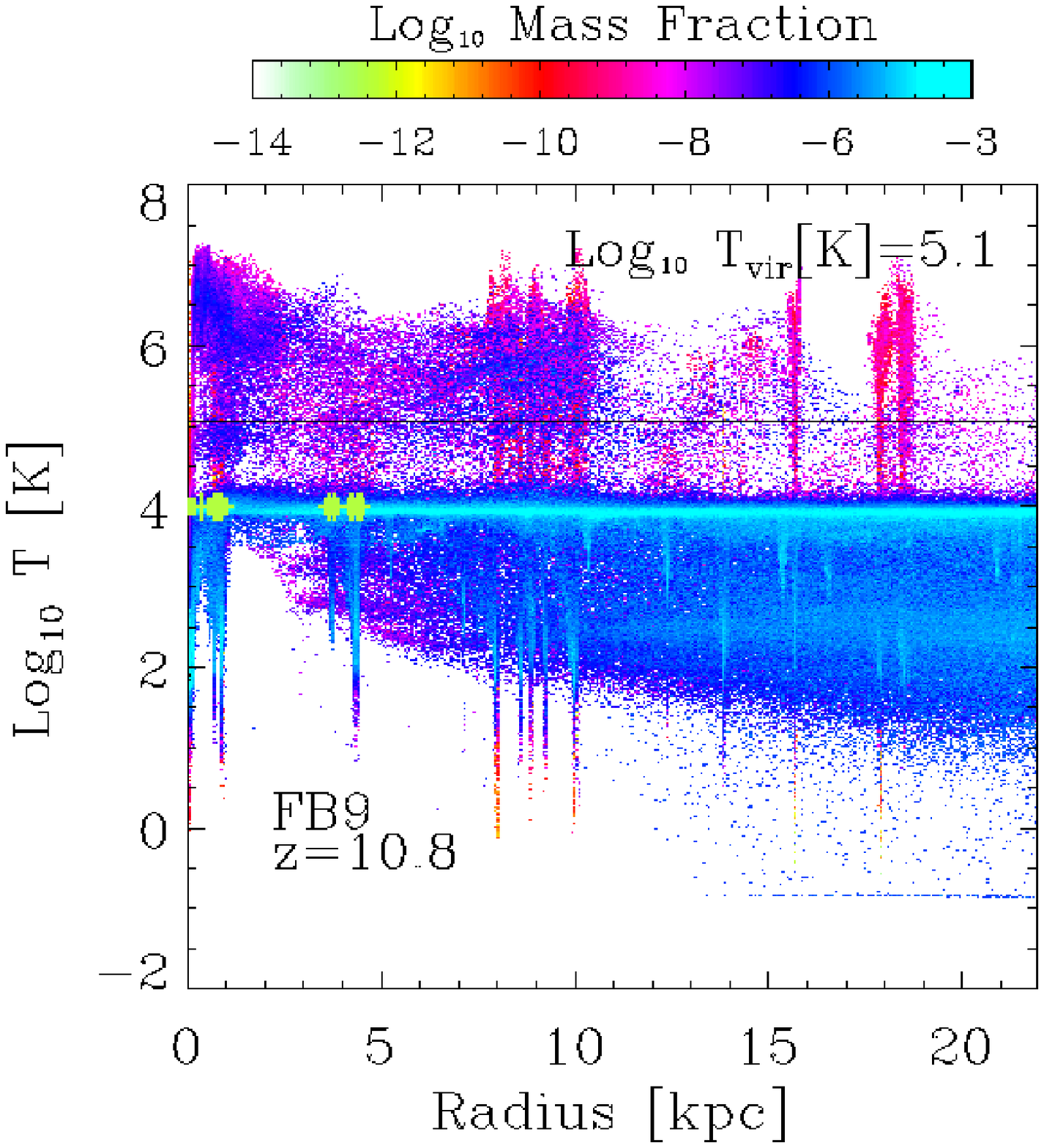}
                  \includegraphics[width=0.24\textwidth,trim = 0mm 0mm
35mm 5mm, clip]{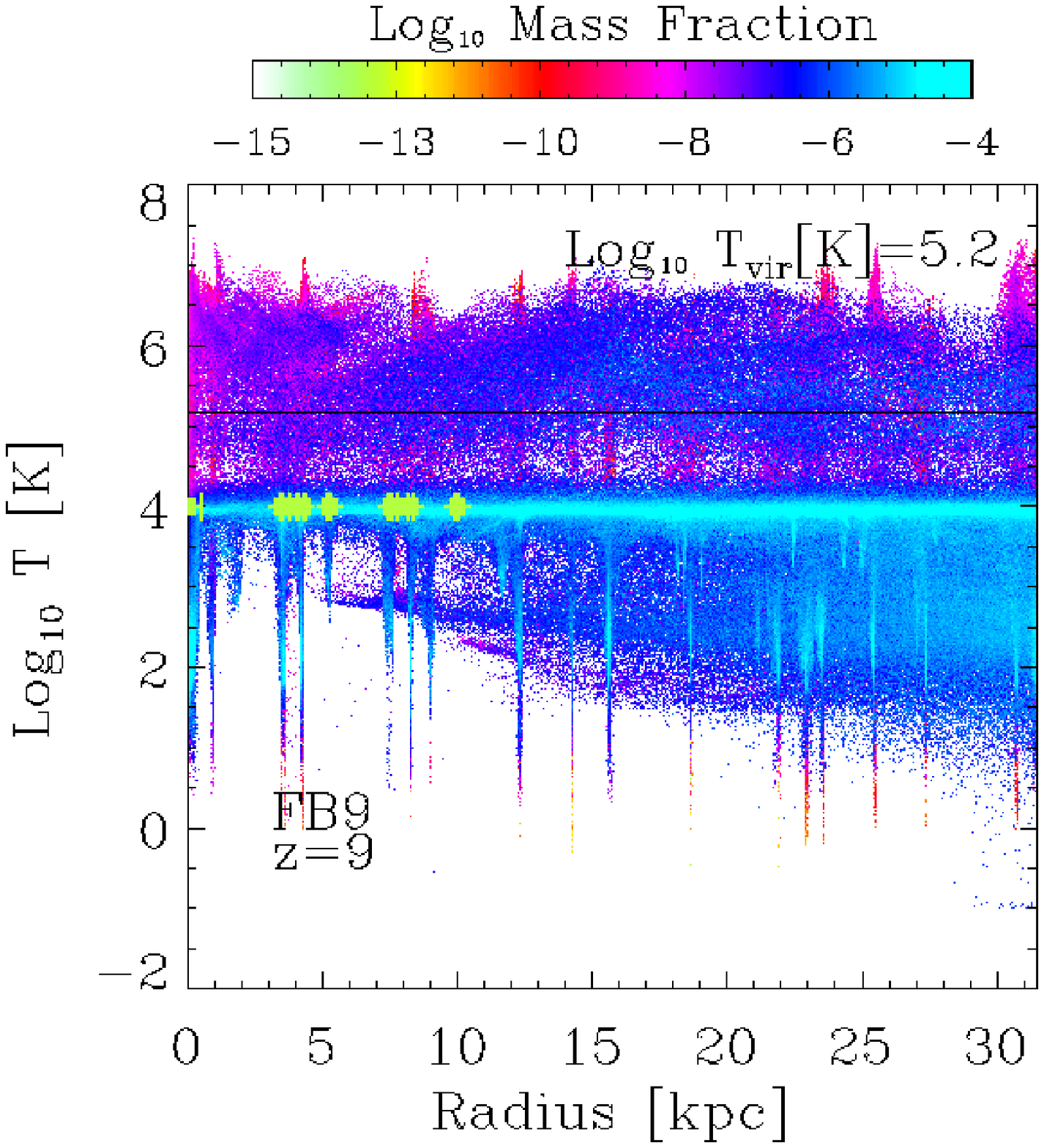}
 \caption{Time evolution of the temperature-radius histograms for all
gas out to $6r_{\rm vir}$ in the cooling (top row) and feedback
(bottom row) runs. The colours show the logarithmic fraction of the total gas mass
within each bin. The position of the main halo (by definition at
radius$=0$) and its subhaloes are overplotted with light green squares
at $T=10^{4}$K, with bars to indicate $0.1 r_{\rm vir}$ for
each. A solid black horizontal line marks the virial temperature,
specified in the top right corner of each
plot.} \label{radhisto_fb9} \end{figure*}

\begin{figure*} \centering
    \includegraphics[width=0.24\textwidth,trim = 0mm 0mm 35mm 5mm,
clip]{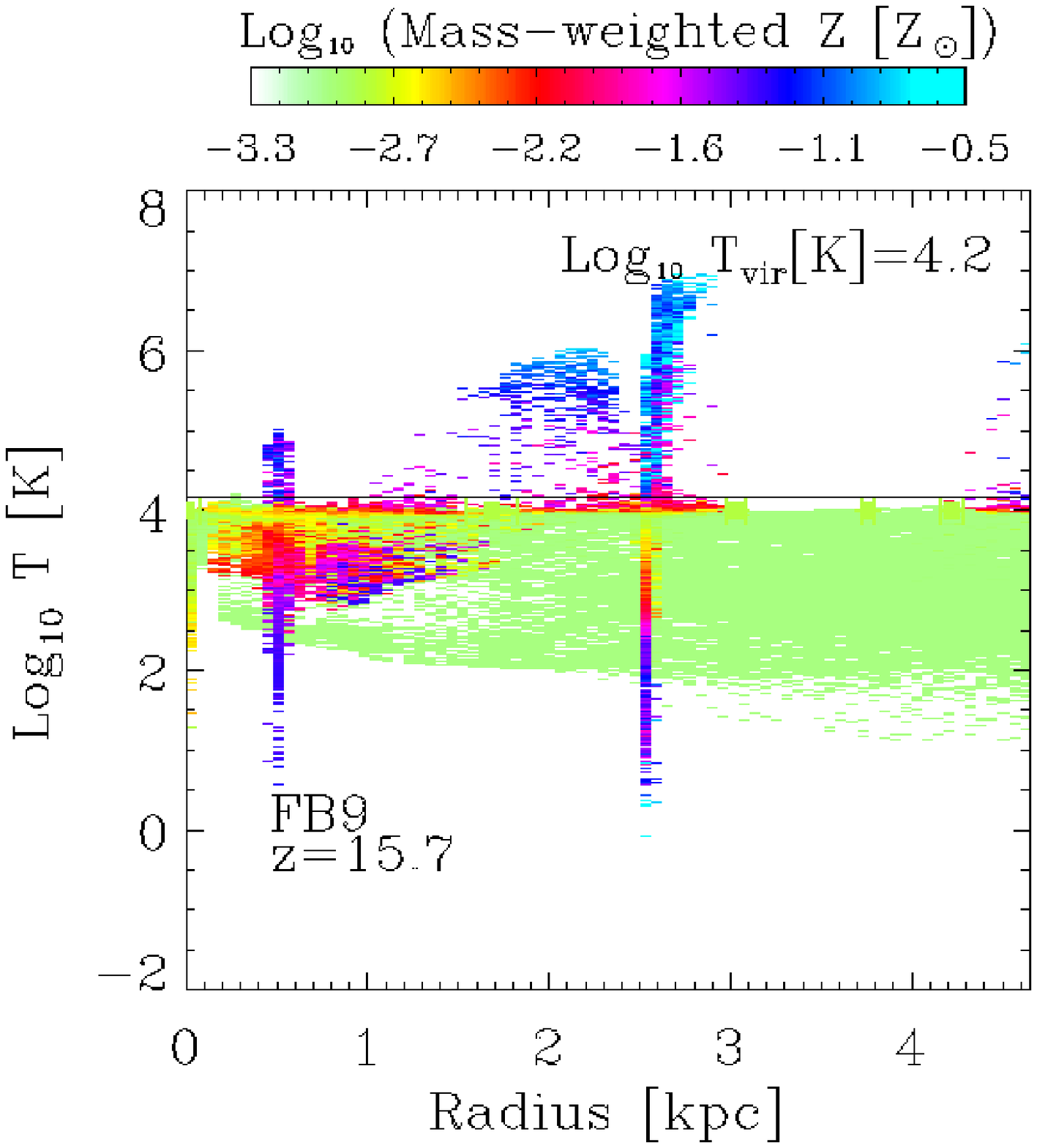}
        \includegraphics[width=0.24\textwidth,trim = 0mm 0mm 35mm 5mm,
clip]{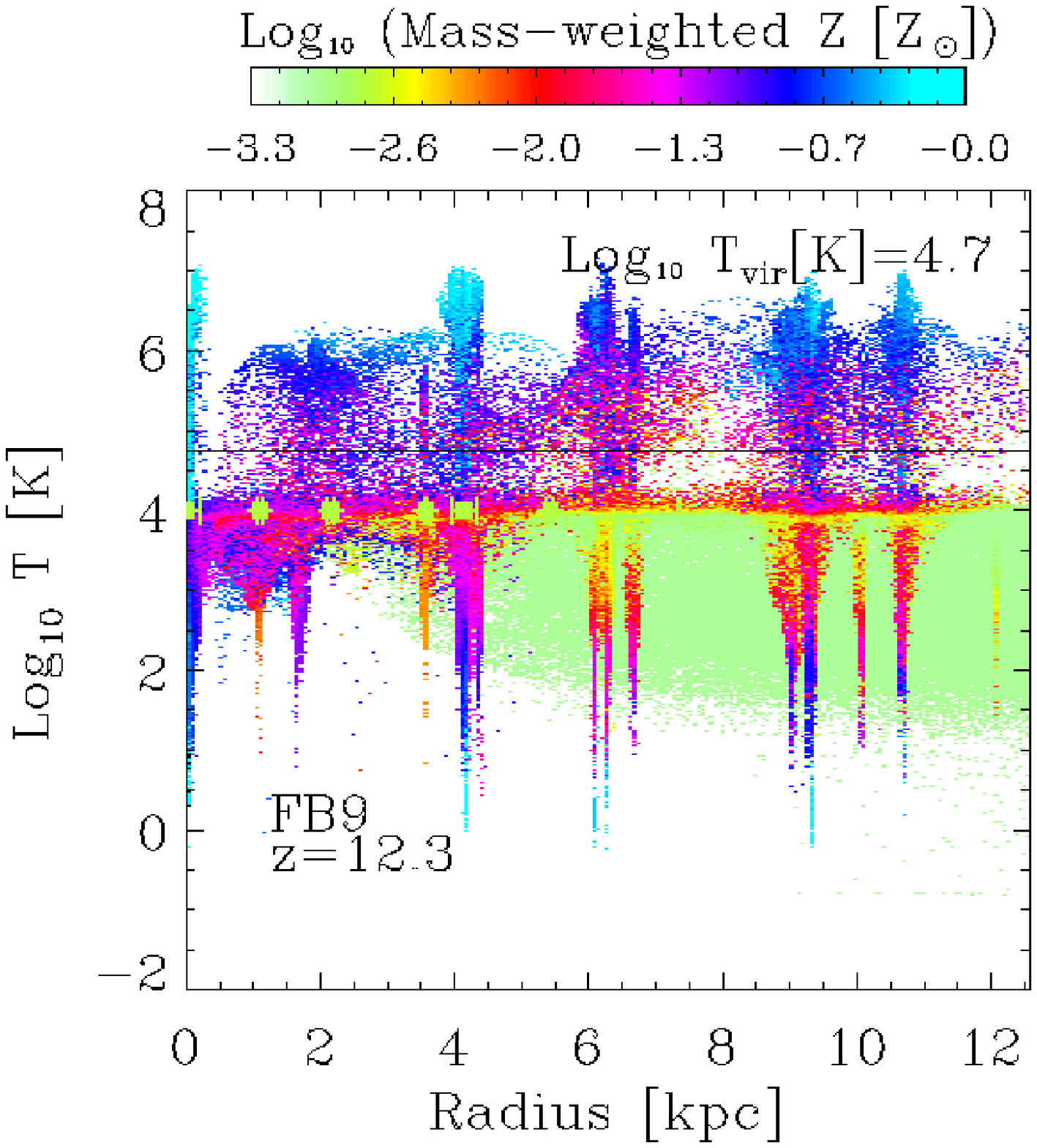}
            \includegraphics[width=0.24\textwidth,trim = 0mm 0mm 35mm
5mm, clip]{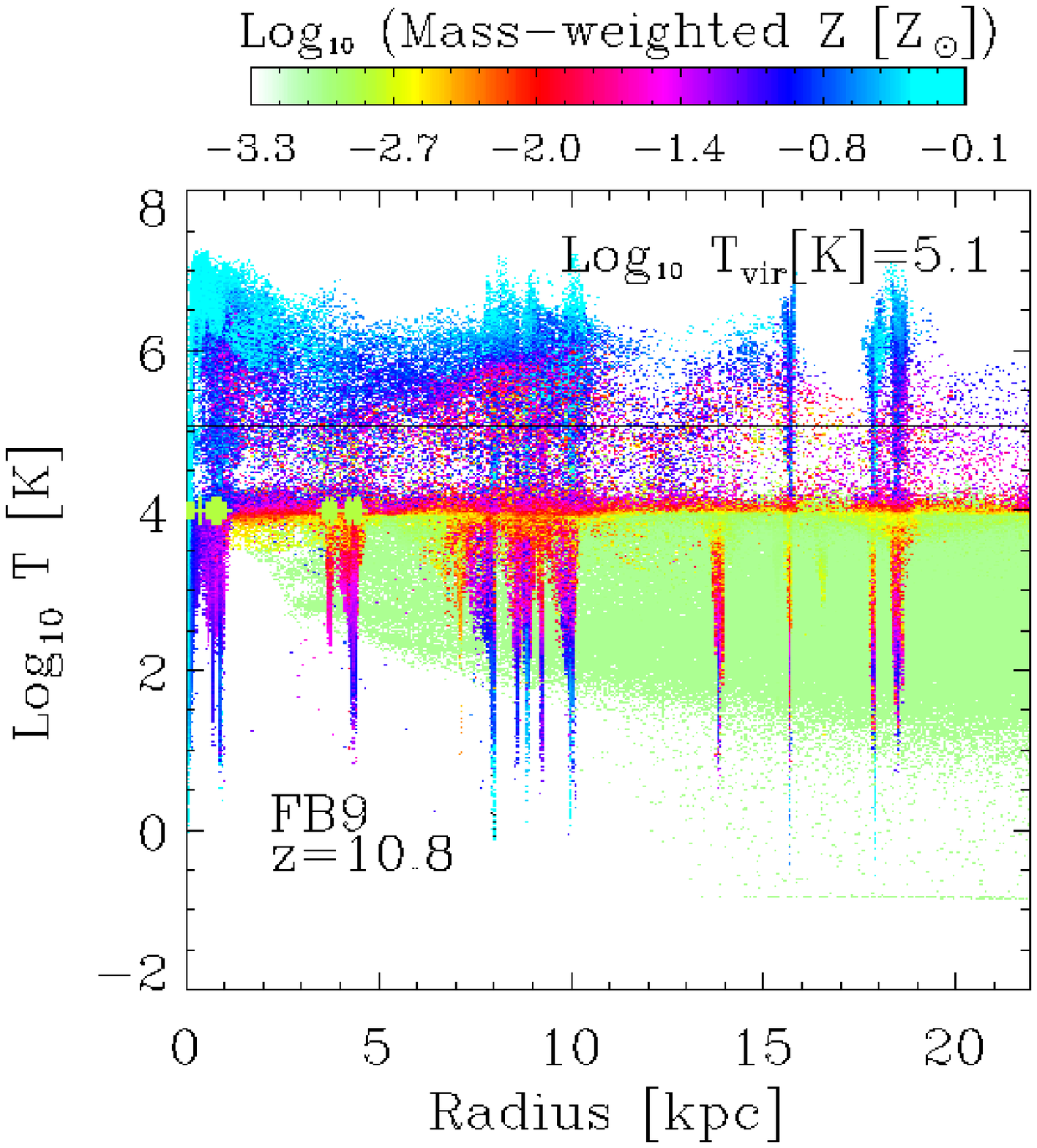}
                  \includegraphics[width=0.24\textwidth,trim = 0mm 0mm
35mm 5mm, clip]{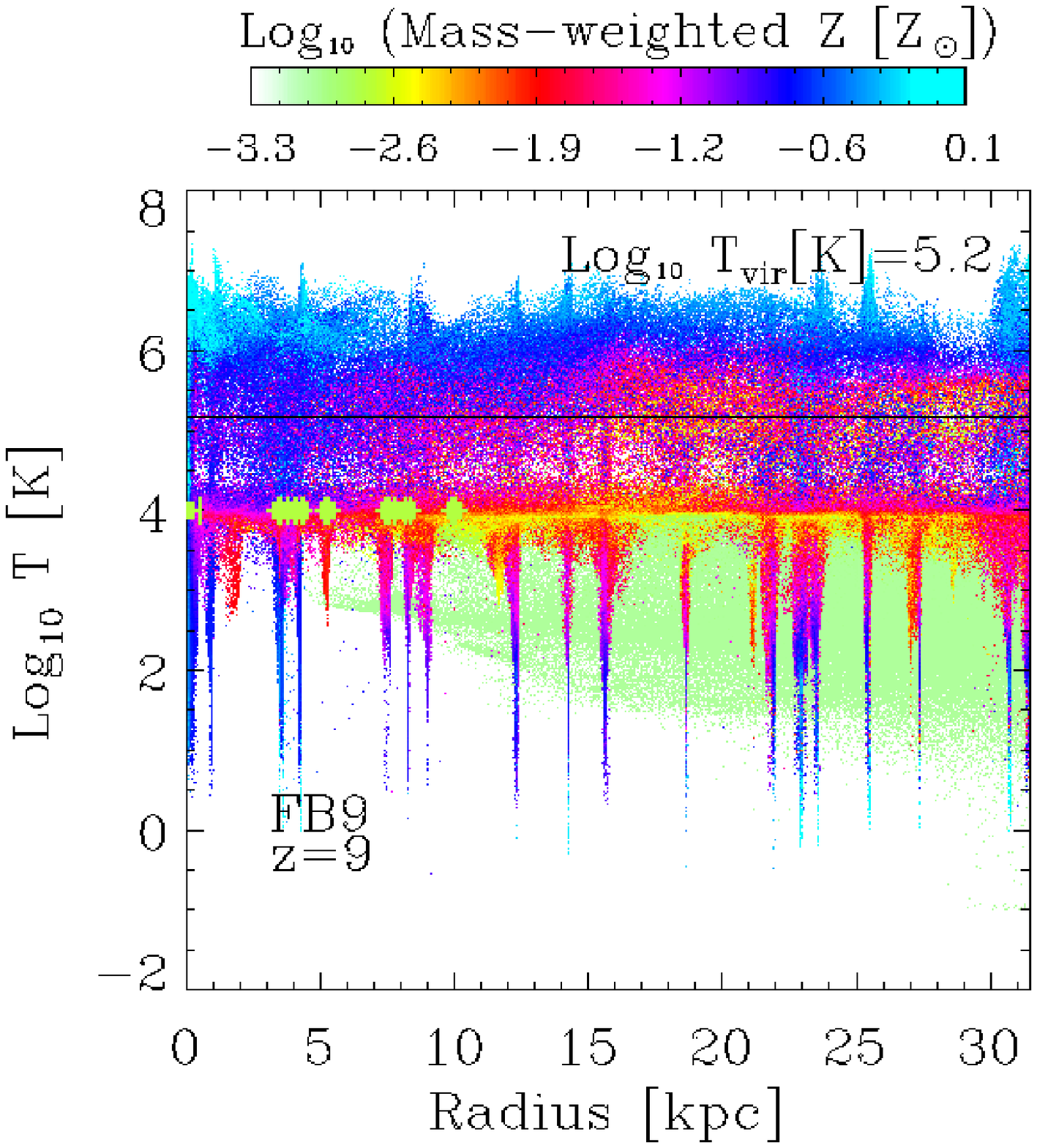}
 \caption{Time evolution of the temperature-radius histograms for all gas  out to $6r_{\rm vir}$
 in the feedback run. The colours show the
average metallicity in each bin. The position of the main halo
(by definition at radius$=0$) and its subhaloes are overplotted with
light green squares at $T=10^{4}$K, with bars to indicate $0.1 r_{\rm
vir}$ for each.} \label{metalradhisto_fb9} \end{figure*}

In order to demonstrate how this far-reaching wind develops and to
corroborate the picture sketched out in the previous section, we examine
how the gas inside $r_{vir}$ is distributed both in temperature and
radius in Fig.~\ref{radhisto_fb9} for the cooling (top row) and
feedback (bottom row) runs. For each redshift output, we measure the temperature distribution in spherical
shells of radial width 10 pc physical centred on the main progenitor. The colour scale shows the mass fraction 
within each temperature-radius bin and the virial temperature at each time is
indicated with a horizontal line.  As we have already seen in Fig.~\ref{histofluxcuts},
most of the gas mass in both the
cooling and feedback runs is concentrated at $T \sim 10^{4}$K at all
radii. Gas at this temperature spans a large range of densities
(approximately $0.01 - 100$ atoms/cc), however the main mass
contribution is from the filaments (see Section \ref{sec:filwind} for
more details). The most striking difference between the runs is the
prevalence of hot gas ($T>T_{\rm vir}$) in the feedback run (bottom row) 
at all radii, in stark
contrast with the cooling run (top row); this gas constitutes the galactic
wind.

There are many `spikes' in temperature in both the cooling and
feedback runs, which correspond to the gas in subhaloes or other
haloes beyond the virial radius of the main halo. We demonstrate this
explicitly for subhaloes within $r_{\rm vir}$ of the main halo, by overplotting squares
at $T=10^{4}$K, to indicate the position of the centres of
the subhaloes with respect to the centre of the main halo (which is at $r=0$, by
design). The horizontal error bars indicate $0.1 r_{\rm vir}$ for
each. The low temperature end of the spike indicates where gas has
cooled and condensed within the subhaloes. 

The high temperature ends of these spikes are of particular interest
in the context of the galactic wind. In Fig.~\ref{metalradhisto_fb9},
we show the $T-r$ diagram from Fig.~\ref{radhisto_fb9}, but now the
colour scale indicates the mass-weighted metallicity. The high
mass-weighted metallicity we measure in the high-temperature spikes
corresponding to (sub)haloes shows that this hot gas results from
SNe exploding in the galaxies hosted by these haloes. By
examining the time sequence of plots for the feedback run
(Fig.~\ref{radhisto_fb9}, bottom row and
Fig.~\ref{metalradhisto_fb9}), we can see that these plumes of hot,
metal-enriched gas (seen particularly clearly at $z=12.3$, second
panel of bottom row of Fig.~\ref{radhisto_fb9} and Fig.~\ref{metalradhisto_fb9}) 
gradually build up into a continuous distribution of hot gas at
all radii, forming the galactic wind.

\begin{figure*} \centering
      \includegraphics[width=0.24\textwidth,trim = 10mm 5mm 10mm 0mm,
clip]{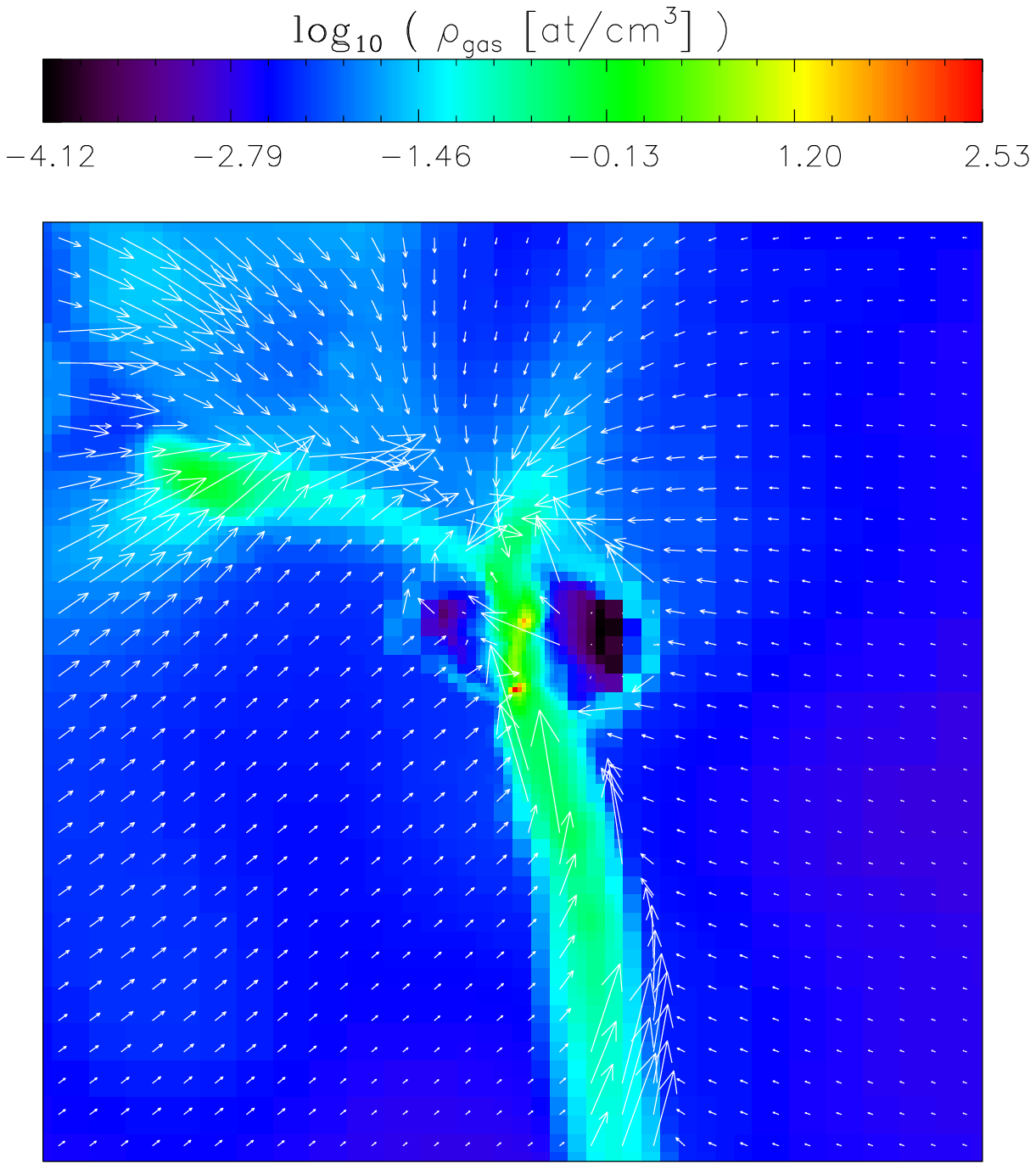}
          \includegraphics[width=0.24\textwidth,trim = 10mm 5mm 10mm
0mm, clip]{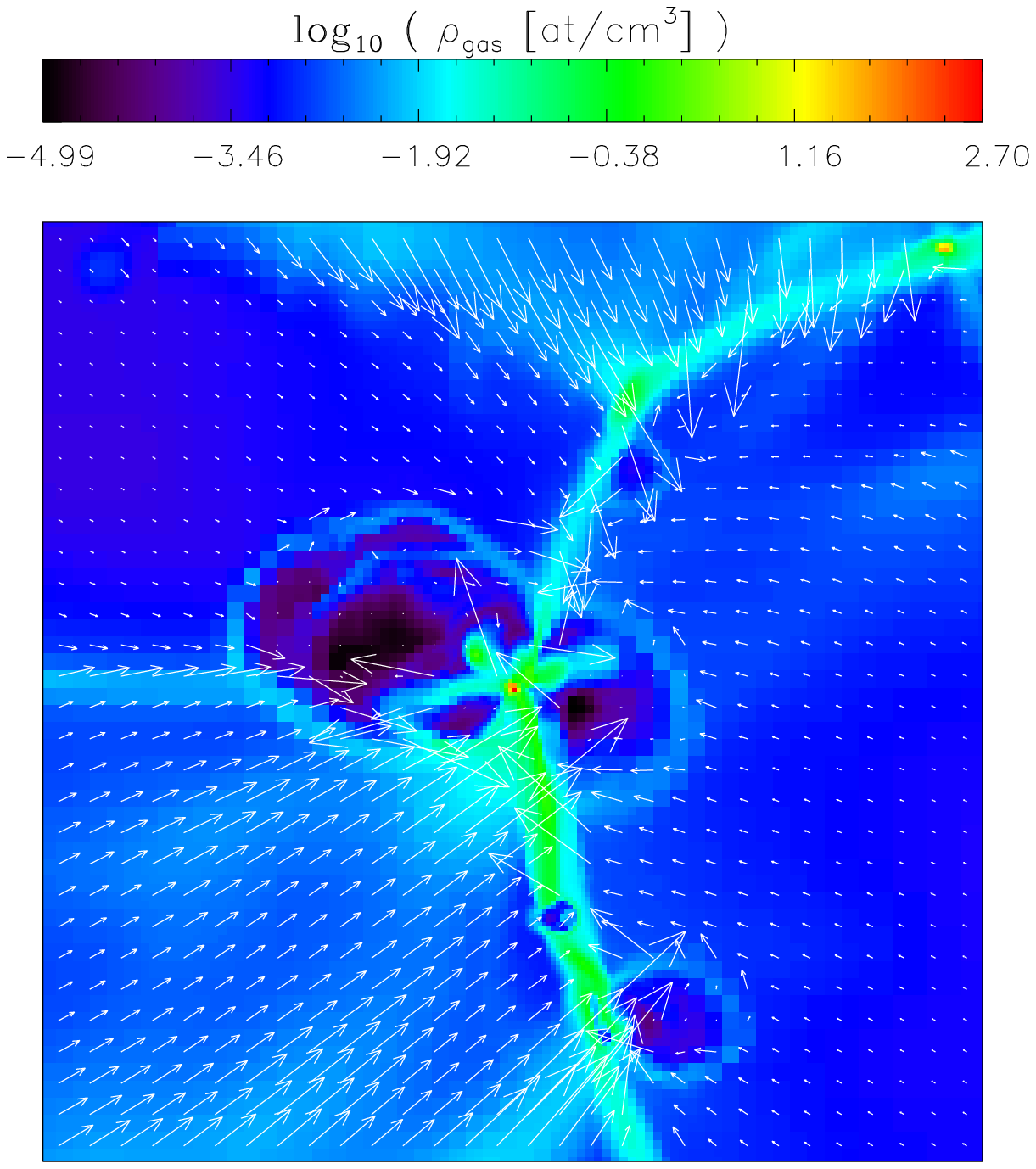}
          \includegraphics[width=0.24\textwidth,trim = 10mm 5mm 10mm
0mm, clip]{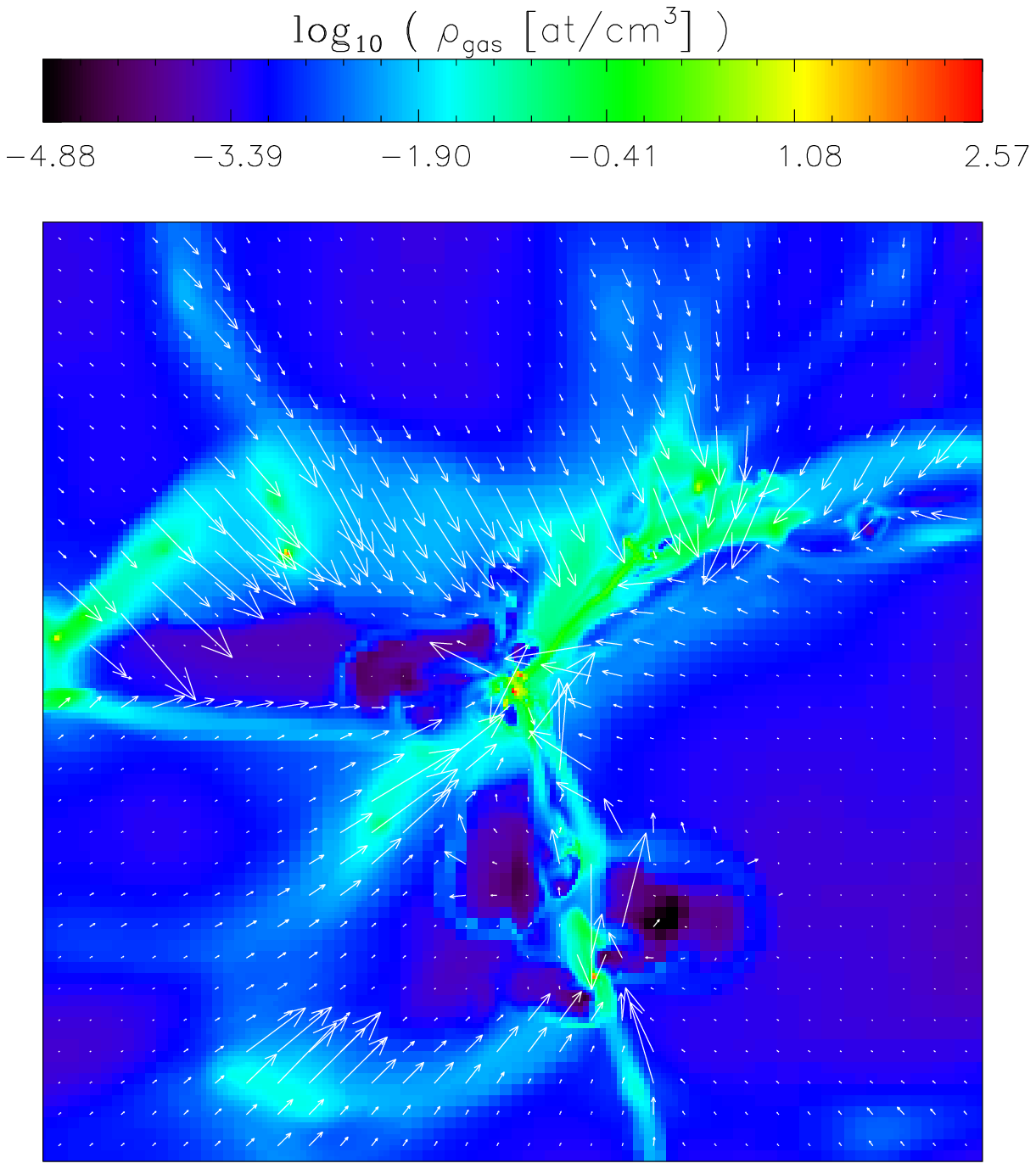}
            \includegraphics[width=0.24\textwidth,trim = 10mm 5mm 10mm
0mm, clip]{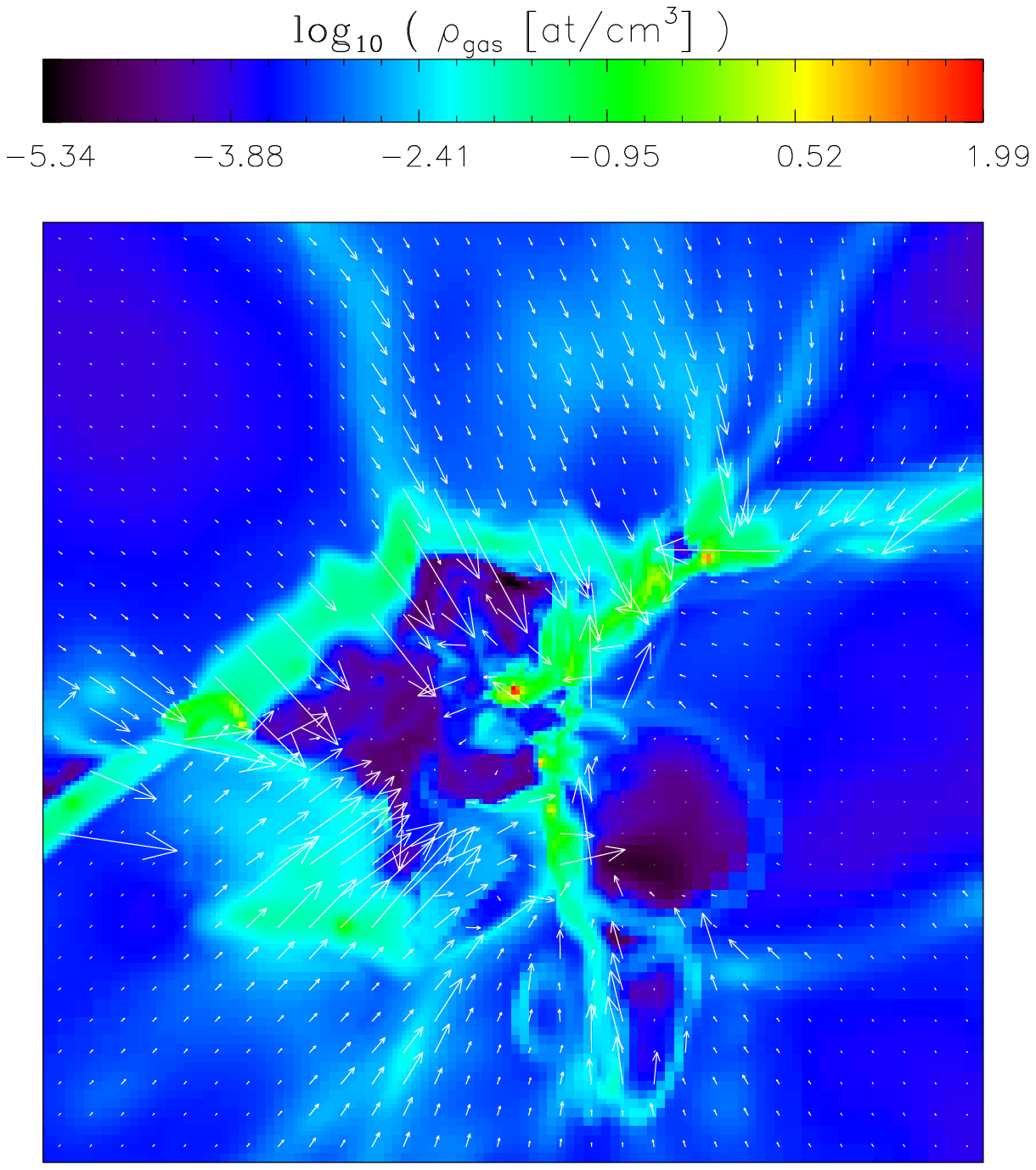}
     \caption{The gas density in a z-slice of thickness $0.05 r_{\rm vir}$ and size
       $(12 r_{\rm vir})^{2}$ centred on the main progenitor for the feedback run at
(from left to right) $z = 15.7$, $12.3$, $10.8$ and $9$. Arrows show
the momentum vectors ($\rho v$). The low density cavities are created
as the SNe blastwaves propagate outwards and eventually
overlap producing a far-reaching galactic
wind. } 
\label{denslice_fb9} \end{figure*}

In Fig.~\ref{denslice_fb9} we show one cell thick ($\sim 0.05 r_{\rm vir}$) slices 
through the gas density in a region of size $12 r_{\rm vir} \times 12 r_{\rm vir}$, centred on the
main halo. Here we can clearly see the low density cavities left
behind as the blastwaves from the SNe propagate outwards. At
higher redshifts (towards the left) these `bubbles' originate only
from the main galaxy, yet as we move to lower redshifts (towards the
right), cavities can also be seen forming further out in the filaments
as galaxies in haloes embedded therein also host SNe explosions. By
$z=9$ (the last panel) we can see the bubbles are overlapping merging
into one outflow with its epicentre located at the main galaxy. 

In previous studies where galactic winds were invoked, often the wind
could not escape the galaxy \citep[e.g.][]{bimodal_marenostrum}, unless it was put in
by hand \citep[e.g.][]{springelhernquist03_sf,oppenheimer_dave_08}. 
We propose it is this mechanism of overlapping bubbles from 
multiple (sub)haloes which (i) significantly increases the reach
  of the wind (ii) more efficiently pushes material outside the
main halo. Indeed, if one considers that the total amount of kinetic
energy imparted to the gas by all SNe (those exploding in the
main halo and its subhaloes) is fixed, but that one is allowed to
distribute it at different points along the path of the outflow, rather
than release it all in the central galaxy, then more energy is available to
drive material out of the main halo's potential well. The reasons are
twofold: (i) the SN driven winds of satellite galaxies have to climb out
of shallower potential wells (ii) the (smaller) amount of gas swept from the central
object (SNe exploding in subhaloes will not contribute to the
mass loading) will see its energy 
replenished on the way outwards by interacting with the subhalo
winds. Note that as a result of this effect, galactic winds might
carry more energy and less mass out of the halo.

High temperature spikes coincident with a (sub)halo are also
occasionally evident in the $T-r$ diagrams for the cooling run,
however these are of a similar colour but much thinner than in the feedback run and therefore
contain a much smaller fraction of the gas mass. This hot gas is generated in the shocking  
of the diffuse circum-galactic material of satellite galaxies (see Powell
et al, in prep for a detailed discussion of shock-heating in
the cooling run).

Another feature of the $T-r$ diagrams (Figs.~\ref{radhisto_fb9} and~\ref{metalradhisto_fb9}) is a
haze of gas (i.e. gas not associated with one of the spikes) with
$T<10^{4}$K which persists at all radii. This is cold, metal poor, diffuse gas that
fills the space between the filaments. In the feedback run (bottom row of Fig.~\ref{radhisto_fb9}) this gas only
becomes significant beyond about $3r_{\rm vir}$ ($\sim 15$kpc) at
$z=9$, whereas it is present in significant quantities within $r_{\rm
vir}$ in the cooling run (top row). Furthermore if we follow the plots
for the feedback run (bottom row of fig.~\ref{radhisto_fb9}) from high to low redshift (left to
right) we can see this haze of cold diffuse gas recedes over time. The
hot wind from the SNe is sweeping away the cold diffuse gas
which surrounds the filaments, reducing the resistance to future
outflows from SNe. We substantiate this hypothesis in Section
\ref{sec:acc} with detailed measurements of the outflow.

\subsection{Wind properties}

\begin{figure*} \centering
   \includegraphics[width=0.33\textwidth,trim = 0mm 0mm 0mm 10mm,
clip]{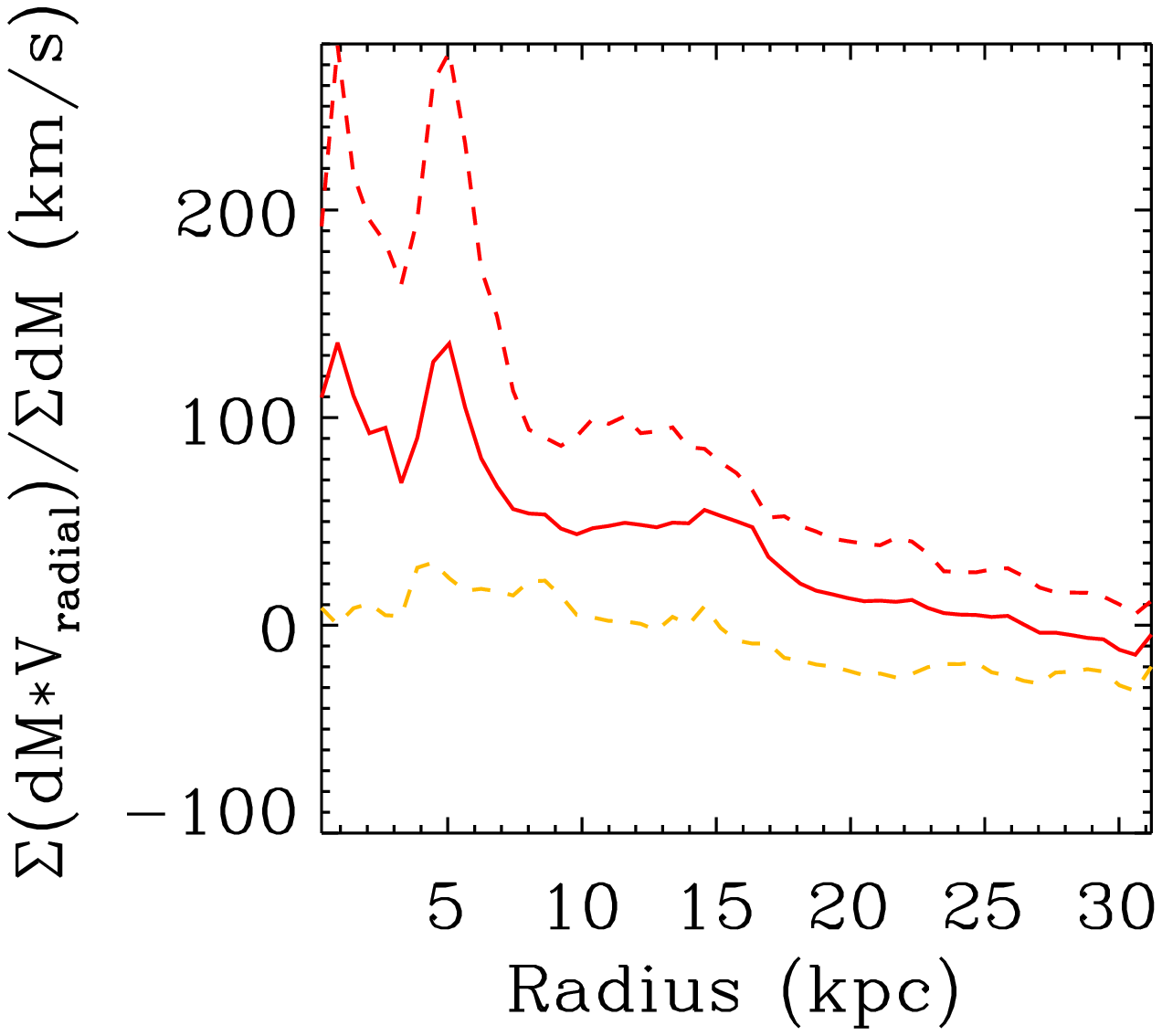}
    \includegraphics[width=0.33\textwidth,trim = 0mm 0mm 0mm 0mm,
clip]{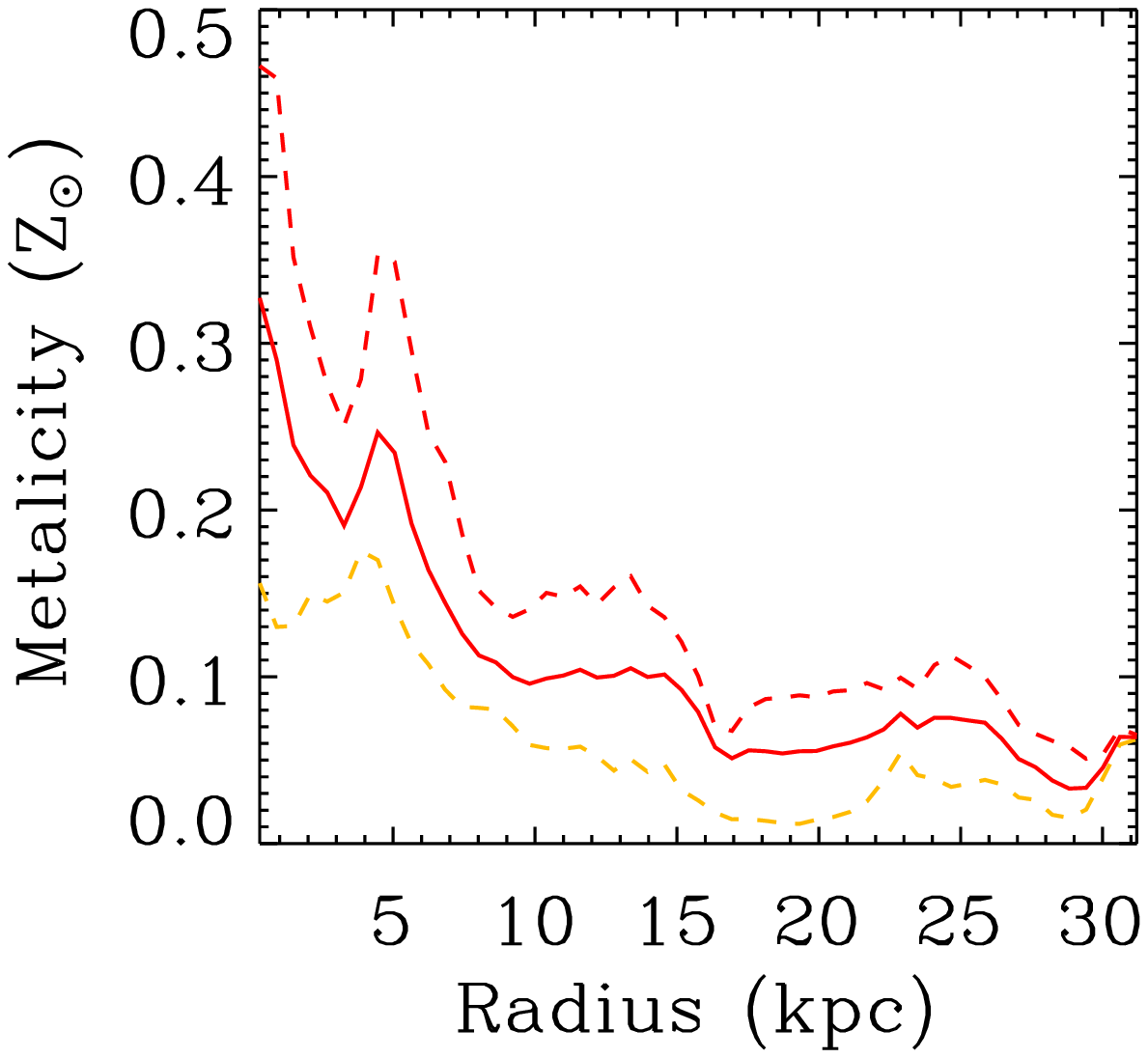}\\
  \includegraphics[width=0.33\textwidth,trim = 0mm 0mm 9mm 5mm,
clip]{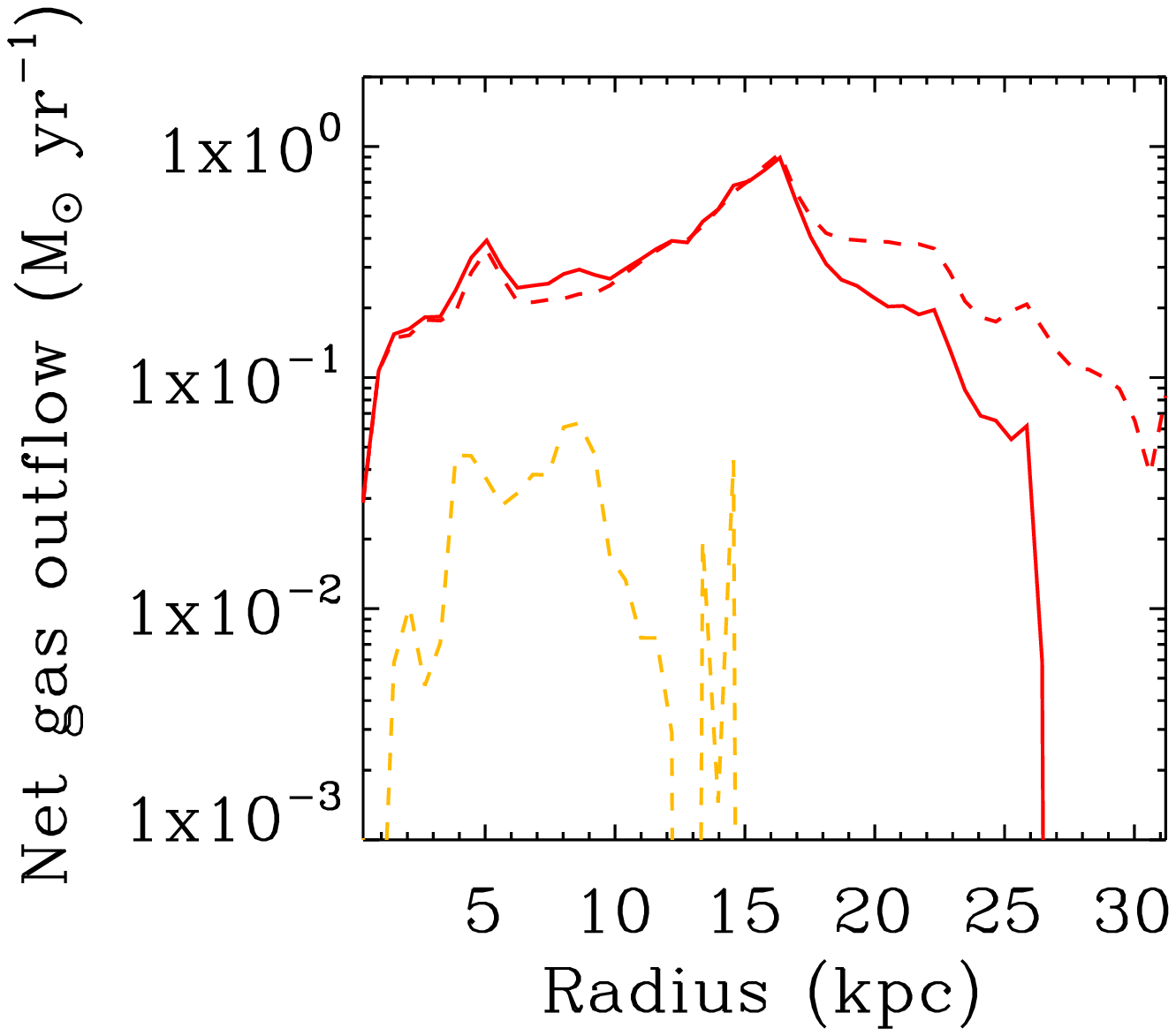}
   \includegraphics[width=0.33\textwidth,trim = 0mm 0mm 9mm 5mm,
clip]{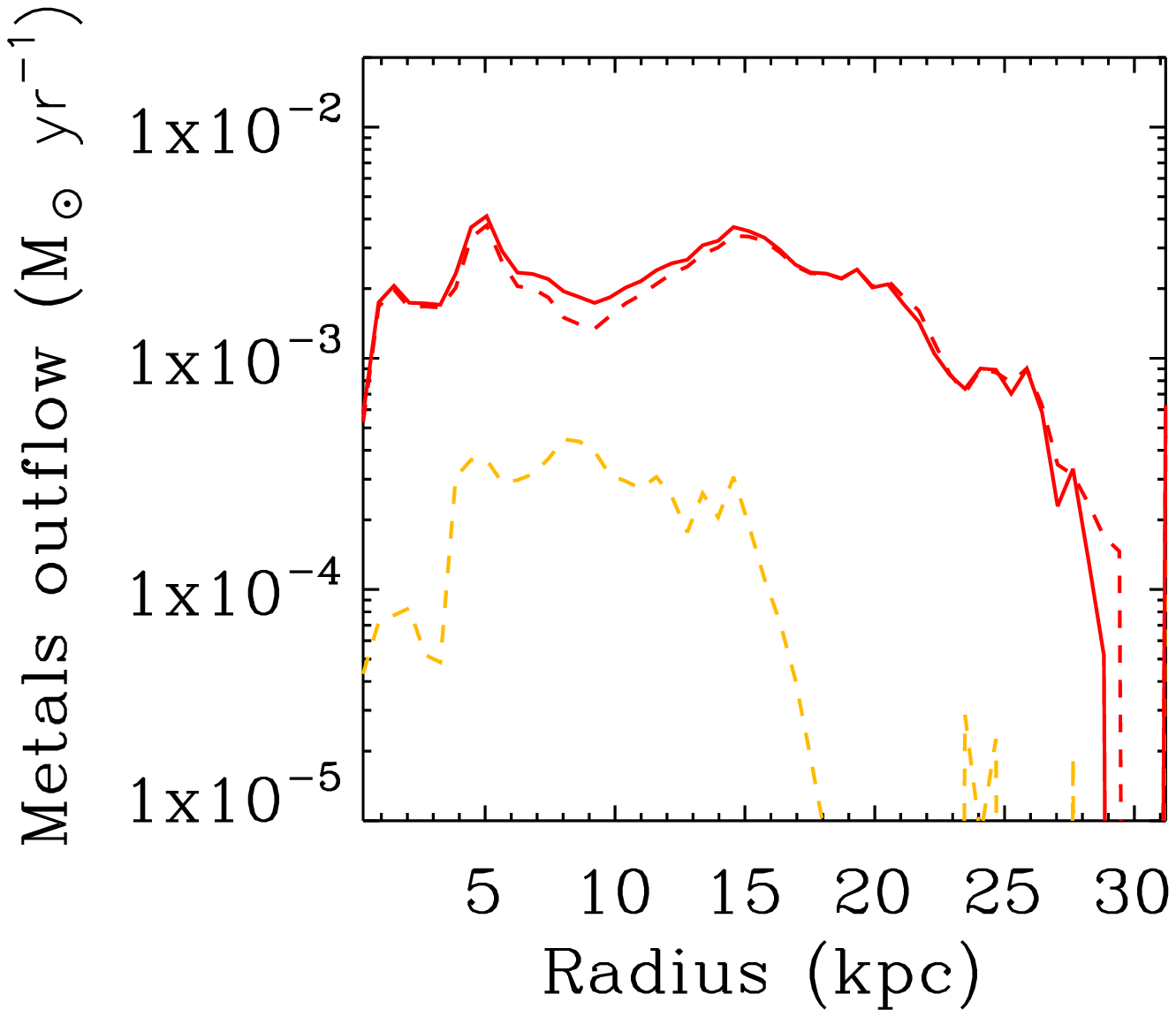}
 \caption{Mass-weighted radial velocity profile averaged (top left), metallicity profile (top right), net mass outflow
rate (bottom left) and net mass outflow of metals (bottom right) in $100$
pc (physical) spherical shells out to $6r_{\rm vir}$ for the feedback
run at $z=9$. Solid red line shows quantities warm + hot gas, which is then
split into warm (orange dashed line) and hot (red
dashed line).} \label{mwvrvr_big} \end{figure*}

In Fig.~\ref{mwvrvr_big} we present the properties of the galactic
wind as a function of radius (within $r_{\rm vir}$ of the main halo) at $z=9$. We show
the mass-weighted velocity (top left), the metallicity profile (top right), the net mass outflow rate (bottom left) and
the net mass outflow rate of metals (bottom right). The solid red line
shows these quantities for the warm and hot gas combined, which is
then split into warm (orange dashed line) and hot (red dashed line) components
(see Table \ref{catsummary} for definitions of `hot' and `warm').

The velocities we measure (top left panel of Fig.~\ref{mwvrvr_big}) compare well to the semi-analytic
  model of \citet{furlanetto_loeb_2003}, both for launching
  (their equation (11) for $v_{\rm w}$) and average expansion velocities
  (equation (12) and Fig.~2 top right panel for $v_{\rm exp}$) of the
  wind. Agreement on $v_{\rm exp}$  is not too surprising because the wind
  bubble in our simulation is able to avoid zones of high pressure (filaments have a negligible
halo volume filling fraction), and therefore it mostly sweeps gas with the
average density (and pressure) of the IGM at the redshift of interest. This
makes Furlanetto and Loeb's approximation of a spherically symmetric expansion in a
homogeneous background of average density a good one, even though 
it is not very consistent with the spherically symmetric accretion of
dense gas which should naturally occur in their model if they did not
wait until all this gas has piled up in the centre of the halo before triggering their wind. 

More remarkable is the agreement with $v_{\rm w}$, considering that the
fraction of the total energy of SNe available to power the wind 
in their model ($f_{\rm esc}$) is calibrated on the idealized simulations of high
redshift starbursts by \citet{mori_ferrara_madau_2002} who quote a value
of about 25\%. Indeed, using the same estimator as Mori and
collaborators (see Fig.~12 of their paper), i.e. the ratio of
kinetic energy flux at the virial radius to the total supernova
luminosity, we find that $f_{\rm esc}$ is around $\sim 2.5$\%. 
This remains the case even when we vary the radius at which we measure the
kinetic flux (down to 0.1 $r_{\rm vir}$) or the time delay ($10-100$ Myr) between the
measurement of the kinetic flux at the virial radius and the supernova luminosity 
to account for the amount of time it takes the central wind to reach
the growing virial radius in our non-idealized feedback simulation. 
This factor 10 lower for the efficiency to power winds in our
simulation compared to that measured by Mori and collaborators can be
attributed to (at least) two differences: (i) the central density of the gas in
\citet{mori_ferrara_madau_2002} is severely underestimated (they use
10 at/cc which is at the bottom end of our definition of clumpy gas
and more representative of the density of filaments at these
redshifts) (ii) the star formation rate (and hence SNe
explosions) is spread out in time (and space) in our simulation rather than 
instantaneous. These two effects, and especially the former, 
mean that radiative cooling affects our supernova remnants to a much
greater level, sapping their energy much more efficiently. 
Finally, we note that the value of $f_{\rm esc}$ in our simulation does not seem to be a strong function of
 the mass of the halo, as the progenitor of our $5 \times 10^9 M_\odot$ halo 
at $z=9$ reaches a similar mass as that of the halo used by Mori and
collaborators ($2 \times 10^8 M_\odot$) at $z \sim 13$ without
$f_{\rm esc}$ changing much. Thus, the agreement with \citet{furlanetto_loeb_2003} for $v_{\rm w}$ (the wind launching velocity) is recovered 
because they assume that the the efficiency of the wind mass loading, $f_{\rm sw}$, is related to the star formation efficiency on galactic scales, $f_\star$, by $f_{\rm sw}=2 f_\star$.  Although the values for 
the $f_\star$ that we
measure are close to the the value of $10$ \% that these authors adopt
(see section~\ref{sec:sfr} for details),
the value we measure for $f_{\rm sw}$,  
is approximately a factor 10 lower than theirs (see section~\ref
{sec:acc} for the exact value). 
Since we adopt an initial mass function (IMF) very similar to theirs (Salpeter with similar mass cut-offs), the
order of magnitude discrepancies in $f_{\rm esc}$ and $f_{\rm sw}$ cancel out in their formula (11) for $v_{\rm w}$ and
we recover a value close to their $300$ km/s, as shown in the top left panel of our Fig~\ref{mwvrvr_big}.

The metallicity profile of the wind out to $6 r_{\rm vir}$ in Fig.~\ref{mwvrvr_big} (top
right panel) shows that the metallicity ranges from 
$0.5 Z_{\odot}$ to $0.1 Z_{\odot}$ at $z=9$. In terms of mass, $75$ per cent of the gas with $T>2 \times 10^4$K within $r_{vir}$ 
has $Z>0.1Z_{\odot}$, with quite narrow scatter since only $10$ per cent of this gas has
$Z>0.5Z_{\odot}$.  Whilst we cannot currently compare measurements of the metallicity of a high-redshift galaxy directly with observations, we can
draw parallels with observations of the metallicity of winds in local
galaxies. Determining the wind metallicity using X-ray observations of
the hot gas is difficult due to various uncertainties and degeneracies
\citep[see][for a discussion]{galacticwindreview05}.  Bearing these
caveats in mind, \citet{martinetal02} find that the
metallicity of the wind in NGC 1569, a local dwarf starburst galaxy,
must be greater than $0.25 Z_{\odot}$ and that the wind seems to carry most
of the metals produced in the current starburst. Our measurements of
the metallicity of the hot diffuse gas are broadly consistent with
these observations, suggesting that metal enrichment 
away from the main site of star formation is plausible at such a
level. 

If we extrapolate our results using Press --Schechter theory
 to calculate the average number density of $5 \times 10^9$ M$_\odot$ haloes at
 $z=9$, we find a value of $\approx$ 0.15 Mpc$^{-3}$ comoving for the 
cosmological model we use. This translates into a mean inter-halo 
distance of about $190$ physical kpc, so that a small but non
negligible volume fraction ($0.4 \%$) of the IGM 
can be filled by metal rich hot gas wind bubbles blown by haloes in this
mass range at those redshifts.  Assuming that the result of 
\citet{furlanetto_loeb_2003} holds in this regime, such that the extent of the wind does not depend sensitively on 
the virial mass of the galaxy host halo at a given redshift, then the volume fraction filled by hot gas will increase if we
start including lower mass haloes. Based on the measurement of the wind metallicity in our simulation (see
  Fig.~\ref{mwvrvr_big} top right and bottom right panels), it seems that
galactic winds may indeed be very efficient at polluting gas in the
the IGM at very high redshifts up to a level comparable to
  that observed in the Ly-$\alpha$ forest at $z \approx 2$ \citep{bouche_etal_2007}.

\section{Measuring gas accretion and outflow} \label{sec:acc}

In Section \ref{sec:filwind}, we defined $5$ gaseous components
(clumpy, filaments, cold diffuse, warm diffuse and hot diffuse) which
are summarised in Table \ref{catsummary}. We now measure the rate of
mass inflow and outflow in each component in order to quantify how
they contribute to the overall accretion/outflow of material onto/from
the central galaxy. In particular, we focus on establishing the impact
(if any) of the galactic wind (which we identify with the warm and hot diffuse phases) on these mass fluxes by comparing the
feedback and cooling runs.

We measure the flux in spherical shells centred on the centre of the dark matter halo of the main galaxy, 
which is taken to be the position of the densest dark matter particle. The shells are of width $100$ pc physical 
and extend out to the halo's virial radius. In order to compute gas velocities relative to the galaxy, we subtract
 the peculiar velocity of the dark matter halo from all gas velocities. We compute the mass flux, $m_{\rm flux}$, 
for each grid cell in a shell using,

\begin{equation}
m_{\rm flux}=\rho_{\rm cell} v_{\rm radial} dx^{3}\\
\end{equation}

\noindent where $\rho_{\rm cell}$, $v_{\rm radial}$ and $dx$ are the cell's gas density, radial gas velocity, 
and physical size respectively.

By definition, gas with negative $m_{\rm flux}$ (i.e. negative $v_{\rm radial}$) is flowing inwards and gas with positive 
$m_{\rm flux}$ (i.e. positive $v_{\rm radial}$) is flowing outwards. We compute the total inflow, $F_{-}$, and total 
outflow, $F_{+}$, separately for each shell as follows,

\begin{eqnarray}
F_{+}= \frac{\Sigma m_{\rm flux, +}}{\Sigma dx^{3}} \; 4\pi r_{\rm shell}^{2}, & F_{-}= \frac{\Sigma m_{\rm flux, -}}
{\Sigma dx^{3}} \;   4\pi r_{\rm shell}^{2}
\end{eqnarray}

\noindent where $ r_{\rm shell}$ is the distance between the halo centre and the shell midpoint and the $+/-$ 
subscript on $m_{\rm flux}$ denotes whether it is positive or negative. This calculation is repeated for each of the gas 
components listed in Table \ref{catsummary}, the only difference being that the terms $m_{\rm flux, +}$ and 
$m_{\rm flux, -}$ only include contributions from cells that meet the temperature and density criteria for the particular
component. The flux we compute is, therefore, an instantaneous flux at the time of the output we are analysing. 

  \subsection{Inflow}

\begin{figure*} \centering

    \includegraphics[width=0.35\textwidth,trim = 4mm 3mm 12mm 12mm,clip]{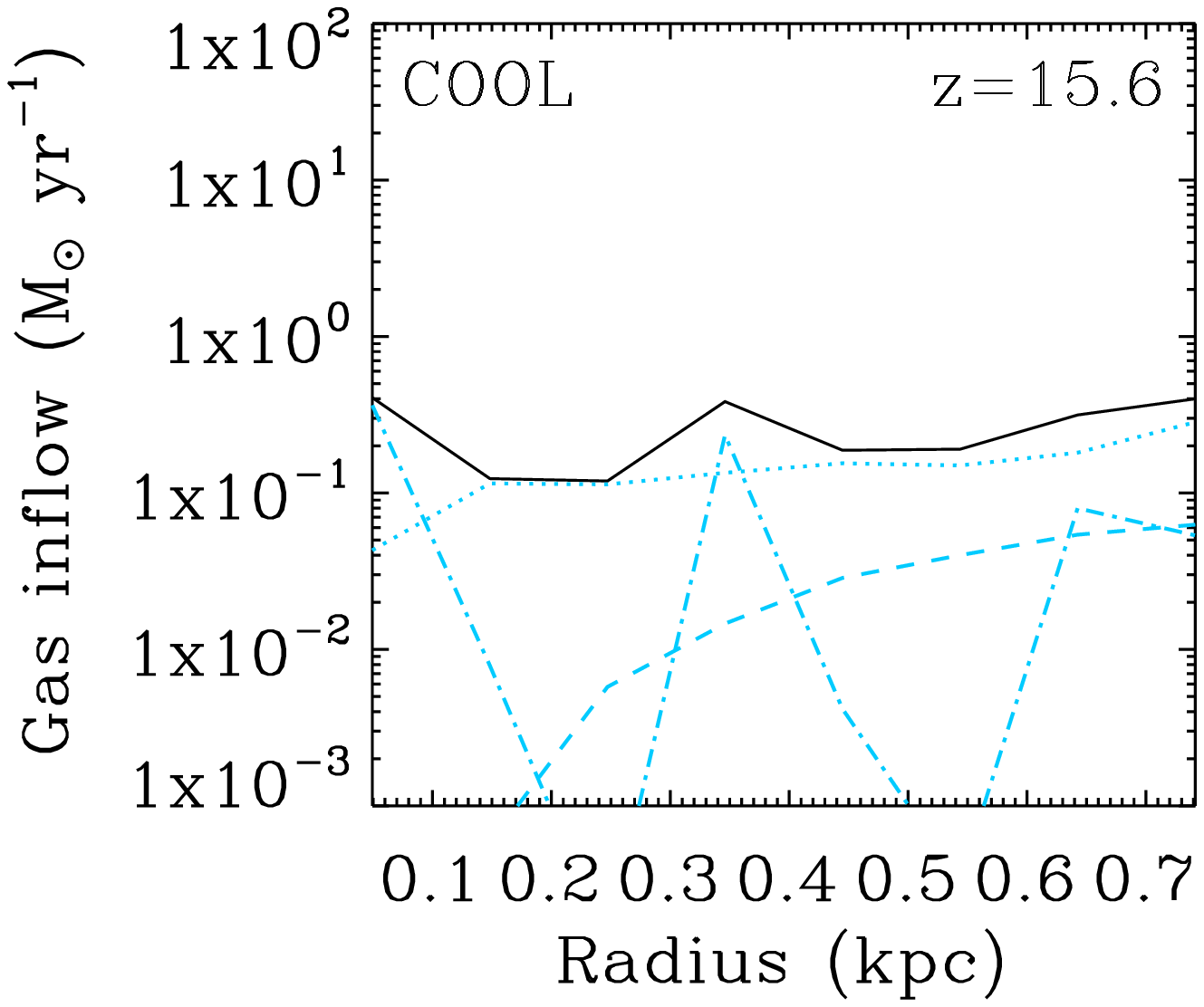}
  \includegraphics[width=0.35\textwidth,trim = 4mm 3mm 12mm 12mm,clip]{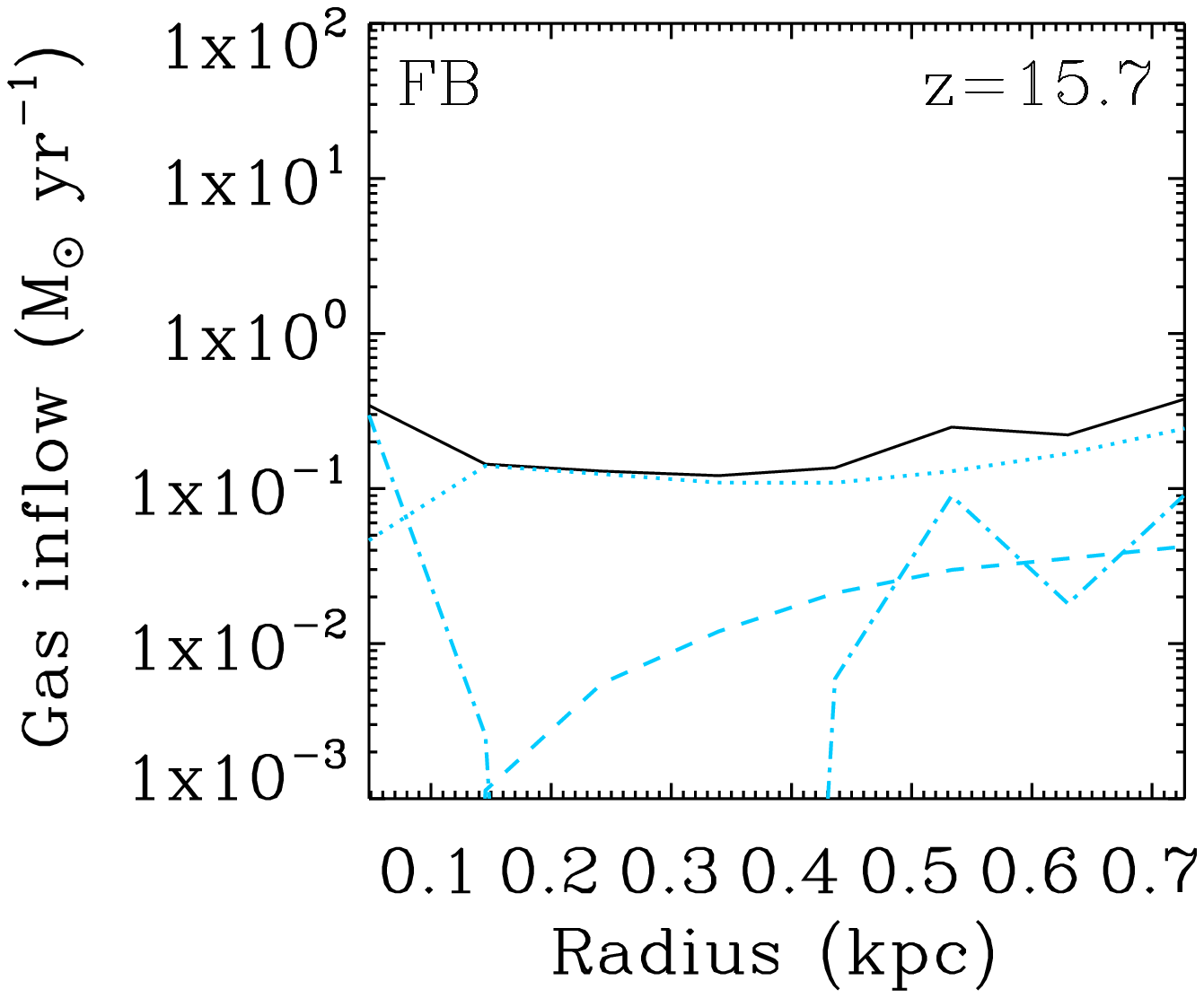}\\

        \includegraphics[width=0.35\textwidth,trim = 4mm 3mm 12mm 12mm,clip]{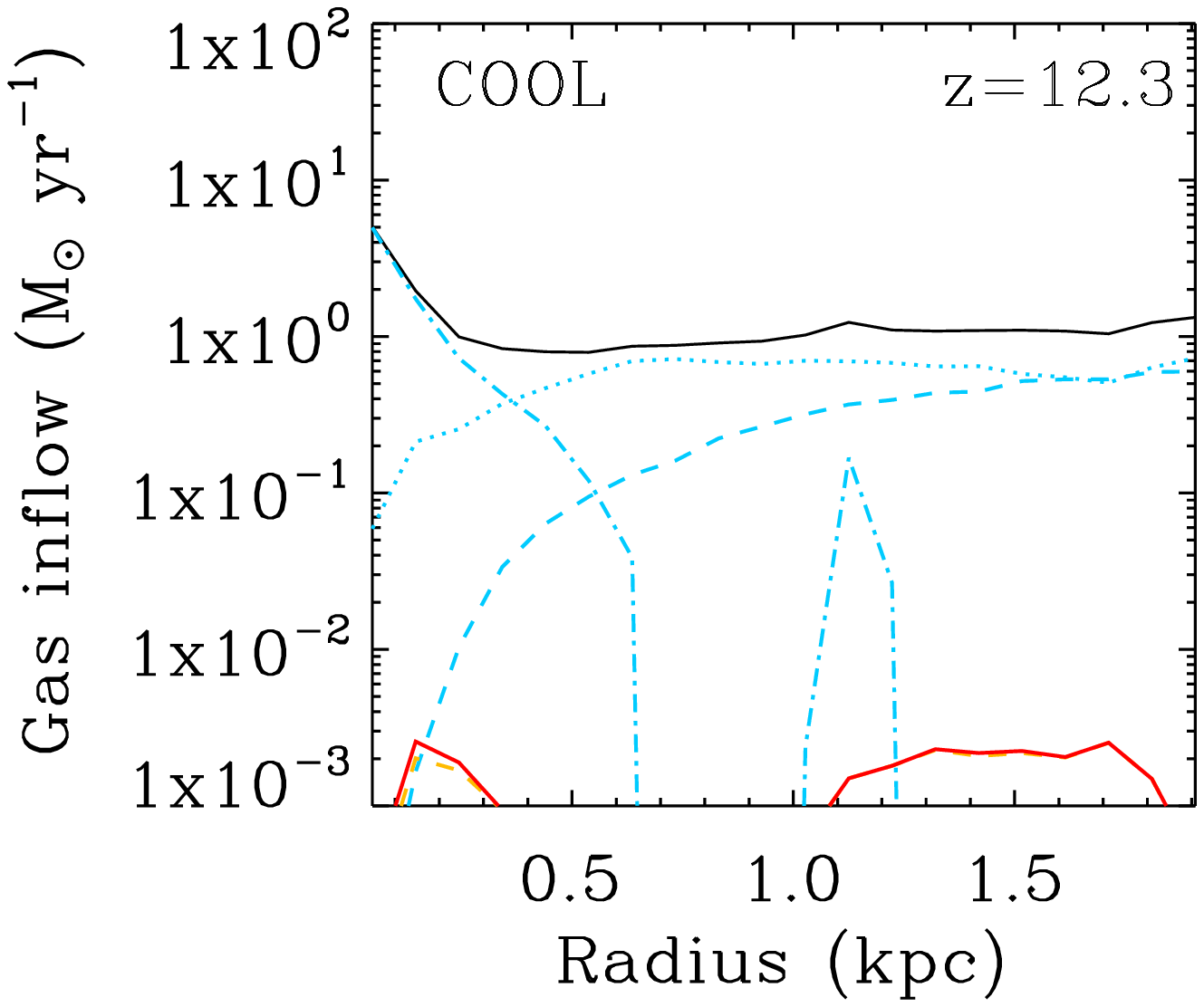}
  \includegraphics[width=0.35\textwidth,trim = 4mm 3mm 12mm 12mm,clip]{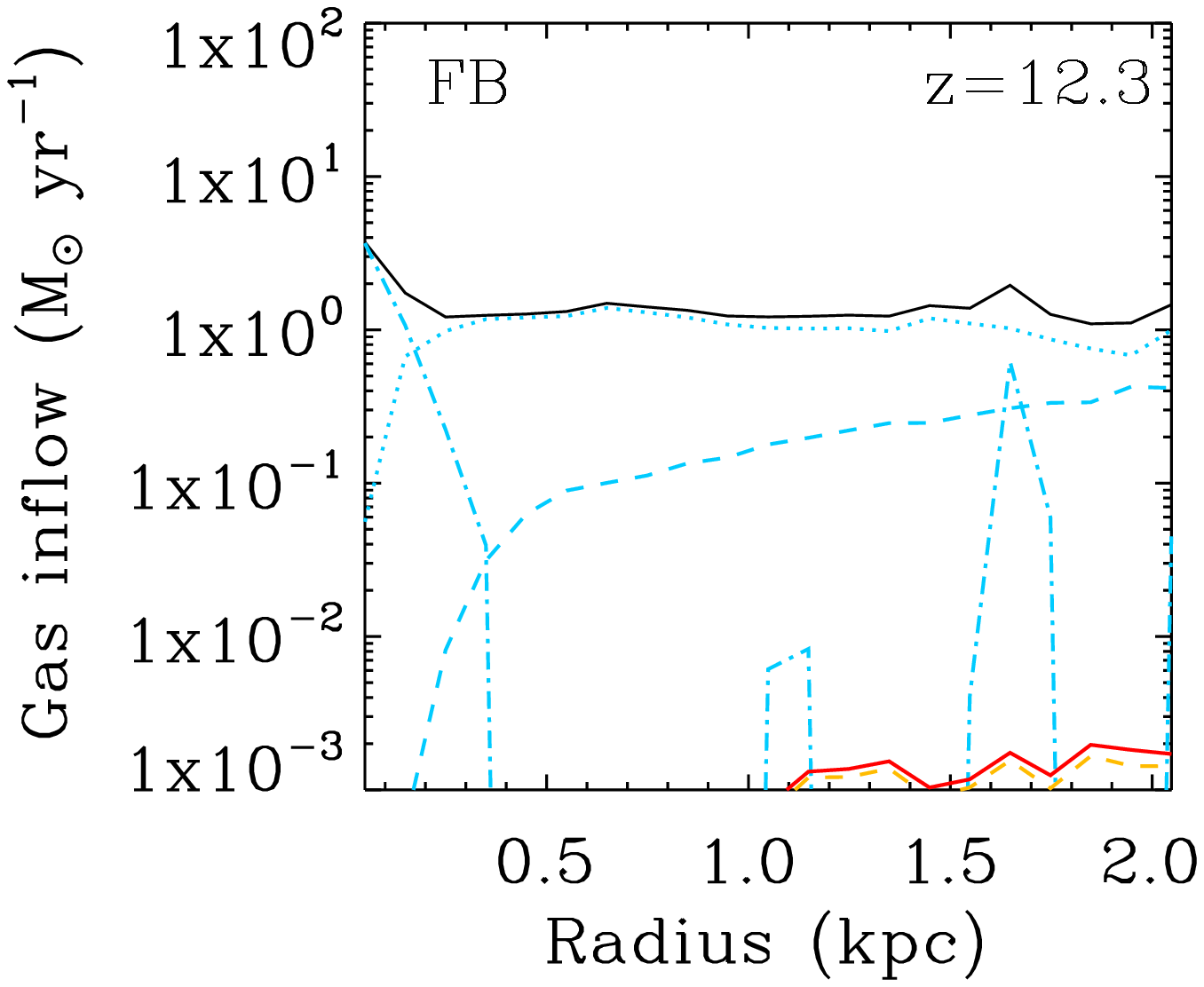}\\

            \includegraphics[width=0.35\textwidth,trim = 4mm 3mm 12mm 12mm, clip]{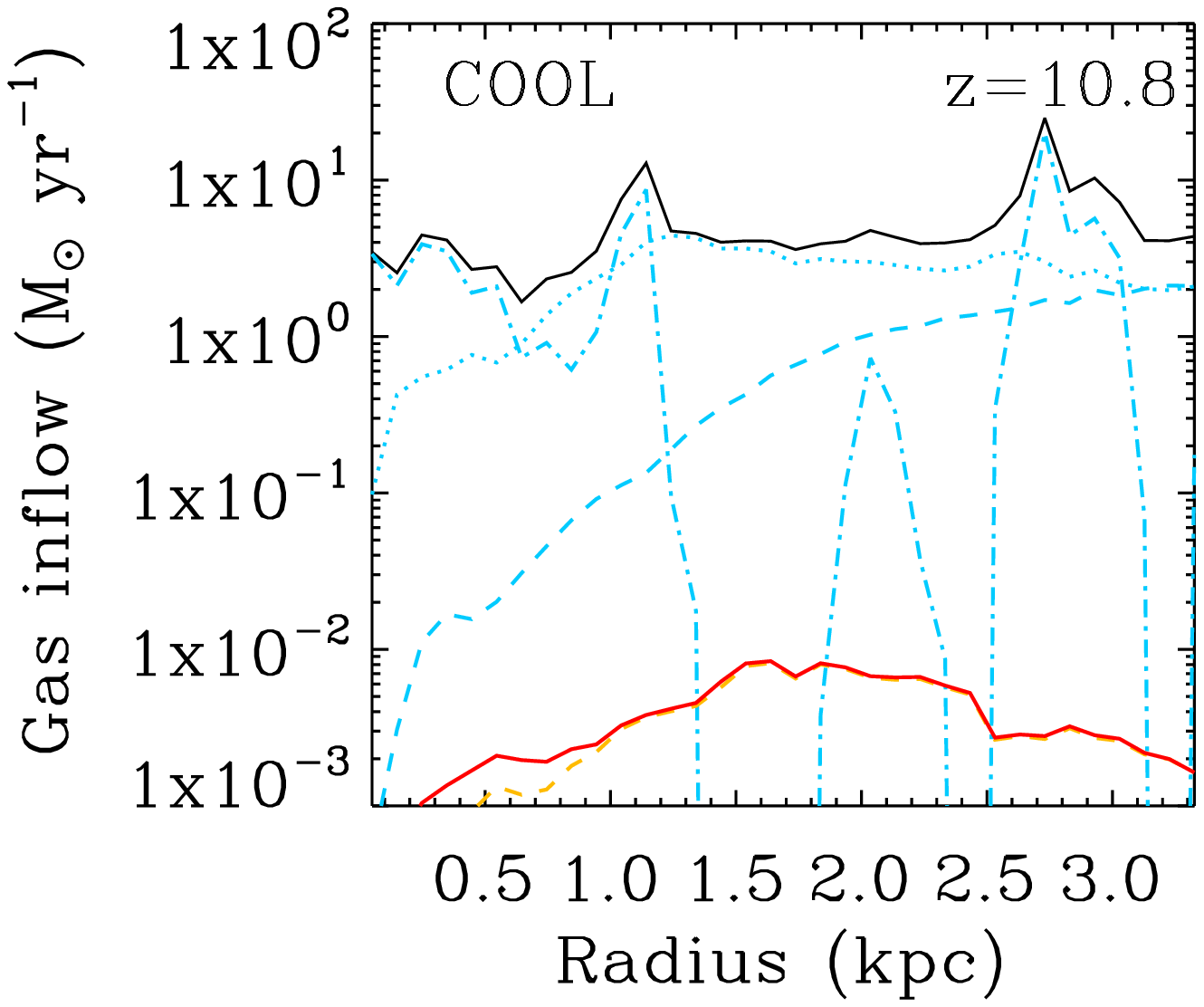}
   \includegraphics[width=0.35\textwidth,trim = 4mm 3mm 12mm 12mm, clip]{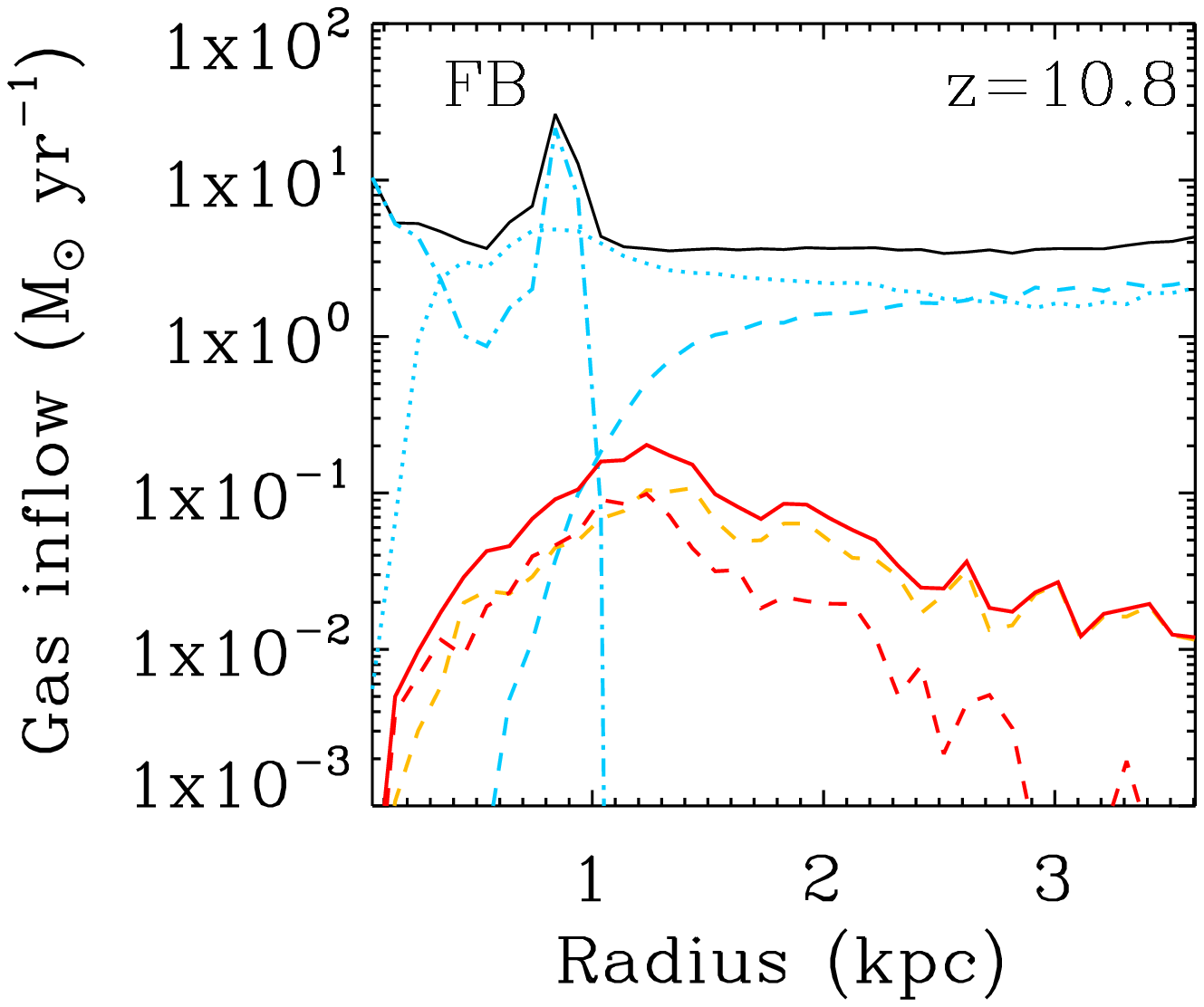}\\

                  \includegraphics[width=0.35\textwidth,trim = 4mm 3mm 12mm 12mm, clip]{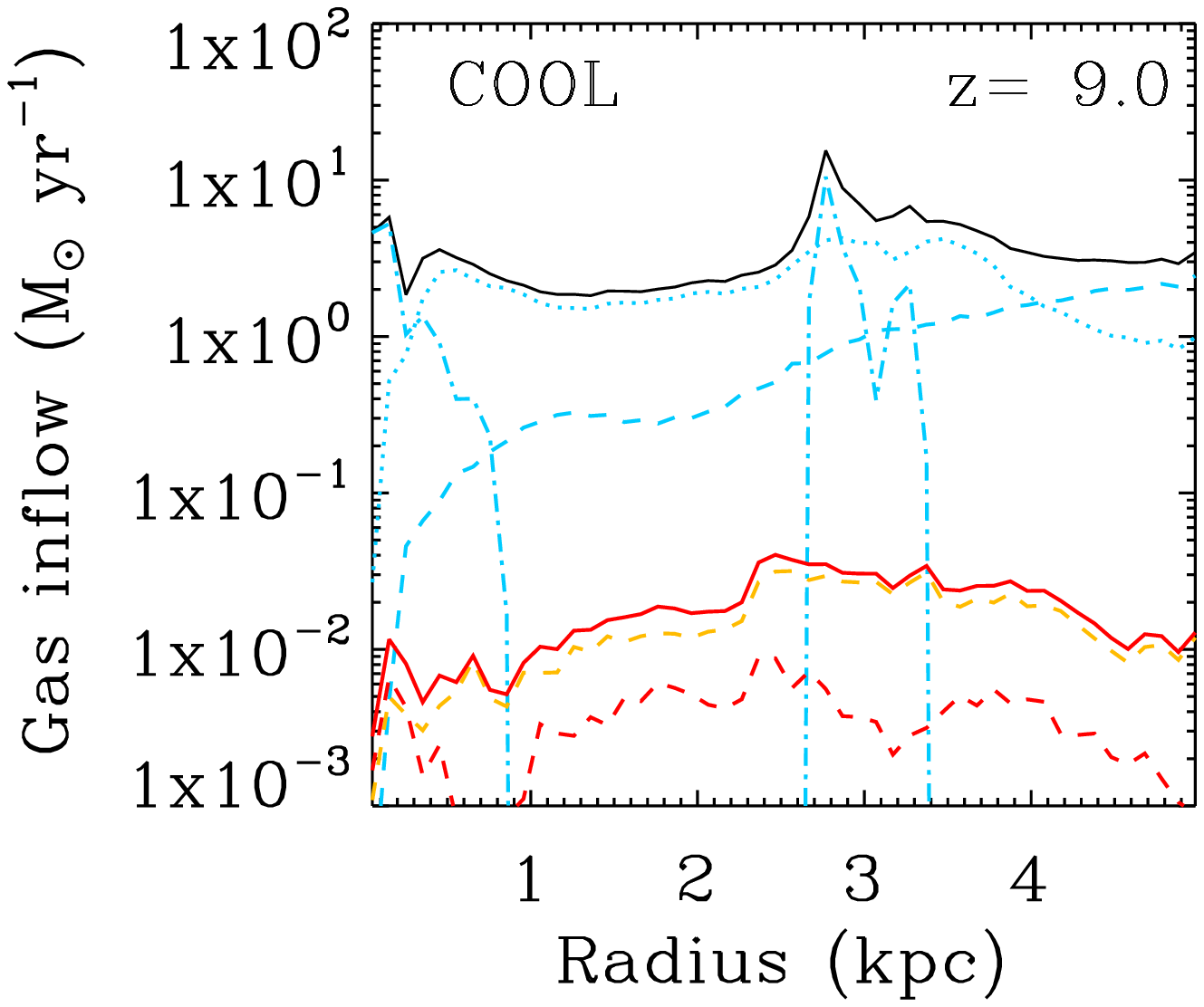}
    \includegraphics[width=0.35\textwidth,trim = 4mm 3mm 12mm 12mm, clip]{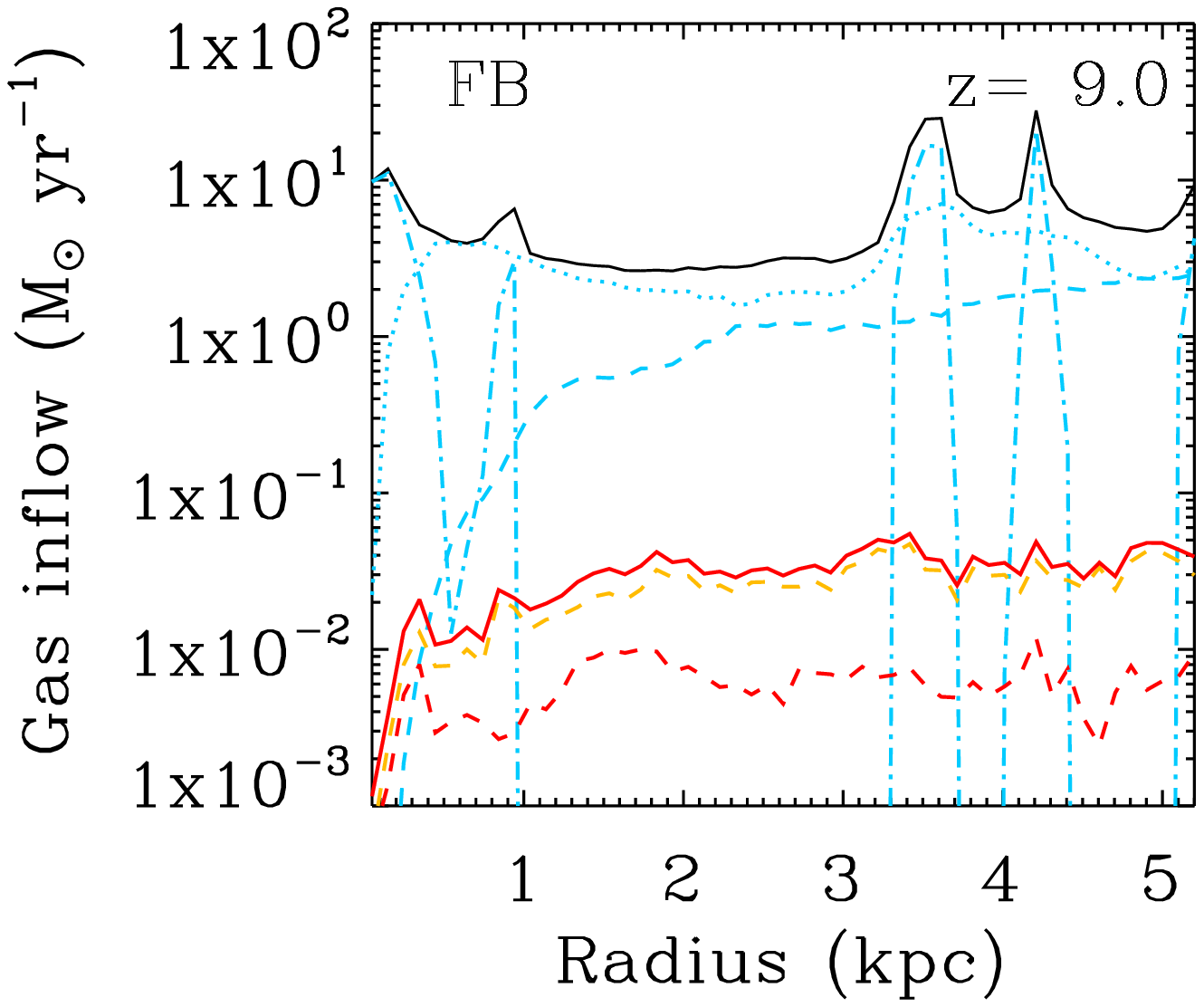}
                          
 \caption{Gas inflow rates averaged in $100$ pc (physical)
spherical shells out to $r_{\rm vir}$ for the cooling (first column) and feedback (second column) runs. Black solid lines show the mass inflow rates of all gas,
which is then split into dense (blue dot-dash line), filamentary (blue
dotted line), cold diffuse (blue dashed line), hot diffuse (red dashed line) and warm
diffuse (orange dashed line) categories as summarized in Table 1. The solid
red line shows the warm diffuse and hot diffuse components
combined.} 
\label{fluxinvr_fb9} 
\end{figure*}

\begin{figure*} \centering

   \includegraphics[width=0.35\textwidth,trim = 4mm 3mm 12mm 12mm,clip]{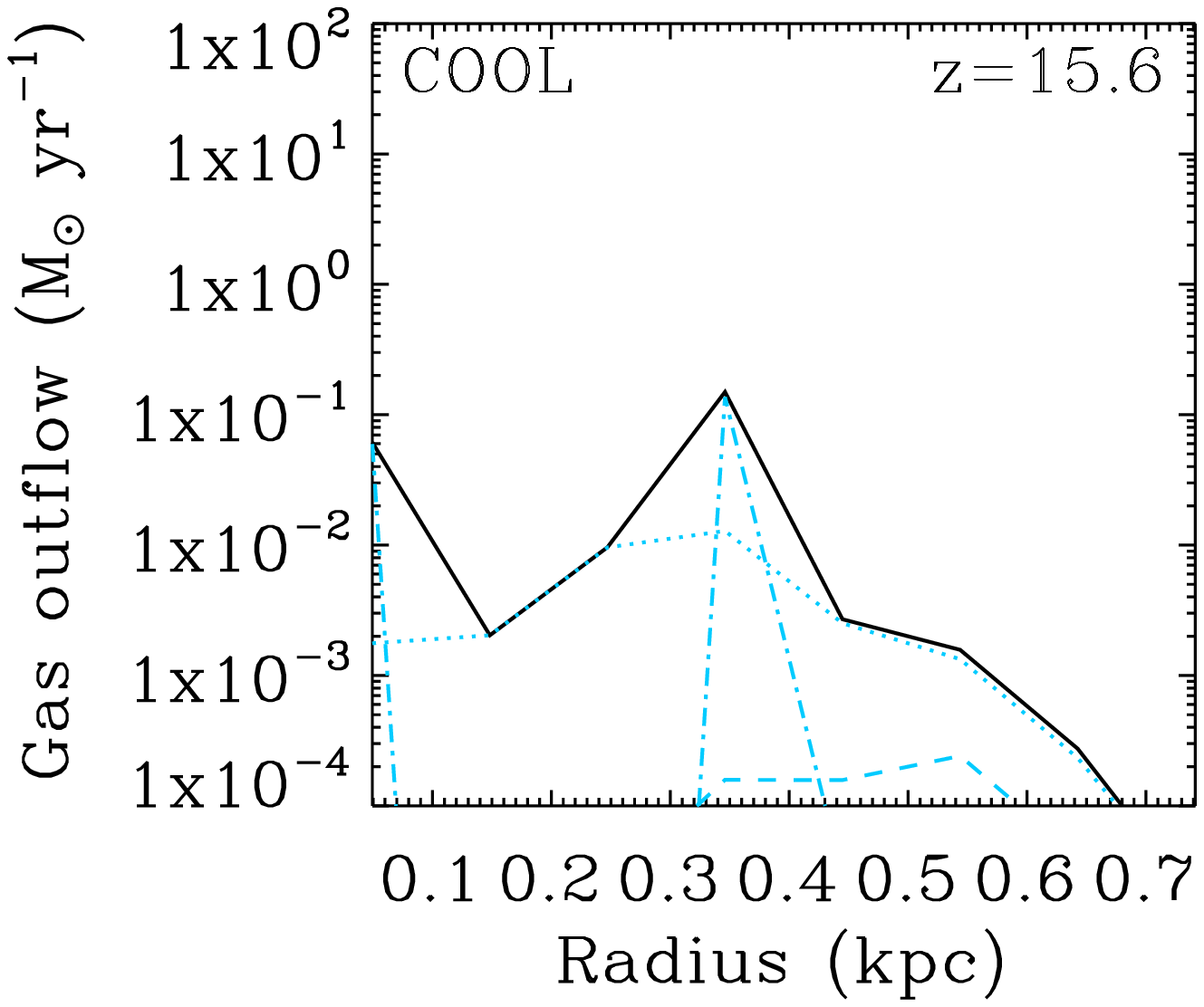}
 \includegraphics[width=0.35\textwidth,trim = 4mm 3mm 12mm 12mm,clip]{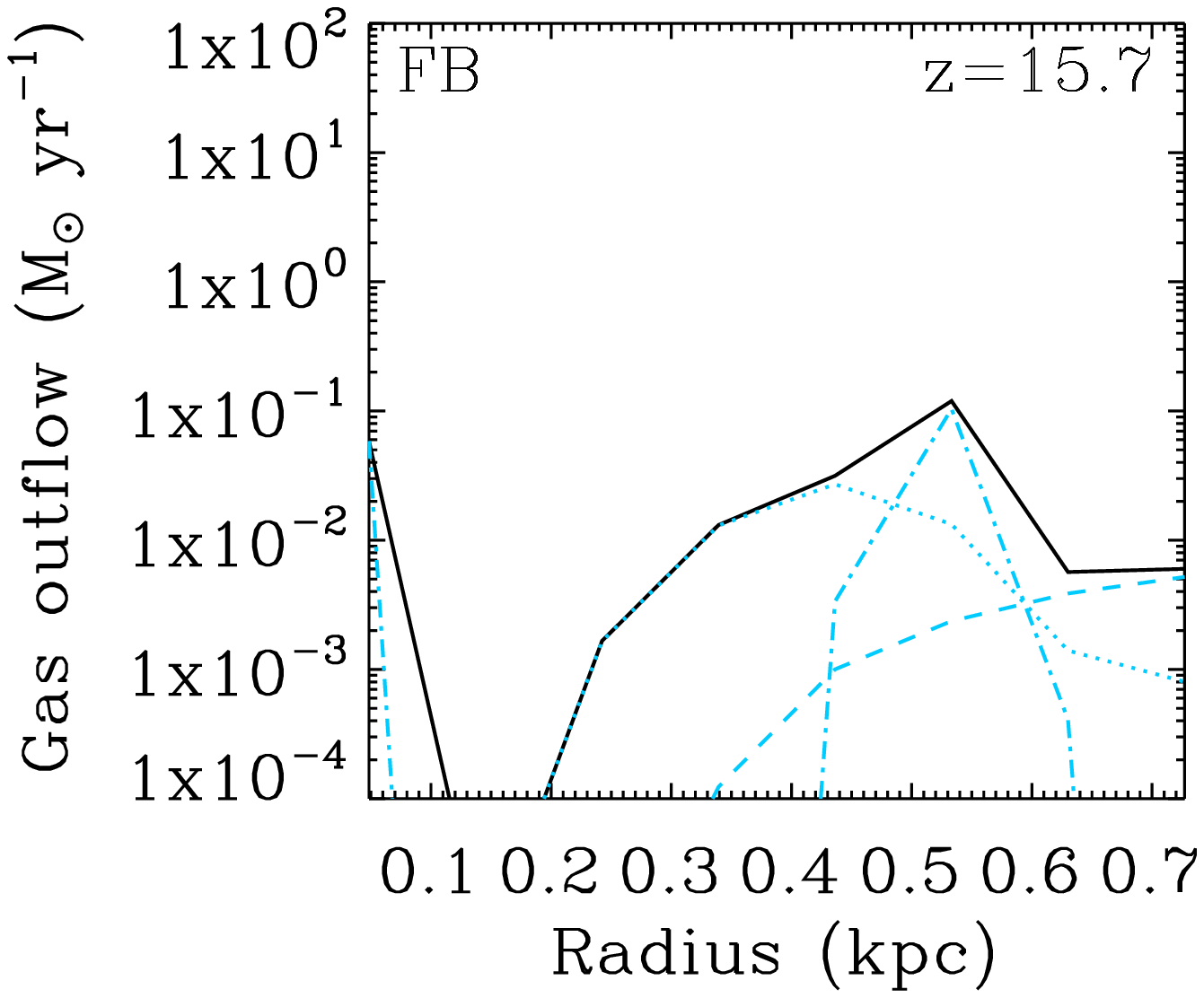}\\

 \includegraphics[width=0.35\textwidth,trim = 4mm 3mm 12mm 12mm,clip]{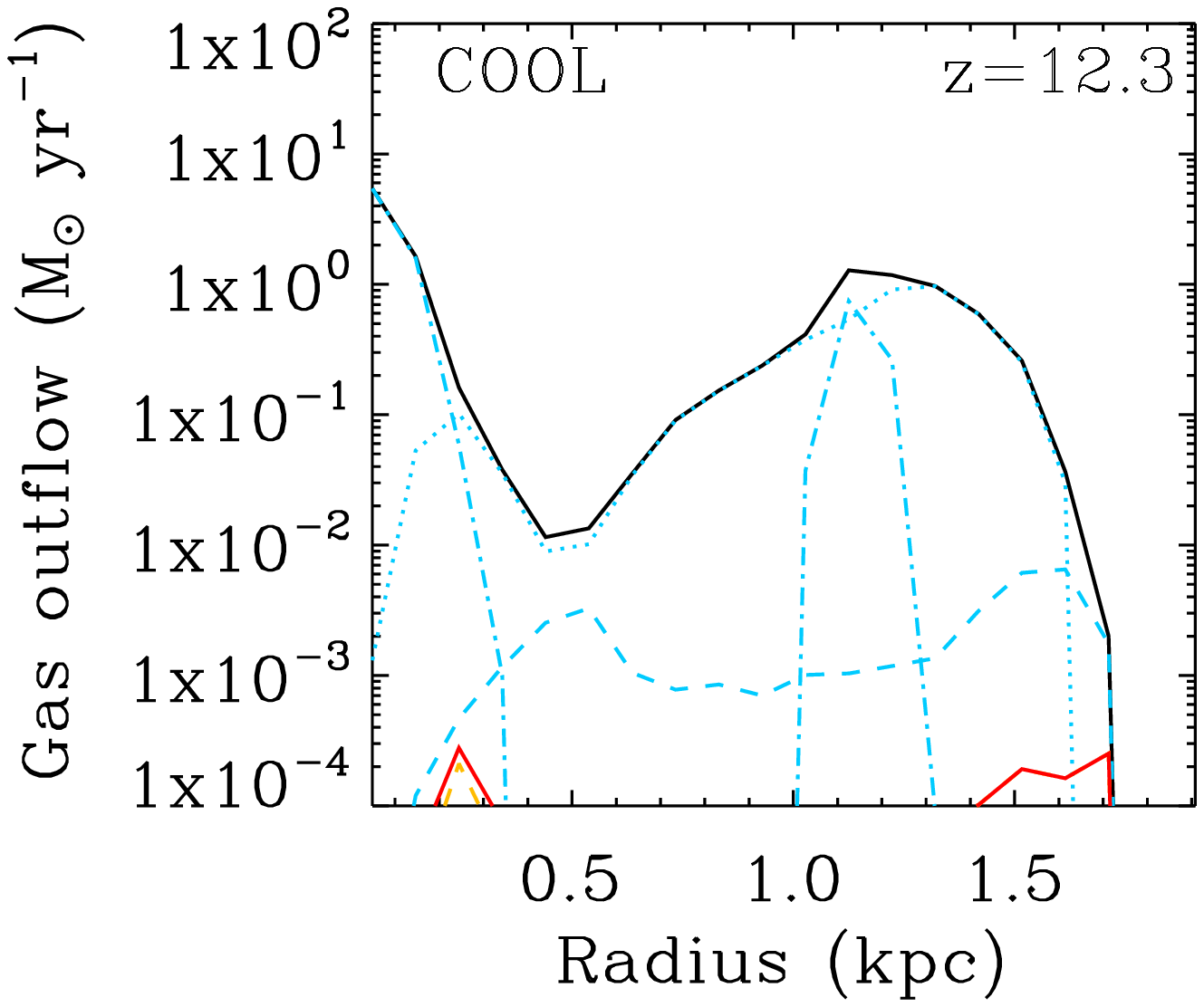}
  \includegraphics[width=0.35\textwidth,trim = 4mm 3mm 12mm 12mm,clip]{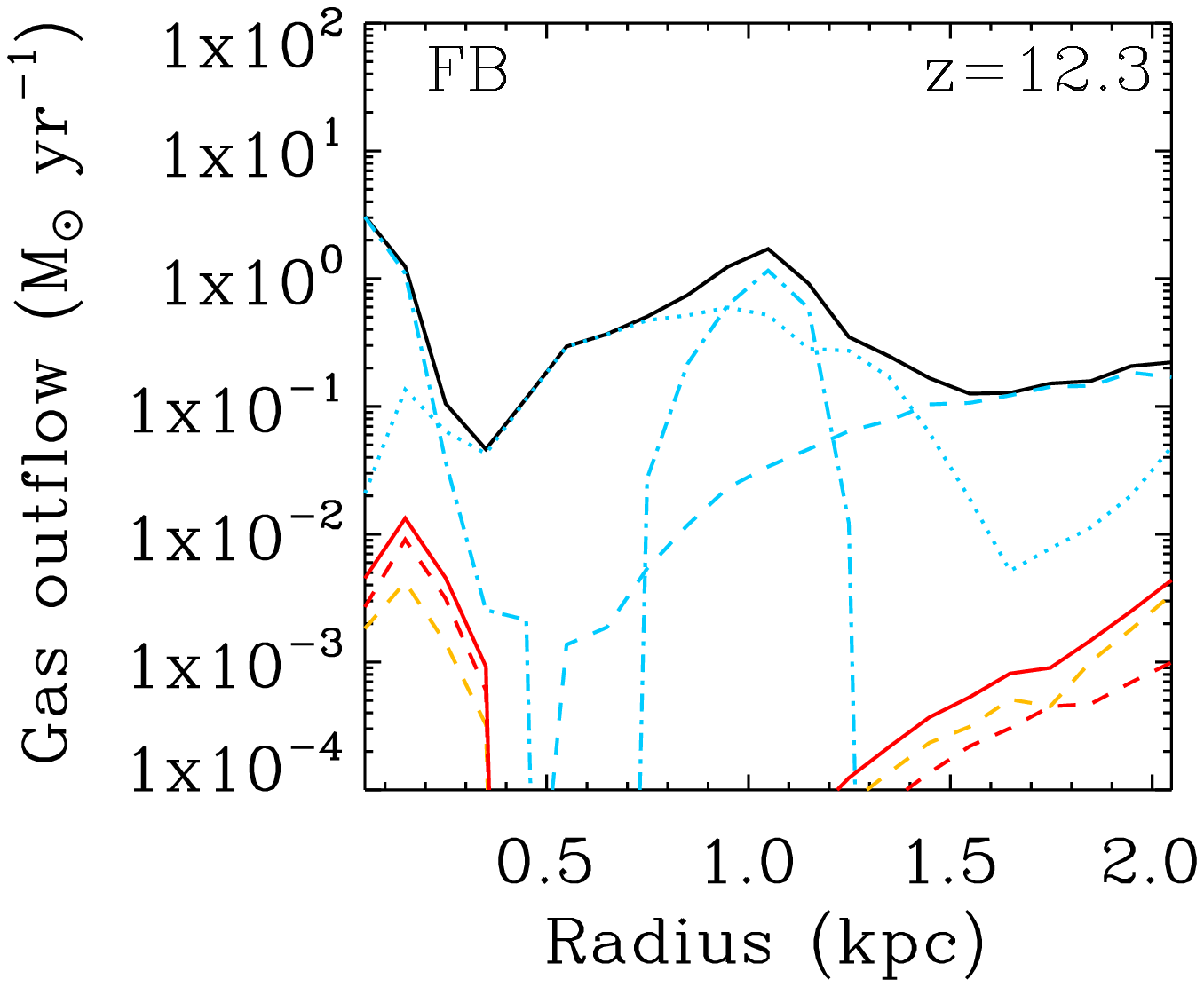}\\

          \includegraphics[width=0.35\textwidth,trim = 4mm 3mm 12mm 12mm, clip]{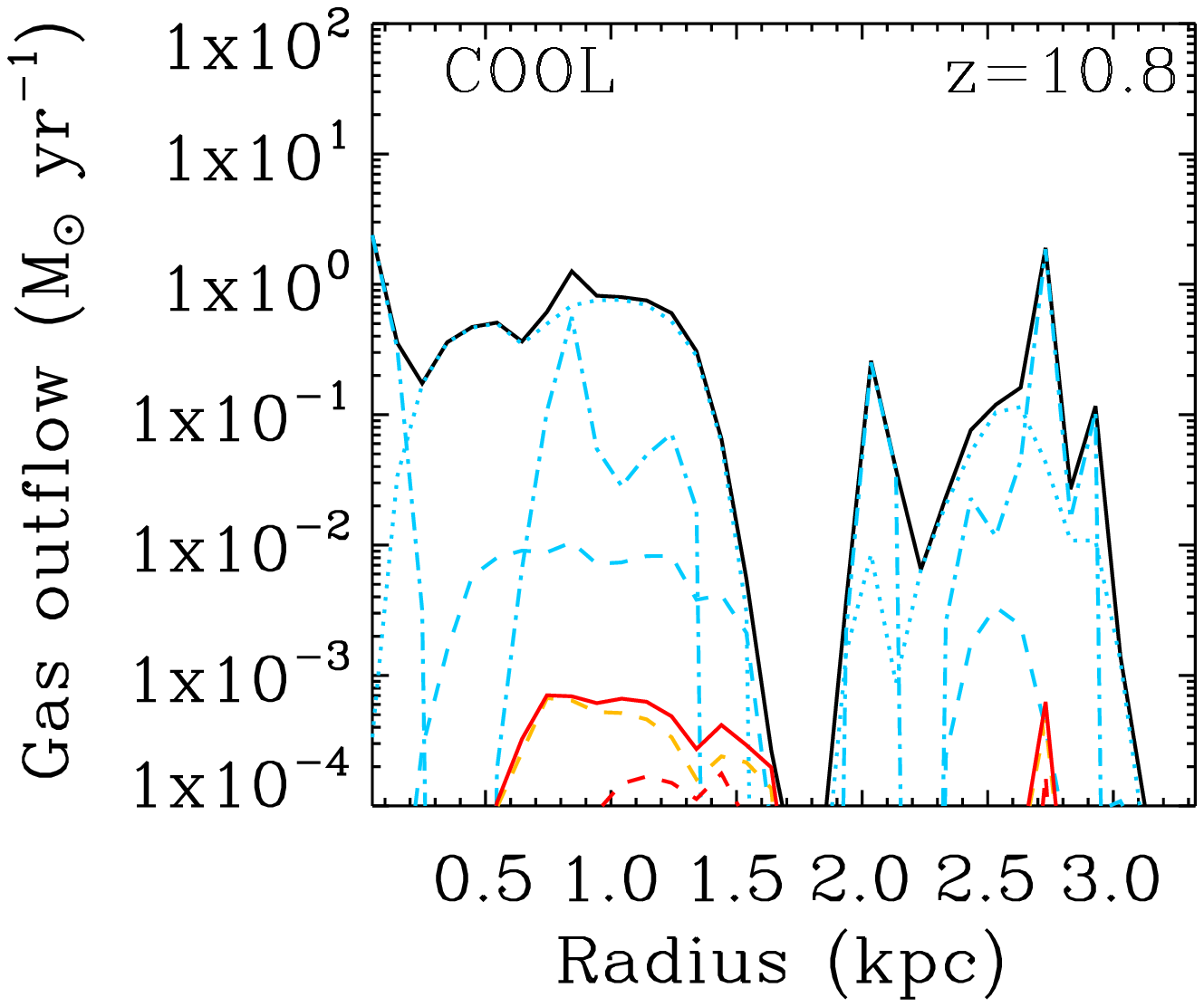}
  \includegraphics[width=0.35\textwidth,trim = 4mm 3mm 12mm 12mm, clip]{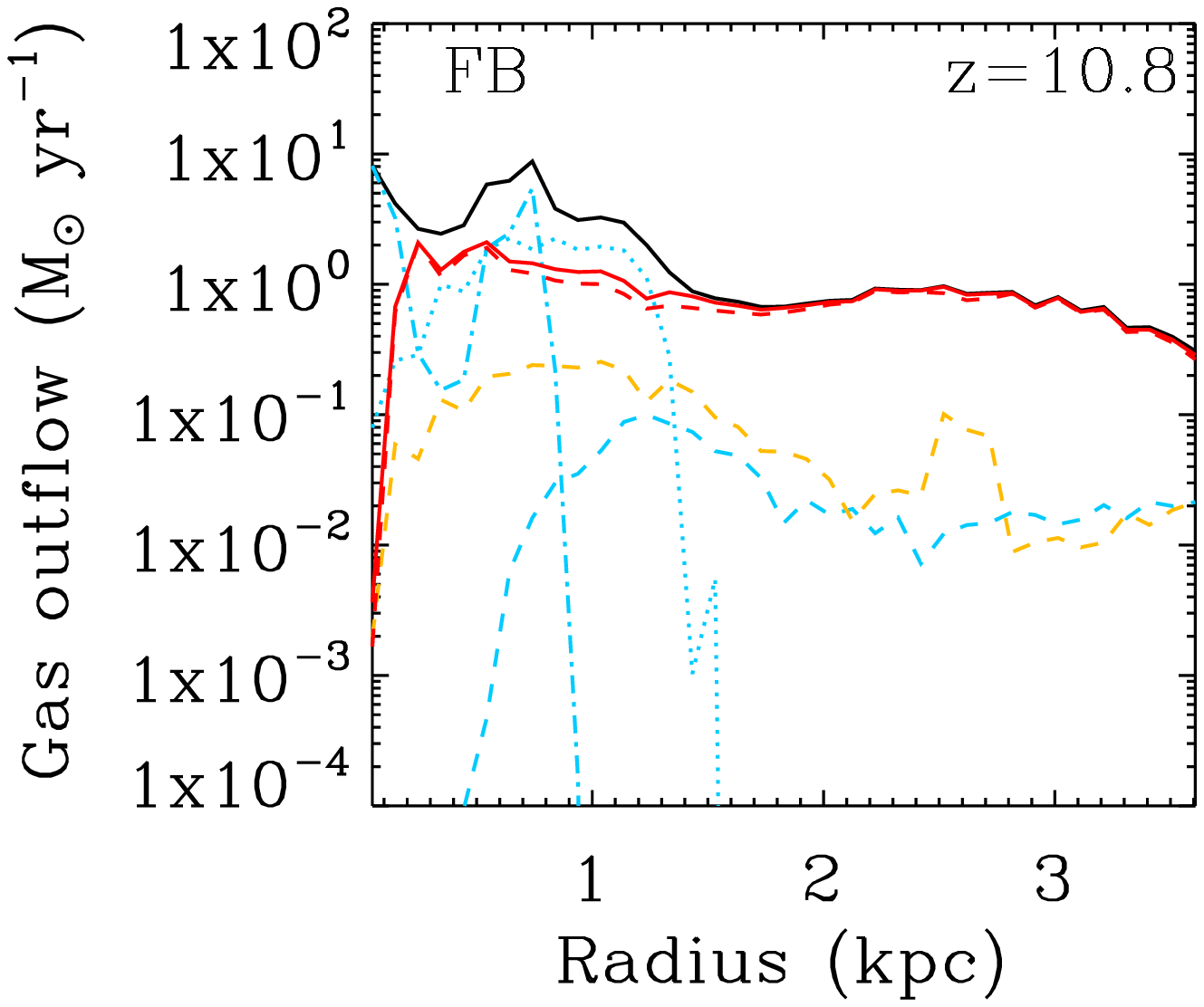}\\

              \includegraphics[width=0.35\textwidth,trim = 4mm 3mm 12mm 12mm, clip]{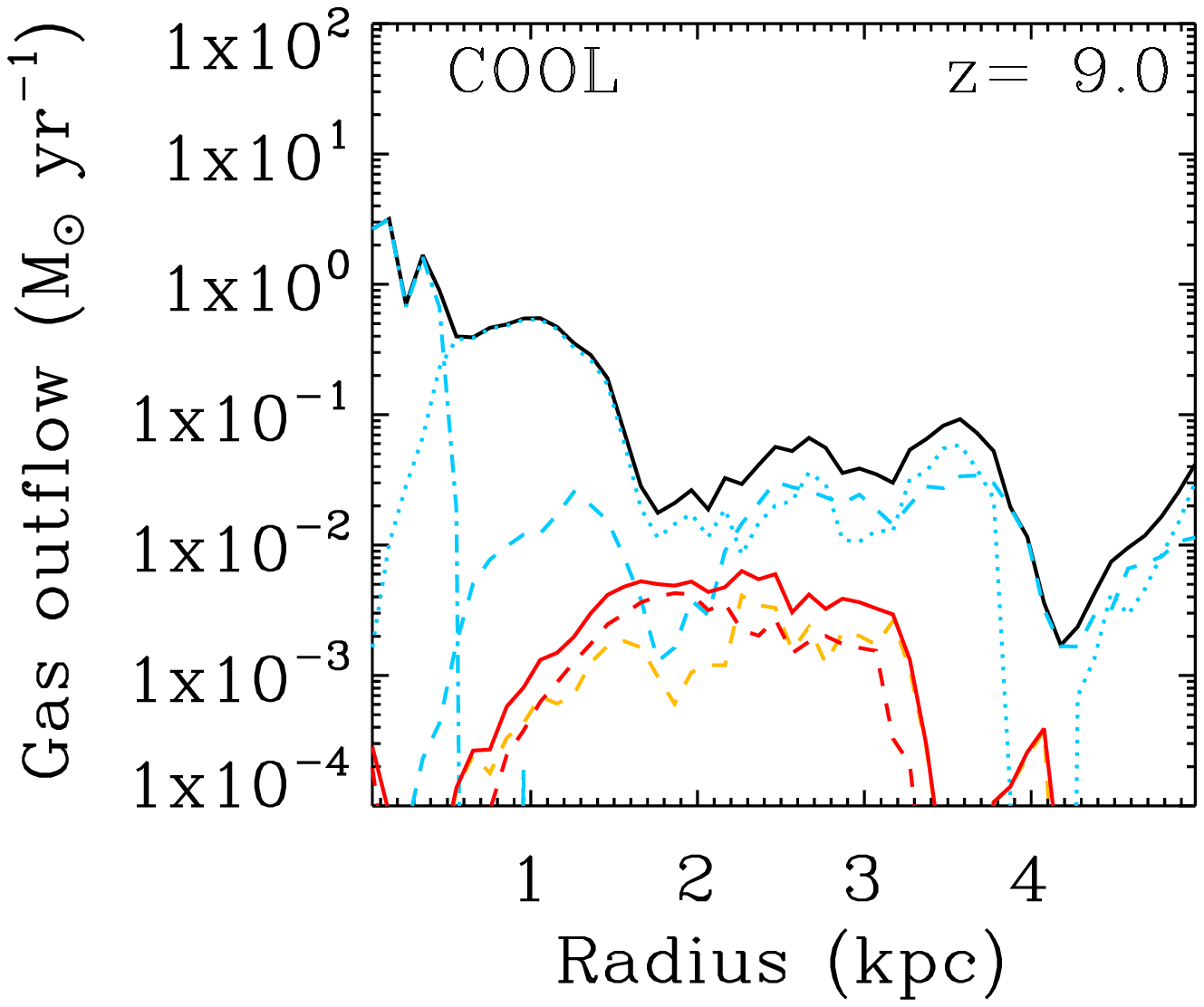}
   \includegraphics[width=0.35\textwidth,trim = 4mm 3mm 12mm 12mm,clip]{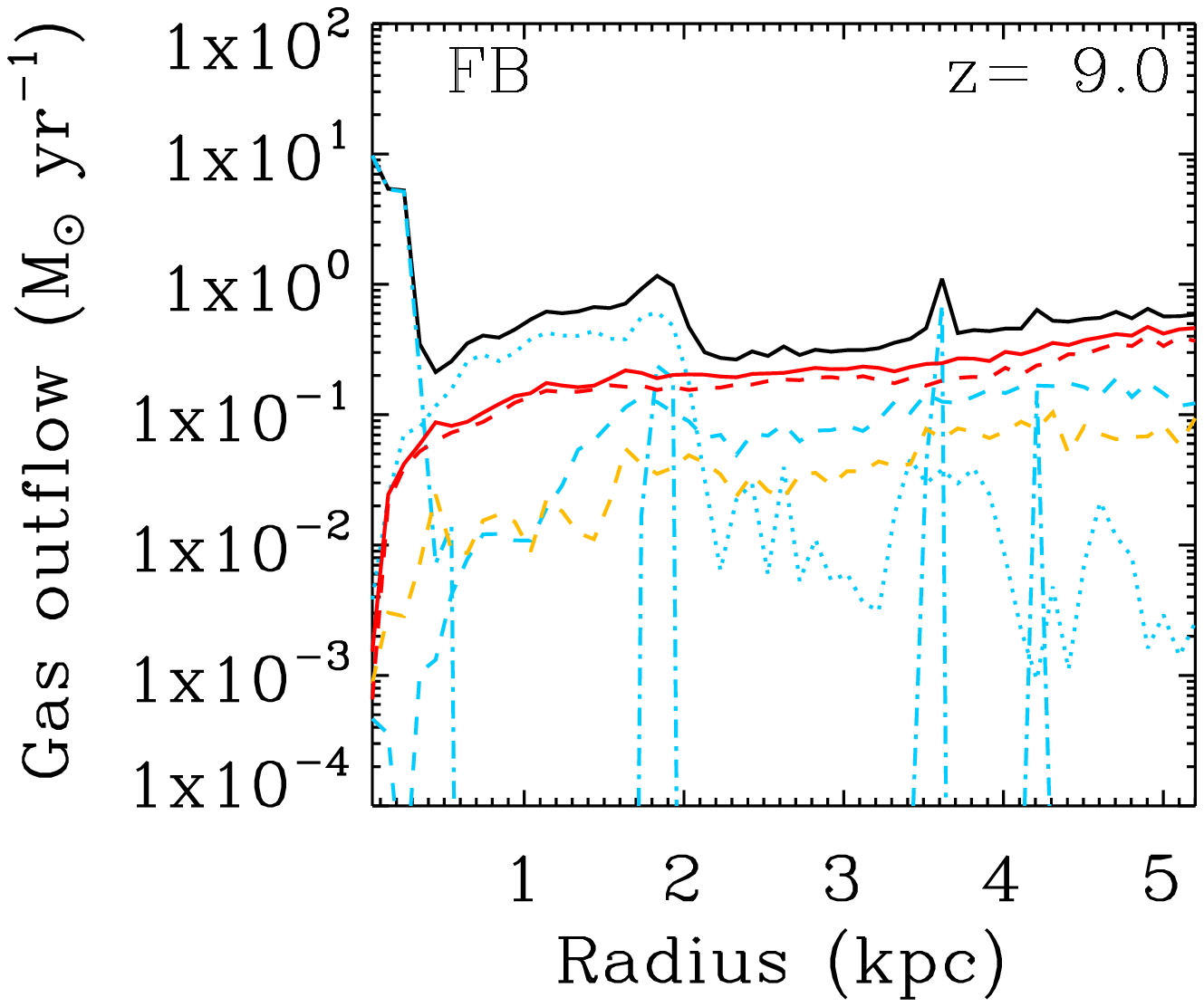}

 \caption{Gas outflow rates averaged in $100$ pc (physical)
spherical shells out to $r_{\rm vir}$ for the cooling (first column) and feedback (second column) runs. Black solid lines show the mass outflow rates of all gas,
which is then split into dense (blue dot-dash line), filamentary (blue
dotted line), cold diffuse (blue dashed line), hot diffuse (red dashed line) and warm
diffuse (orange dashed line) categories as summarized in Table 1. The solid
red line shows the warm diffuse and hot diffuse components
combined.} 
\label{fluxoutvr_fb9} 
\end{figure*}

\begin{figure*} \centering
    \includegraphics[width=0.45\textwidth,trim = 0mm 0mm 5mm 10mm,
clip]{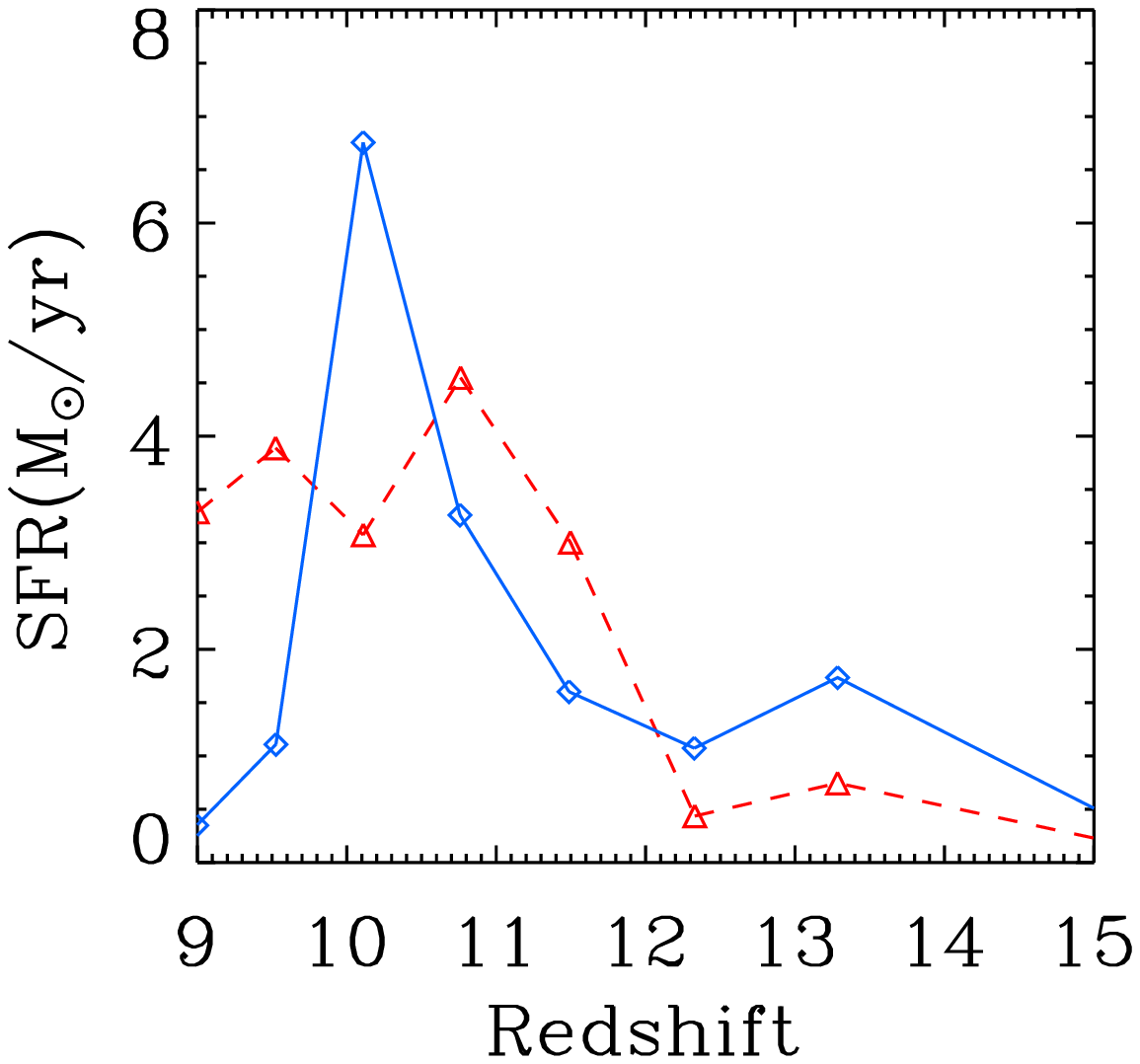}
     \includegraphics[width=0.45\textwidth,trim = 0mm 0mm 5mm 10mm,
clip]{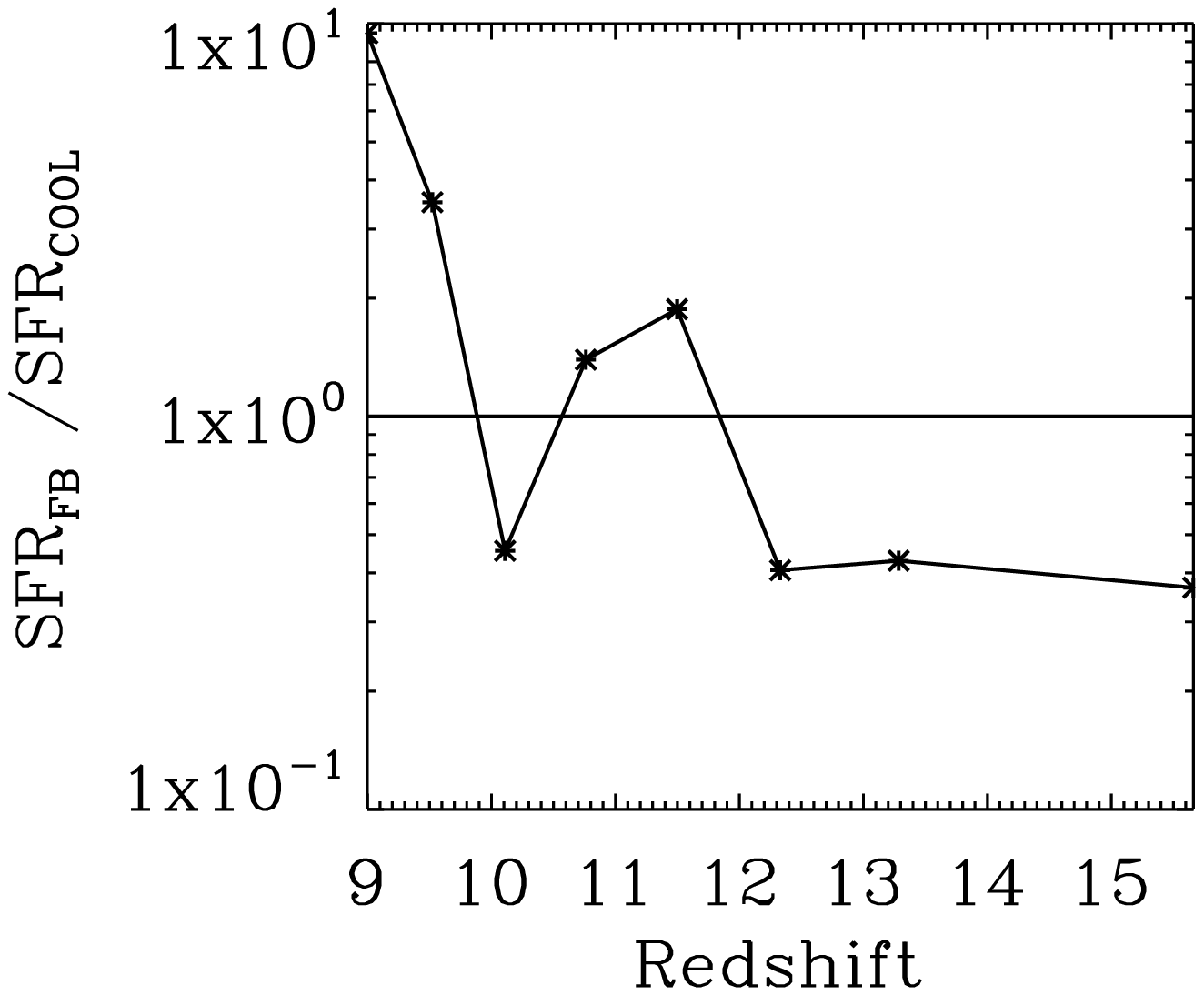}
 \caption{{\bf Left:} Star formation rate averaged over 10 Myr intervals versus
redshift for the feedback run (dashed red line) and cooling run (solid blue line). This is calculated
using the stars within the virial radius of the main progenitor at
each redshift and indicates the in-situ star formation rate
for the main progenitor itself. {\bf Right:} Star formation of main
progenitor only (averaged over 10 Myr intervals) for the feedback run divided by that
for the cooling run, as a function of redshift. The horizontal line indicates
when the SFR in the feedback run equals that in the cooling
run.} \label{sfrmainprog} \end{figure*}

Fig.~\ref{fluxinvr_fb9} shows the gas inflow versus radius (up to the
virial radius) for the cooling (first column) and feedback (second column)
runs. The mass flow is measured for each gas phase in
spherical shells of width $100$ pc physical, centred on the main
progenitor halo, out to its virial radius. 

The total mass
inflow rate (solid black line) is approximately constant with radius,
reaching values above $1 {\rm M}_{\odot}{\rm yr}^{-1}$ by
$z<12$, in both runs and is dominated by the contribution from the
filaments (blue dotted line).This level of inflow compares well with the estimate of the baryonic accretion rate 
at the virial radius derived by simply painting baryons onto dark matter using 
extended Press-Schechter theory \citep{neisteinetal06}. 
For $\Lambda$CDM, this yields:

\begin{equation} \dot{M}\simeq 6.6 (1+z)^{2.25}\left(\frac{M_{\rm
vir}}{10^{12}{\rm M}_{\odot}}\right)^{1.15} \left(\frac{f_{\rm
b}}{0.165}\right){\rm M}_{\odot} {\rm yr}^{-1} \end{equation}

\noindent \citep{dekeletal_nature}. In this formula, $f_{\rm b}$ is
the universal baryon fraction which is $0.17$ in our simulation, $z=9$ and
${\rm M}_{\rm vir}\approx 5 \times 10^{9} {\rm M}_{\odot}$, so we get
$\dot{M}=2.8 {\rm M}_{\odot} {\rm yr}^{-1}$, which is within 50 percent
of our measured values in the cooling and feedback runs (cf. top two rows of
Fig.~\ref{fluxoutvr_fb9}). At $z  = 15.6$, the virial mass of our dark
matter halo has dropped by  a factor $60$, to ${\rm M}_{\rm vir}\approx 8 \times 10^7{\rm M_\odot}$.
Using the same formula we therefore predict an accretion rate of  $\dot{M}=0.073 {\rm M}_{\odot} {\rm yr}^{-1}$
at this redshift, which is lower, but of the same order of magnitude
to that which we measure  ($\dot{M}=0.3 {\rm M}_{\odot} {\rm yr}^{-1}$)
in the simulation.  This confirms that our resimulated halo is a fairly
`typical' halo and that gas, even in high resolution simulations 
is being accreted proportionally to dark matter, at least at the virial radius.

While the similarity of the gas accretion rate to the prediction for
the dark matter accretion rate at the virial radius is not very surprising,
the near constancy of the filamentary gas accretion rate all the way
down to the disc is more remarkable. This is caused by the radial velocity of the gas 
in the filaments remaining constant around $\approx 50$km/s, independently of radius.
Indeed, whereas the dark matter filaments are
dynamically heated when they penetrate the virial radius, the gaseous filaments 
radiate away the extra gravitational energy (recall their temperature is stable around $\sim 10^{4}K$) 
to continue their journey supersonically (at about Mach 5) deep into the centre of the dark matter
halo, all the way to the central disc.

On the other hand, the mass inflow rate in the cold diffuse component (blue dashed lines Fig~\ref{fluxinvr_fb9})
increases with radius, until it becomes comparable with that in the
filaments at the virial radius.  This trend with radius is a
geometrical effect; as the shell surface area decreases with decreasing radius, more and more diffuse
gas is compressed and joins the filamentary phase, and thus the inflow
of diffuse cold gas per unit time is reduced as the radius shrinks. 
As the filaments have a roughly constant cross-sectional
area (as illustrated in the middle row of Fig.~\ref{3dfluxcuts}),
no such effect occurs and thus they rapidly become the dominant supply of gas in
the inner regions of the halo. The accretion rate is very similar for the feedback and cooling runs
and so it seems that the geometry of the supply of cold gas to the
central object means that the SNe feedback is unable to
significantly diminish it. The main source of such gas is the dense,
collimated filaments, which present more pressure resistance to the
galactic wind than the more symmetrically distributed diffuse
material.

The impact of SNe feedback on accretion is also considered in
  \citet{vandevoort_etal_2010}, in which the authors examine
  accretion onto a large sample of galaxies and their haloes in a suite of cosmological
  simulations with different physics. We note that their recipe for SNe feedback involves adding in a galactic wind `by hand' (i.e. by applying a velocity kick to particles near an exploding star particle) and so their technical approach is very different to ours. A direct comparison to our
  results is challenging, not least because we are considering a very
  different halo mass and redshift range ($\approx 5\times10^9 {\rm M}_{\odot}$ halo
  at $z>9$, compared to haloes with $M \ge 10^{10} {\rm M}_{\odot}$ at
  $z<5.5$). In the discussion that follows, we will compare to their results for haloes with $M_{\rm vir}\ltsim 2\times 10^{11} {\rm
    M}_{\odot}$ at $z=2$, as this is the closest match to the halo in our study. \citet{vandevoort_etal_2010} find that including SNe feedback reduces the total smooth
  accretion rate by a factor of $\approx 2.5$  for haloes in this range and by a factor of $\approx 10$ for the
 central galaxies hosted by these haloes. This suggests that accretion from the IGM has been considerably impeded by SNe feedback; this is in stark contrast to our results.

A clue to the possible origin of this 
difference can be found in their Fig.~7 (last row) where the lowest
mass halo illustrated (which has $10^{11.5} {\rm M}_{\odot}$) is
embedded {\it within} a filament. This means there 
is no lower density material for a SNe-driven wind to preferentially
flow into. This is very different to our lower mass, but higher
redshift halo, which is fed by 3 filaments that are thin 
compared to its virial radius. It is interesting to note that for their  $10^{12} {\rm M}_{\odot}$ halo, which {\it is} fed by 3 filaments (see their Fig.~7, middle row) the total accretion rate onto the halo is not altered by SNe feedback. We would therefore argue that the {\it
  geometry} of the accretion must be considered, not just the halo 
mass (although the two are interlinked) in order to accurately predict/explain the effects of galactic winds on gas accretion.

The spikes in the inflow in the clumpy component (blue dot-dash
lines in Fig~\ref{fluxinvr_fb9})
correspond to satellite galaxies, or other over-dense clumps of
gas. Note that there is always a spike in the clumpy inflow rate at
the centre (at $r=0$); this is due to the presence of a rotating gas
disc, which also gives rise to similar central spikes in the
corresponding outflow plots (Fig.~\ref{fluxoutvr_fb9}. Indeed, since gas discs are generally not axisymmetric, as different rotating
  elements of gas of the same disc pass through a spherical 
shell they can contribute either negatively or positively to the radial velocity 
component of the shell, therefore leading the gas in the disc to be detected both
as inflowing and outflowing.

There is both warm inflow (orange dashed line) and to a lesser extent, hot
(red dashed line) inflow, in the cooling run (Fig.~\ref{fluxinvr_fb9}, first column). While this is around 3 orders of magnitude less than the
contribution from the filaments it indicates the possibility for gas
to be heated in low mass haloes even without
feedback. Fig.~\ref{3dfluxcuts} revealed that the origin of this
heated material is a shock front, where cold diffuse material is
heated as it impacts the denser filamentary material surrounding the
disc (this, and the properties of gas inflow in general, is discussed 
in more detail in Powell et al,  in prep). There is a similar rate 
of mass inflow in the hot and warm components
in the feedback run, but it is dwarfed by
the rate of outflow.

\subsection{Outflow}

Fig.~\ref{fluxoutvr_fb9} shows the gas outflow versus radius (up to
the virial radius) for the cooling (first column) and feedback (second column) runs, respectively. The rate of mass outflow in the feedback run
is dominated by the contribution from the hot component. Despite
velocities reaching several $100$ km/s (see Fig.~\ref{mwvrvr_big},
left) for this component, the mass outflow rate is only
around $10-30$ percent of the total inflow. This is simply because the hot
gas has very low densities; Fig.~\ref{histofluxcuts} (bottom panel) shows
that the greatest mass fraction of hot gas occurs at densities of
$\sim 10^{-3}$ atoms/cc.

The cold diffuse outflow in the feedback run is a factor of
$10$ higher than that in the cooling run (see particularly $z=9$),
reinforcing our interpretation of Figs~\ref{radhisto_fb9} and \ref{metalradhisto_fb9} 
that hot outflows from SNe sweep away some of the cold diffuse gas 
falling onto the central galaxy. This fraction of swept-up material is about  
$10$ percent of the amount of cold diffuse gas which flows in,
suggesting that the material being swept out is the low density ($\sim
10^{-3}$ atoms/cc) cold ($10^{3}$K $<T< 10^{4}$K) tail of the distribution seen in the
$\rho-T$ phase diagram for the cooling run (Fig.~\ref{histofluxcuts},
top panel). Indeed this gas is clearly missing in the feedback run
(compare the sections labelled `CD' in Fig.~\ref{histofluxcuts}, for the cooling (top) and feedback (bottom) runs). As with the filamentary
component, the impact of feedback on the cold diffuse accretion
rate is surprisingly small even though a high velocity galactic wind
has been successfully launched.

The large spikes in the mass outflow rate of the clumpy component
(blue dot dash line) in both runs are due to the presence of rotating gas
discs in satellite galaxies. The outflow seen in the filamentary
component (blue dotted line) is a result of the `looping' of the filaments
(particularly pronounced in the central regions of the halo in the
cooling run) and, to a lesser extent, material associated with the
outer region of satellites which is not distinguished from the
filament in which the satellite is embedded due to the smooth density
gradient.

\section{Correlating accretion, outflows and star
formation}\label{sec:sfr}

One of the main reasons why the cold-mode of accretion has received so much
attention, is because of its implications for the star formation histories of galaxies. 
In the traditional picture, star formation is
delayed as the gas is first shock heated to the virial temperature of the dark matter halo and 
requires time to cool and condense isotropically onto the central galaxy.
If, on the other hand, gas comes in cold and fast along dense filaments, 
star formation can commence earlier, be much more rapid and localized 
in space. In order to quench such
  starburst episodes, the supply of fuel for star formation has to be
  shut off by e.g. SNe feedback. However, since galactic winds are
  expected to take the path of lowest pressure to propagate in the
  halo,  the filaments could be resistant to destruction and the 
  starburst prolonged. Therefore, in this
section we investigate how star formation rates correlate
with net mass inflow rates, comparing the cooling and feedback runs
to quantify the impact of SNe.

\begin{figure} \centering
    \includegraphics[width=0.45\textwidth,trim = 0mm 0mm 5mm 10mm,
clip]{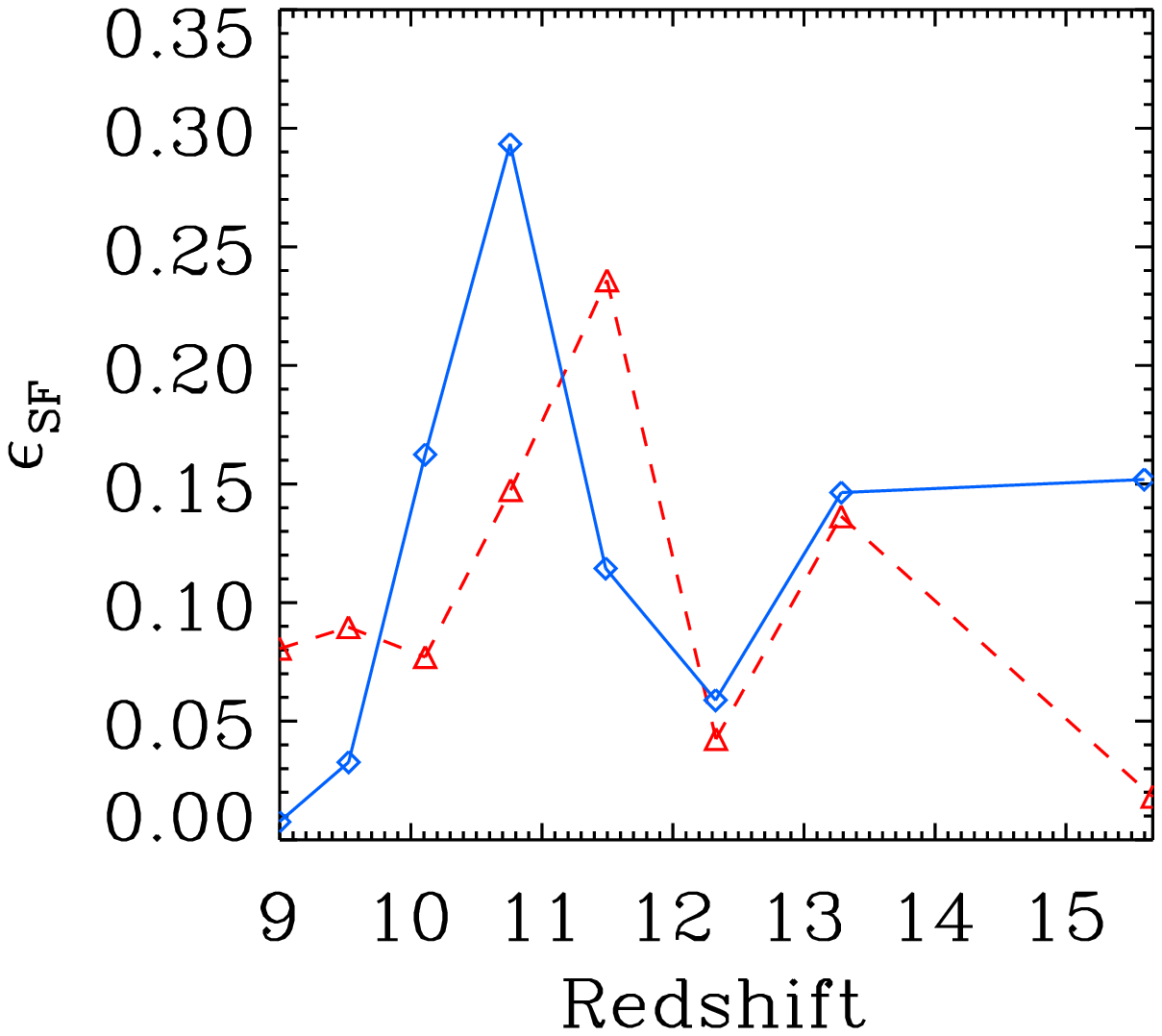}
 \includegraphics[width=0.45\textwidth,trim = 0mm 0mm 5mm 0mm,
clip]{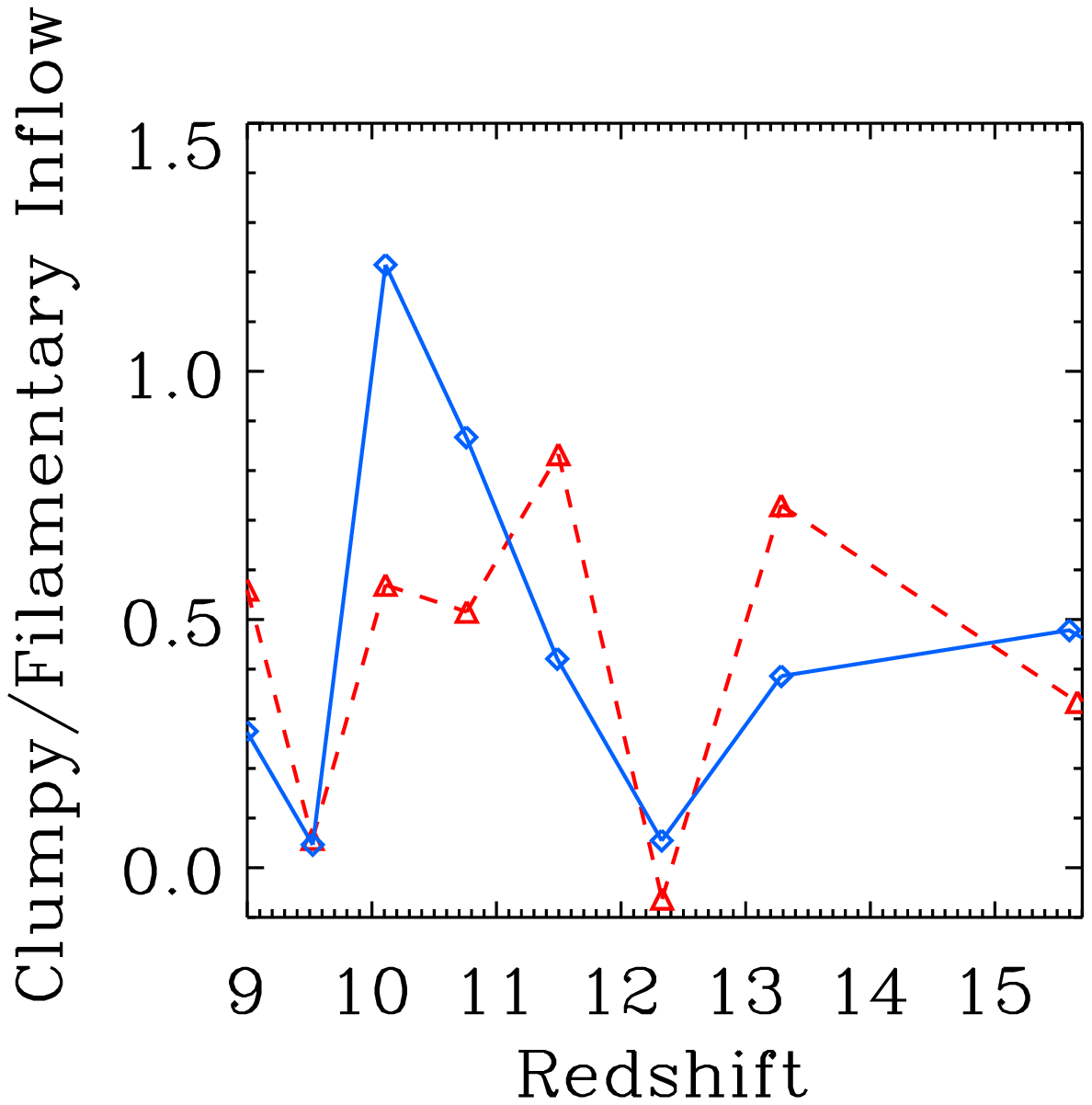}
 \caption{{\bf Top:} The Kennicutt-Schmidt star formation efficiency versus
redshift for the cooling (blue solid line) and feedback (red dashed line)
runs. These quantities are computed for the main progenitor, within a
sphere of radius $0.1 r_{\rm vir}$. {\bf Bottom:}  Ratio of net clumpy inflow to net filamentary inflow versus redshift for the cooling (blue solid line) and feedback (red dashed line) runs. The value of the inflow rate at each $z$ is obtained by finding the net inflow per radial bin and averaging over the radial bins within $r_{\rm vir}$ (full data with radial dependence is presented in Figs.~\ref{fluxinvr_fb9} and \ref{fluxoutvr_fb9}).} \label{ks} \end{figure}

To examine the star formation rate (SFR) of the main progenitor in isolation we
analyse the stars within the virial radius at each available output
and compute the SFR in a small time interval only (10 Myrs), thereby
excluding stars formed outside the main progenitor and accreted during mergers
(left panel of Fig~\ref{sfrmainprog}). The right panel of 
Fig.~\ref{sfrmainprog} shows the ratio of these in-situ star 
formation histories in the feedback and cooling runs. Whilst the SFR in the cooling run is a factor 2 larger than 
that in the feedback run for $z > 12$, the situation is reversed at lower
redshifts ($z < 10$) where the SFR of the feedback run exceeds that in the 
cooling run by up to a factor 10. This is intriguing as
it suggests that SNe are a source of positive
feedback, i.e. they enhance star formation rather than inhibit it. 
Although in order to confirm the origin of this positive feedback effect
it would be necessary to perform another {\sc nut} simulation with supernova
feedback turned on but metal enrichment switched off, we attribute 
most of it to the extra cooling provided by metal lines in the 
feedback run rather than, for example, compression of the ISM by blast waves.
Indeed it is obvious from Fig~\ref{histofluxcuts} that such an extra channel for cooling 
enables a larger mass fraction of cold gas to reach the star forming density threshold
in the feedback simulation (bottom panel).

To assess how these differences affect the global efficiency of star formation, we estimate the 
`effective' star formation efficiency, $\epsilon_{\rm sf}$, in
both runs by writing a Kennicutt-Schmidt law for the entire central galaxy which gives,

\begin{equation}
  \epsilon_{\rm sf}={\rm SFR} \frac{t_{\rm dyn}}{m_{\rm gas}}
   \label{effeqn}
    \end{equation}

\noindent where $m_{\rm gas}$ is the mass of gas measured inside a sphere centred on the galaxy, 
with radius $0.1r_{\rm vir}$ and we take $t_{\rm dyn}\approx t_{\rm ff}$ as the dynamical time. 
Here $ t_{\rm ff}$ is the usual free-fall time, i.e. the time for the mass enclosed in the sphere 
to collapse under its own gravity i.e.,

\begin{equation} t_{\rm dyn}\sim t_{\rm ff} = \sqrt{ \frac{3\pi}{32 G
\rho}} \end{equation}

\noindent The dynamical time varies by a factor of around $2$, over the redshift
range examined (z $\approx 9-15$), from $\approx 5$ Myr at high redshift
to $10$ Myr at low $z$.

\begin{figure*} \centering
  \includegraphics[width=0.45\textwidth,trim = 0mm 0mm 5mm 0mm,
clip]{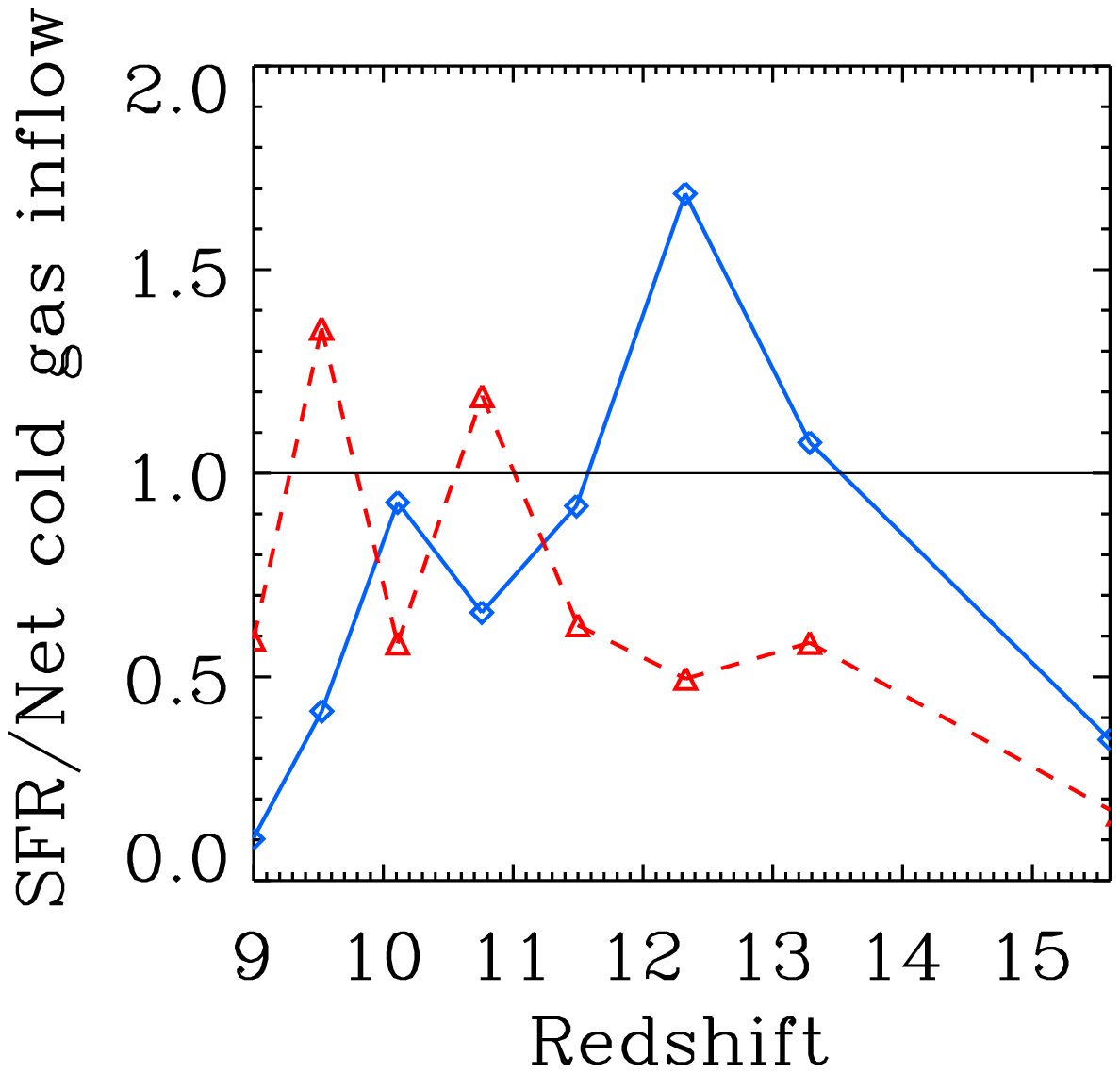}
  \includegraphics[width=0.45\textwidth,trim = 0mm 0mm 5mm 0mm,
clip]{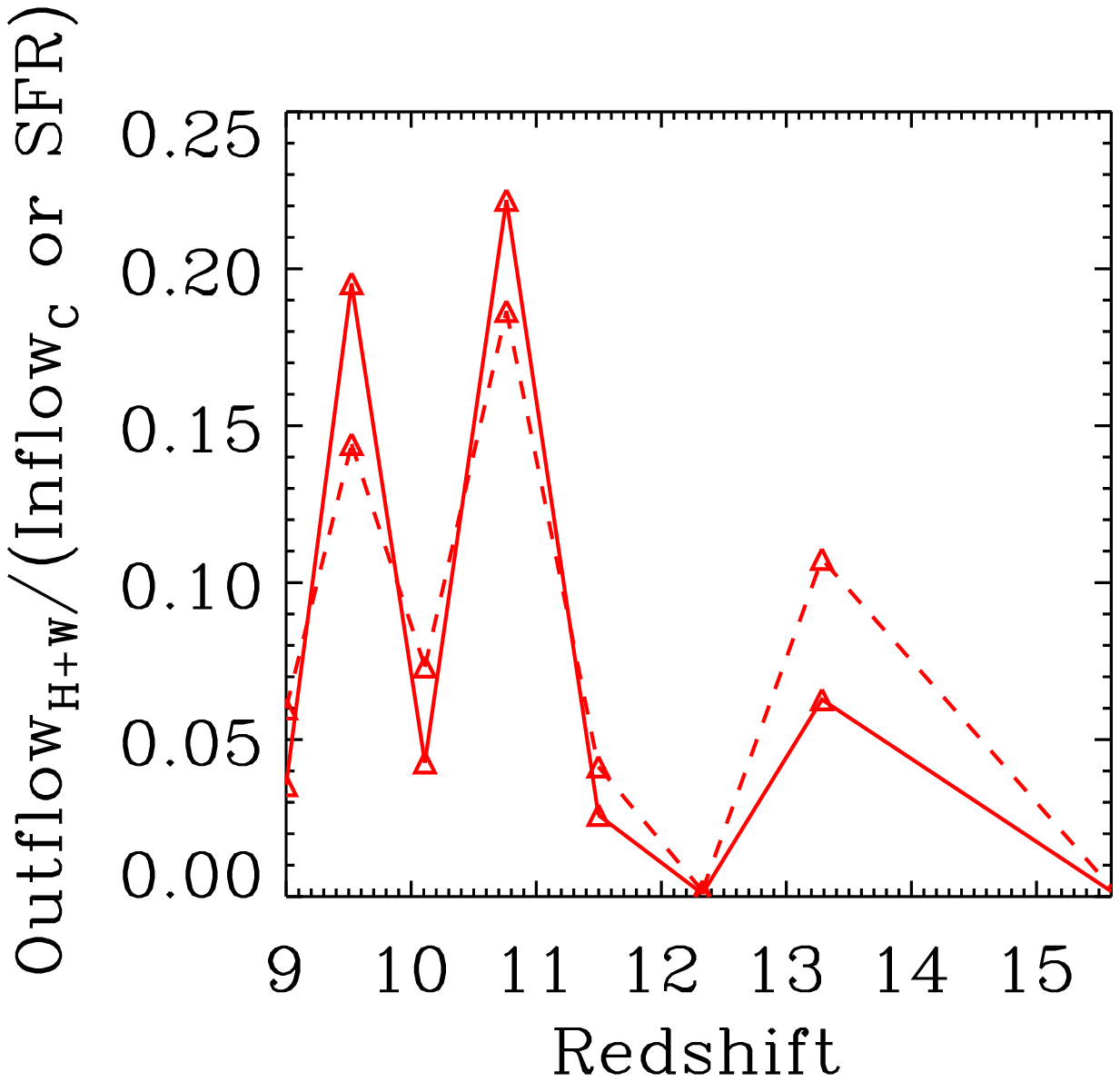}
    \caption{ {\bf Left:} Star formation of main progenitor only
(averaged over 10Myr) divided by net cold gas mass inflow rate versus
redshift for the cooling (blue solid line) and feedback (red dashed
line) runs. The horizontal line indicates when the cold gas inflow
rate equals the SFR. {\bf Right:} Net mass outflow rate of warm + hot
gas divided by the net mass inflow rate of cold gas (solid line) or
SFR (dashed line) for the feedback run. The value of the inflow or
outflow rate at each $z$ is obtained by 
finding the net flow per radial bin and averaging over the radial bins within $r_{\rm vir}$ (full data with radial dependence is presented in Figs.~\ref{fluxinvr_fb9} and \ref{fluxoutvr_fb9}).}
\label{ratio} 
\end{figure*}

Fig.~\ref{ks} (top) shows the efficiency, $\epsilon_{\rm sf}$, as
defined in equation \ref{effeqn} for the cooling (solid line) and
feedback (dashed line) runs. Note that this effective (global) star formation
efficiency is, on average, an order of magnitude higher than both the
typical star formation efficiency of $0.01 - 0.02$ determined from
observations of the local Universe \citep{krumholz_tan07} and the
local efficiency parameter in {\sc ramses}, which we set to $0.01$ 
in both cooling and feedback simulations (for details of
the star formation implementation see the Appendix). Such low efficiencies of
this order of magnitude are only measured in the cooling run for
$z \approx 9$.

This effect can be understood as follows. Equation \ref{effeqn} represents an averaging of the star formation
properties of the whole region we are considering, whereas the
efficiency parameter in the simulation can only control the star
formation rate in individual cells which have exceeded the required
density threshold. If star formation occurs in a clumpy medium where
many regions exceed the density threshold for star formation, the
overall efficiency of the object will be larger than if the medium is
more homogeneous. Higher densities in a clumpy medium also give rise
to shorter star formation time-scales on a cell by cell basis. It
seems appealing to conclude that, at very high-redshift, star
formation on galactic scales is more efficient, not because it is more
efficient on local cloud core scales, but because a larger fraction of
the ISM gas mass is locked in dense clumps (clumps are evident by eye in the gas disc in the
feedback run: see Fig.~\ref{tempmaps}, bottom right), than at low ($z \approx 0$) redshift.

This view receives some support from observations of clumpy galaxies at $z>1$ \citep[e.g.][]{clump_cluster}.
These clump masses are estimated to lie around $M\sim10^{8}{\rm
  M}_{\odot}$, which are consistent with the findings of high resolution hydrodynamical simulations. 
Indeed \citet{agertz_clumpygal} and \citet{ceverino_clumpygal} report massive clumps of order $10^{8}{\rm M}_{\odot}$ at $z=2-3$
The clumps we see in our simulated disc at $z=9$ are less massive ($\sim 10^{5-7}{\rm
M}_{\odot}$), but we believe they are the high redshift
counterparts of the observed (and simulated) clumpy galaxies.
We defer a detailed study of these stellar clumps to a
companion paper (Slyz et al, in prep.), but we note that their
  typical masses ($M_{\rm clump}\sim 10^6 {\rm M_{\odot}}$) are 
consistent with the Jeans mass values one expects from the gravitational 
fragmentation of the gaseous disc.

In order to understand how the cold gas components fuel star formation
and how they yield the high global star formation efficiencies
measured, we evaluate how the SFR correlates with the net mass flow
averaged over all radii.
In Fig.~\ref{ratio} (left panel), we examine the ratio of the SFR to the cold gas
inflow rate versus redshift for the cooling (blue solid line) and
feedback (red dashed line) runs. In the interval $9\leq z\leq10$ the
cold gas supply far exceeds the SFR in the cooling run. In the feedback run during the same
time period, the dashed line oscillates around $1$, showing that a
amount of gas comparable to the amount of incoming material is 
converted into stars. However, the high
efficiency of star formation that we measure 
on the scale of the galaxy (Fig~\ref{ks}) seems uncorrelated with the epoch of high
inflow rates, whereas the global star formation rate is, as shown by 
the lines on the right panel of Fig.~\ref{ratio}. Once again, we interpret this as 
evidence that the global efficiency is more sensitive to the local 
microphysics than the global properties of mass accretion. Indeed, the
epochs of high global efficiency correlate quite well with the times at
which the radially averaged clumpy component of cold gas
accretion is high (see Fig.~\ref{ks}, bottom), i.e. mergers/strong interactions
with neighbouring galaxies take place. These are thought to trigger violent instabilities 
which will compress the ISM (or increase turbulence and therefore increase fragmentation) and allow a larger fraction of it to reach
the threshold for star formation, resulting in higher global
star formation efficiencies  and a merger induced starburst. Note
  that when the ratio of clumpy to filamentary 
inflow reaches its peak, we also see a corresponding peak in the SFR
(left panel, Fig.~\ref{sfrmainprog}).
Compared to these violent events the 
impact of smooth filamentary accretion on the global star formation
efficiency is a second order effect, even though it maintains 
it at a higher level (between $5-10$\%) than the $\approx 1$ \% expected from 
our choice of efficiency parameter on small scales, which should
remain unchanged by diffuse accretion alone.

Finally, Fig.~\ref{ratio} (right panel) shows that the hot outflow rate is on average one order
of magnitude below the SFR. This is contrary to observations of lower
redshift galaxies in \citet{cmartin99_outflows}, and \citet{heckman_etal_2000} who estimate the mass-loss rate to be comparable to the
SFR. This means that the hot gas outflow rate is also about a factor of $10$ lower
than the rate of mass entrained by SNe ejecta in our modelling
(see the Appendix for details of SNe feedback), 
suggesting that a significant proportion of this gas mass does reform 
stars on very short timescales. Whether this is caused by a lack of implementation
of physical processes known to occur in the interstellar medium
(e.g. radiative transfer, stellar winds) or this is an intrinsic problem of (high-redshift)
SNe  driven winds remains to be established. 

In light of this study, we conclude that SN explosions
alone are very inefficient at driving massive galactic winds from
typical $L_\star$ galaxies at high redshift. We stress that this finding is resolution-dependent; at
lower resolutions the Sedov-Taylor solution necessarily
represents multiple SN remnants and a more massive wind may be driven. Since
such a solution has only been proven to accurately describe an {\it
  individual} SN remnant, we are confident that the low mass wind we
produce is a robust result. However, there are additional 
processes that could affect the mass-loading of the wind which are not
included in our simulations, such as stellar winds, ionizing photon heating,  cosmic
rays and magnetic fields. Furthermore, as in all cosmological
simulations, we use a (simple) subgrid model for star formation and have
assumed an IMF, both calibrated on local
Universe observations. As the rate and energy of SNe are
dependent on these assumptions so, in turn, should be the mass outflow
rate in the wind.

\section{Conclusions}\label{sec:conc}

We have undertaken a detailed analysis of a suite of subparsec
resolution simulations (the {\sc nut} simulations) with the aim of
understanding the mechanism via which a far-reaching galactic wind
can arise in a protogalaxy at high redshift 
 ($z\approx9$) and how this impacts the surrounding IGM and the galaxy's evolution. 
In particular we examine the relationship between cold-mode, filamentary
accretion and star formation and explore the impact that the
introduction of SNe feedback and the resulting wind has on these processes. \\

\noindent Our main findings are:

\begin{enumerate}

\item A far-reaching SNe driven wind (illustrated in
  Fig.~\ref{tempmaps}, top right) does develop; this is particularly
  significant since we do not directly model the wind in our feedback
  scheme, but instead we resolve the Sedov blastwaves of individual
  SNe and the wind arises naturally. We find that SNe explosions occur
  not only in the main progenitor, but also in subhaloes and
  neighbouring dark matter haloes (demonstrated in the  
temperature-radius histograms, Fig.~\ref{radhisto_fb9}). These create
overlapping bubbles which are critical for creating the far-reaching
wind and for facilitating the escape of the gas from the halo
potential. Since subhaloes are key, we speculate that similar winds
can also be produced by higher mass haloes at lower redshifts, and we
will certainly investigate this issue in the very near future. The wind
is enriched with metals to the order of $0.1 Z_{\odot}$ (see the
metallicity profile in Fig.\ref{mwvrvr_big}, top right). While one
must be careful when comparing high-redshift simulations 
results to local observations, we conclude that it seems feasible 
for such a wind to enrich the IGM to the level required by 
observations at $z \approx 2$ . On the other hand, we find the
ejection of mass by the wind is very inefficient: the outflow rate in 
the hot phase is only $10-30$ per cent of the total mass inflow rate. 
We also find that the mass outflow rate is about an order of magnitude
lower than the SFR, in disagreement with local observations 
\citep[e.g][]{cmartin99_outflows, heckman_etal_2000}, perhaps
indicating an important difference between low and high redshift
SNe driven winds or a missing physical process in the simulation 
which drastically alters the mass loading factor of the wind (stellar
winds, radiative transfer). We plan to address this issue in a
forthcoming paper. 

\item The total gas accretion rate at the virial radius of the
  main progenitor (see Fig.~\ref{fluxoutvr_fb9}, top two rows) ranges
  between $1-10 {\rm M}_{\odot} {\rm yr}^{-1}$ depending on redshift,
  and is in fair agreement with analytical predictions based on
  extended Press-Schechter theory, indicating that the gas accretes
  like dark matter at the virial radius (note that the accretion is cold). 
 However, in contrast to the accretion of dark matter, this rate
 persists for the gas filaments right down to the central object and
 thus these filaments are the dominant supply of cold gas to the
 central disc (compared to clumpy or diffuse, spherical cold accretion).
 While the filaments appear more perturbed in the inner halo in our 3D 
 visualisations of the feedback run than in those of the run without
 SNe feedback (Fig.~\ref{3dfluxcuts}, second row), we measure similar 
 mass accretion rates in both runs, suggesting that the majority of
 the gas survives its journey to the central object intact and gets
 replenished by cold gas coming in from larger distances. So, despite 
 the development of a galactic wind, the cold, filamentary accretion
 is not significantly altered. This suggests that, at least at $z \approx 9$, 
 SNe driven winds cannot reduce the amount of cold, accretion onto 
 dwarf-mass protogalaxies and therefore cannot significantly reduce
 star formation at this epoch, as is commonly assumed.

\item At the lower end of the redshift range studied ($z<12$), the SFR in the feedback run is
  more often than not greater than that in the cooling run 
  (see Fig.~\ref{ratio}, left for the ratio of SFRs versus redshift). 
  An extreme example is at $z=9$ when the SFR in the feedback run
  is $\approx 3 {\rm M_{\odot} {\rm yr}^{-1}}$, yet only  $0.3 {\rm  M_{\odot} {\rm yr}^{-1}}$ 
  (the minimum SFR over the redshift range considered) in the cooling
  run. We attribute this `positive' feedback effect to the metal
  enrichment active only in the former (we have already 
  demonstrated that the gas accretion rates are very similar in both
  runs). Increased metallicity means more efficient cooling and this 
  combined with the wind which can spread the metals, means a 
  greater mass of gas will cool and condense enough to form stars. 
  In fact, we find that the mass outflow rate is significantly lower than the rate of input of ejecta and entrained ISM into the SNe blastwaves, suggesting much of this material is recycled for next generation star formation.

\item Star formation is globally very efficient in both the cooling and
feedback runs. Even though in the star formation implementation we set
the efficiency to $0.01$ when the gas density in a cell goes over
$10^{5}$ atoms/cc, we measure average efficiencies of $\sim 0.2$
within $0.1r_{\rm vir}$. This indicates that star formation at high
redshift proceeds in dense clumps that represent a larger fraction of
the total gas mass of the galaxy than they do at low redshift. The
global efficiency is well correlated with the average inflow 
rate of clumpy gas (i.e. gas that is condensed in
satellites/subhaloes) (see Fig.~\ref{ks}, bottom) i.e. with the
occurrence of mergers or strong interactions between neighbouring 
galaxies. When mergers occur the ISM is compressed and/or 
turbulence triggers increased fragmentation, allowing a larger
fraction of the gas to reach high enough densities for star formation, 
resulting in higher global star formation efficiencies.

\end{enumerate}

\section{Appendix: Gas physics} \label{sec:appendix}

\subsection{Cooling and heating}

The cooling and heating of gas is included in {\sc ramses} by adding
an additional source term to the energy equation. Above $T = 10^{6}$K
the cooling is dominated by bremsstrahlung radiation whereas in the
range $10^{4} \le T \le 10^{6}$ K, collisional and ionization
excitation and recombination processes dominate and can cool the gas
to $T= 10^{4}$ K. To cool below $T=10^{4}$ K, molecular cooling
and/or fine structure metal cooling need to be implemented; in
the simulations presented here, cooling down to 1K is possible via metal line cooling. 
 We note that collisional ionization equilibrium is adopted. 

Although 
it does not matter for the current paper, we turn on a 
\citet{uvbackground} like UV radiation field (which varies with
$z$, but is uniform in space) instantaneously at $z = 8.5$ to model
the reionization of the Universe.

\subsection{Tracking the metals}

The metallicity of the gas is followed as a passive scalar and is advected with the flow of gas.
For both simulations with and without SNe feedback, we force
the pristine gas to have a constant floor metallicity $Z = 0.001
Z_{\odot}$. This value provides a fairly good approximation to the
real ${\rm H}_2$ cooling rates at work when the Universe is still devoid of
heavy elements (Yohan Dubois, priv. comm.). For the simulation with
SNe feedback, this floor is soon superseded by higher values
because of the injection of heavy elements produced in the explosions
of massive stars. 
When a SN explosion occurs, a fixed percentage of newly synthesized metals in
  the star particle (the `yield', which we set to $0.1$) is returned to the gas when 
creating the Sedov blastwave (see the section on supernova feedback). 

\subsection{Star formation}

The implementation of star formation in {\sc ramses} is outlined in
\citet{rasera_teyssier_sf} and \citet{dubois_teyssier_sn} and we
recall the main details here.  Given observational evidence for a
universal relation between the gas surface density and the star
formation rate over many orders of magnitude at low redshift
\citep{kennicutt98}, we assume star formation proceeds according to a
Schmidt law (a volumetric representation of the Kennicutt law). Gas is
therefore converted into stars in cells in which the gas density,
$\rho$, exceeds a threshold value, $\rho_{0}$, at a rate given by,

\begin{equation} \dot{\rho}= - \frac{\rho}{t_{*}} \end{equation}

\noindent where $t_{*}$ is the star formation scale, given by,

\begin{equation}
t_{*}=t_{0}\left(\frac{\rho}{\rho_{0}}\right)^{-1/2} \end{equation}

\noindent and is proportional to the free-fall time.

To set the star formation parameters, we refer to modern observations
which have shown that star formation occurs at a constant efficiency
of $\sim 1-2 \%$ of gas per free-fall time over $5$ orders of
magnitude in density, up to $10^{5} \rm{cm^{-3}}$
\citep{krumholz_tan07}. While it is not possible to achieve stellar
densities in our simulations, we set the density threshold as high
as our spatial resolution allows us, i.e. $\rho_J$, the density at which 
 the Jeans length associated with gas at 100 K becomes equal to 
 $\Delta x_{\rm min} \approx 0.5$ pc. This allows us to identify potential sites 
of star formation as realistically as possible. In the simulations described in this paper
this corresponds to a density threshold at the
top of the range currently probed by observations, $\rho_{0}= 10^{5}
\rm{cm^{-3}}$, and a compatible star formation efficiency of
0.01. Note that such a high density threshold for star formation
allows us to bypass the issue of having to decide which fraction of
the gas is molecular and thus eligible for star formation as virtually 
all the gas is in molecular form at such densities.
Since the free-fall time for gas at $\rho_{0}= 10^{5}
\rm{cm^{-3}}$ is $\sim$ 0.2 Myr, we then set $t_{0} = 20$Myr to obtain
our chosen efficiency. Finally, in order to completely avoid artificial gas
fragmentation on scales of the order of the Jeans length the following 
polytropic equation of state is also enforced:

\begin{equation} T =
T_{0}\left(\frac{\rho}{\rho_{0}}\right)^{\gamma_{0}-1} \end{equation}

\noindent where $T_0$ is set to $100$K to prevent the Jeans length
from becoming shorter than $\Delta x_{\rm min}$ and $\gamma_0$ to 4/3
to stabilize the gas against gravity.

When the conditions for star formation are met in a cell, the number
of star particles to be created, $N$, is drawn from a Poisson
distribution,

\begin{equation} P(N)=\frac{\lambda_{\rm p}}{N!}e^{-\lambda_{\rm
p}} \end{equation}

\noindent where

\begin{equation} \lambda_{\rm p}=\left( \frac{\rho\Delta x^{3}}{m_{\rm
*, min}} \right)\frac{\Delta t}{t_{*}} \end{equation}

\noindent where $\Delta x^{3}$ is the cell volume and $\Delta t$ is
the timestep evaluated at the current refinement level. The minimum
star mass, $m_{\rm *, min}$, is related to the resolution of the
simulation by,

\begin{equation} m_{\rm *, min}=\frac{\rho_{0}\Delta x_{\rm
min}^{3}}{1+\eta_{\rm SN}+\eta_{\rm W}} \end{equation}

\noindent where $\Delta x_{\rm min}^3$ is the volume of a cell on the
finest refinement level of the simulation, and $\eta_{\rm SN}$ and
$\eta_{\rm W}$ are dimensionless parameters that account for the mass
loss of stars through SNe explosions and the entrained gas mass
respectively. Details of $\eta_{\rm
SN}$ and $\eta_{\rm W}$ are given in section \ref{appendix:sn} below. In our
simulations, $m_{\rm *, min} \approx167 {\rm M}_{\odot}$, but note
that this is reduced to $m_{\rm *, min} \approx77 {\rm M}_{\odot}$ in
the feedback run, since $\eta_{\rm SN}$ and $\eta_{\rm W}$ are
non-zero.

In order to limit the number of star particles spawned in a
simulation, the $N$ particles which the Poisson law tells us should be
created in a given cell are then fused together into one, so that each
star particle mass is $m_* = Nm_{\rm *, min}$. 
To preserve numerical stability, no more than 90 \% of the gas in any
given cell is turned into stars in one timestep (this is the same
threshold as used by \citet{dubois_teyssier_sn}).

\subsection{Supernova feedback} \label{appendix:sn}

The scheme for SNe feedback employed in our feedback run is
presented in \citet{dubois_teyssier_sn}. When a star particle is
formed the amount of gas depletion can be broken down into the mass
locked up in stars permanently, $m_{*}$, and the mass which will be
ejected by the supernova, $m_{\rm ej}$, where

\begin{equation} m_{\rm ej} = m_{*}(\eta_{\rm SN}+\eta_{\rm
W}) \end{equation}

\noindent $\eta_{\rm SN}$ is the fraction of the total mass of stars
formed which is recycled and becomes supernova ejecta. We set
$\eta_{\rm SN}\simeq 0.106$, i.e. the standard value for a Salpeter IMF 
truncated between 0.1 and 100 $M_\odot$.  $\eta_{\rm W}$ is a free parameter, the
{\it mass loading} factor which quantifies the mass of gas surrounding
the star that will be swept away by the shock wave which propagates
outwards after the explosion, on scales that we cannot resolve; this additional mass is added to the
star particle when it forms. Setting $\eta_{\rm W}=1$, as we do,
ensures that angular momentum is conserved when we enforce the Sedov
blast wave solution described below for the gas (see
\citet{dubois_teyssier_sn}, Appendix A for details). Note that this
means that the same amount of gas is entrained in the explosion as
that which is locked into stars. Observations of galactic outflows by
\citet{cmartin99_outflows} show that the removal of gas from the
galactic disc occurs at a rate that is comparable to the rate of star
formation, which somewhat supports our choice of parameter, even though 
these measurements are performed on a much larger scale than that at
which we enforce our Sedov solution ( $\sim$1 pc).

Star particles undergo SNe explosions $10$Myr after their
formation, corresponding to the typical lifetime of a 10$M_\odot$ Type
II supernova progenitor. The total energy released, $E_{\rm ejecta}$,
is set using the typical energy, $E_{\rm SN}=10^{51}$ergs, and mass,
$M_{\rm SN}=10{\rm M}_{\odot}$, of a Type II supernova,

\begin{equation} E_{\rm ejecta}=\eta_{\rm SN}\frac{m_{*}}{M_{\rm
SN}}E_{\rm SN} \end{equation}

\noindent Half of this energy is kinetic and produces a Sedov blast
wave with an initial radius of $2$ cells on the highest level of
refinement ($\sim 1$ pc). The other half of the energy is thermal and heats the
surrounding gas. Note that at these very high resolutions, the results
are not very sensitive to the exact form (kinetic or thermal) in which
the energy is deposited (Y. Dubois priv. comm).

\section*{Acknowledgments}

The authors thank R. Teyssier, A. Kravtsov and A. Pontzen for valuable
discussions at various stages of this work and acknowledge the 
hospitality of the Aspen Centre for Physics where part of this paper
was written. Thanks also to F. Bournaud for comments on an earlier version of the manuscript.
The simulations were performed on Jade at the Centre Informatique 
National de l'Enseignement Sup\'erieur (GENCI grant SAP2191) and on the 
STFC/DBIS facility DiRAC facility jointly funded by STFC, the large
facilities capital fund of BIS and the University of Oxford. LCP was supported by
an STFC studentship at Oxford Astrophysics at the time when the
majority of this work was undertaken. JD and AS' research is supported 
by Adrian Beecroft, the Oxford Martin School and STFC. 

\bibliographystyle{mn2e} 
\bibliography{galaxy,numerical,theory,winds}
 \label{lastpage}

\end{document}